\documentclass[10pt,pra,aps,superscriptaddress,twocolumn]{revtex4-1}
\usepackage{amsmath,bbm}
\usepackage{amsthm}
\usepackage{amsmath}
\usepackage{latexsym}
\usepackage{amssymb}
\usepackage{float}
\usepackage{graphicx}           
\usepackage{color}
\usepackage{mathpazo}
\usepackage{comment}
\usepackage[colorlinks=true,linkcolor=blue,citecolor=magenta,urlcolor=blue]{hyperref}
\usepackage{multirow}

\newcommand{\be}{\begin{equation}}
\newcommand{\ee}{\end{equation}}
\newcommand{\bea}{\begin{eqnarray}}
\newcommand{\eea}{\end{eqnarray}}

\def\squareforqed{\hbox{\rlap{$\sqcap$}$\sqcup$}}
\def\qed{\ifmmode\squareforqed\else{\unskip\nobreak\hfil
\penalty50\hskip1em\null\nobreak\hfil\squareforqed
\parfillskip=0pt\finalhyphendemerits=0\endgraf}\fi}
\def\endenv{\ifmmode\;\else{\unskip\nobreak\hfil
\penalty50\hskip1em\null\nobreak\hfil\;
\parfillskip=0pt\finalhyphendemerits=0\endgraf}\fi}

\newcommand{\blk}{\color{black}}

\makeatletter
\newtheorem*{rep@theorem}{\rep@title}
\newcommand{\newreptheorem}[2]{%
\newenvironment{rep#1}[1]{%
 \def\rep@title{#2 \ref{##1}}%
 \begin{rep@theorem}}%
 {\end{rep@theorem}}}
\makeatother

\newreptheorem{thm}{Theorem}

\begin{document}

\title{Loophole free interferometric test of macrorealism using heralded single photons}

\author{Kaushik Joarder}
\affiliation{Raman Research Institute, C. V. Raman Avenue, Sadashivanagar, Bengaluru, Karnataka 560080, India}

\author{Debashis Saha}
\affiliation{Center for Theoretical Physics, Polish Academy of Sciences, Aleja Lotnik\'{o}w 32/46, 02-668 Warsaw, Poland}

\author{Dipankar Home}
\affiliation{Center for Astroparticle Physics and Space Science (CAPSS), Bose Institute, Kolkata 700 091, India}

\author{Urbasi Sinha}
\email{usinha@rri.res.in}
\affiliation{Raman Research Institute, C. V. Raman Avenue, Sadashivanagar, Bengaluru, Karnataka 560080, India}

\begin{abstract}
We show unambiguous violations of the different macrorealist inequalities, like the Leggett-Garg inequality (LGI) and its variant called the Wigner form of Leggett-Garg inequality (WLGI) using a heralded, single-photon based experimental setup comprising a Mach-Zehnder interferometer followed by a displaced Sagnac interferometer. 
In our experiment, the negative result measurements are implemented as control experiments, in order to validate the presumption 
of non-invasive measurability used in defining the notion of macrorealism. Among all the experiments to date testing macrorealism, the present experiment stands out in comprehensively addressing the relevant loopholes. The clumsiness loophole is addressed through the precision testing of any classical or macro-realist invasiveness involved in the implementation of negative result measurements. This is done by suitably choosing the experimental parameters so that the quantum mechanically predicted validity of all the relevant two-time no-signalling in time (NSIT) conditions is maintained in all the three pairwise experiments performed to show the violation of LGI/WLGI. Further, importantly, the detection efficiency loophole is addressed in our experimental scheme by adopting suitable modifications in the measurement strategy enabling the demonstration of the violation of LGI/WLGI for any non-zero detection efficiency. We also show how other relevant loopholes like the multiphoton emission loophole, coincidence loophole, and the preparation state loophole are all
closed in the present experiment. We report the LGI violation of $1.32\pm 0.04$ and the WLGI violation of $0.10\pm 0.02$ in our setup, where the magnitudes of violation are respectively eight times and five times the corresponding error values, while agreeing perfectly with the ranges of the quantum mechanically predicted values of the LGI, WLGI expressions that we estimate by taking into account the non-idealities of the actual experiment. 
At the same time, consistent with quantum mechanical predictions, the experimentally observed probabilities satisfy all the two-time NSIT conditions up to the order of $10^{-2}$. Thus, the non-invasiveness in our implemented negative result measurement is convincingly upper-bounded to   $10^{-2}$. 
\end{abstract}

\maketitle


\section{Introduction}

The notion of realism is central to the classical world view. It assumes that at any instant, irrespective of whether or not actually measured, a system is definitely in one of the possible states for which all its observable properties have definite values. This tenet has been a subject of a variety of experimental tests, like the test of local realism in terms of the Bell-type inequalities for the entangled systems \cite{PhysicsPhysiqueFizika.1.195}. On the other hand, Leggett and Garg suggested a procedure for testing the validity of the concept of realism for the single systems in the macroscopic domain \cite{PhysRevLett.54.857}. For this purpose, they formulated an inequality involving the observable time-separated correlation functions, known as the Leggett-Garg inequality (LGI).  This \blk is based on the notion of realism used in conjunction with the concept of non-invasive measurability (NIM).  Here one \blk assumes the possibility of measurements with only arbitrary small disturbance affecting the subsequent evolution of the system. The quantum mechanically predicted violation of LGI under suitable conditions is then taken to signify repudiation of what has been called the notion of macrorealism.

The quantum mechanically predicted violation of LGI has been tested using various physical systems like the superconducting qubits \cite{Palacios-Laloy2010}, nuclear spins \cite{PhysRevA.87.052102,PhysRevLett.107.130402}, electrons \cite{PhysRevB.86.235447}, quantum dot qubits \cite{PhysRevLett.100.026804}, nitrogen vacancy centre \cite{George3777}, atoms hopping in a lattice \cite{PhysRevX.5.011003}, spin-bearing phosphorus impurities in silicon \cite{Knee2012}, oscillating neutrinos \cite{PhysRevLett.117.050402}, and photonic systems \cite{Xu2011,PhysRevLett.106.040402,Goggin1256,Suzuki_2012,PhysRevA.97.020101}.

Here a point worth noting is that, although the original motivation leading to LGI was for testing realism vis-à-vis quantum mechanics in the macroscopic regime, 
one important use of LGI over the years has been the certification of nonclassicality or ‘quantumness’ pertaining to different types of microscopic systems \cite{Emary_2013,Li2012}. For facilitating such studies, another macrorealist condition called no-signalling in time (NSIT), also called `quantum witness' \cite{Li2012}, was proposed \cite{PhysRevA.87.052115,PhysRevA.91.062103,PhysRevLett.116.150401}.  This \blk stipulates that the choice of measurement at any instant does not affect the statistical results of any measurement at a later instant. While nearly all the experiments so far testing macrorealism have been based on LGI, the latest experiment using superconducting flux qubit \cite{Knee2016} had invoked NSIT for testing macrorealism. Also, to be noted, there is a further macrorealist inequality which has been called Wigner form of LGI (WLGI) \cite{PhysRevA.91.032117}.  This \blk is  obtained from the notion of macrorealism in terms of the assumed existence of overall joint probabilities pertaining to different combinations of measurement outcomes. By appropriate marginalization, the pair-wise observable joint probabilities are then obtained, analogous to the way Wigner’s form of the local realist inequality was derived \cite{wignerajp}.

Our present paper reports the first experimental demonstration of the decisive  violations of both the forms of the macrorealist inequalities, viz. LGI and WLGI  in the same setup.  This \blk has been achieved using single photons by closing all the relevant loopholes. Towards this end, we have devised suitable strategies  in the context of the Mach-Zehnder interferometric setup used in our experiment.  This \blk can also be adapted for applications in the different photonic contexts. While there are a number of well-known tests of nonclassicality of single photons, ours  stands out in providing a robust signature of the nonclassicality of single photons in a way which comprehensively refutes the quintessential classical notion of realism.  Concurrently, our experimental results are found \blk to be fully compatible with the quantum mechanical predictions within the limits of experimental inaccuracies or non-idealities. A particularly notable feature of the design of our experiment  is the way we use the no-signalling condition as a tool for closing one of the key loopholes. Now, before explaining the specifics of this experiment, let us first present an overview of the relevant background in order to make clear the motivation underpinning the way our experiment has been conceived and implemented. 

\section{Background and Motivation}

A key feature of most of the experiments based on LGI has been the use of projective measurements, along with negative result measurement (a measurement in which an outcome is inferred when the detector is not triggered). Negative result measurement is employed to ensure the validity of NIM so that the observed violation of a macrorealist inequality can be attributed solely to the quantum mechanical violation of realism per se (the underlying conceptual justification for using negative result measurement in order to ensure NIM has been discussed by Leggett \cite{Leggett_2008}). However, the non-idealness in the empirical implementation of negative result measurement can result in the classical disturbance affecting the measured system, thereby contributing to the observed violation of the macrorealist condition being tested. This has been referred to as the clumsiness loophole \cite{Wilde2012}. Different strategies have been adopted for tackling this issue. For example, in the experiment testing macrorealism based on NSIT using a single superconducting flux qubit \cite{Knee2016}, the classical disturbance resulting from measurement is determined using appropriate control experiments. In the ``IBM quantum experience" (IBM QE) study of the extension of such a test for two or more qubits, the NSIT condition per se has been reformulated in terms of an ``invasive-measurement bound", which is estimated in the context of the relevant circuits in the IBM QE. This bound, thus, takes into account the clumsiness loophole, and its violation is regarded as signifying the violation of ``clumsy macrorealism" in such tests \cite{Ku2020}. 

On the other hand, an entirely different strategy for testing macrorealism has been explored in terms of what has been called “ambiguously measured LGI” formulated by avoiding the use of NIM and hence not requiring negative result measurement \cite{PhysRevA.96.042102}. But, for this procedure to work, one has to consider two sets of measurements, one projective and the other ambiguous measurements realized by a class of POVMs, and then, importantly, invoke an additional assumption called equivalently invasive measurement (EIM), which equates the invasive influence of ambiguous measurements with that of the unambiguous ones. It has been argued that EIM leads to a testable condition and by violating the ambiguous version of LGI while satisfying this condition, the clumsiness loophole can be avoided. Further, It has been pointed out \cite{PhysRevA.96.042102} that an assumption similar to EIM is also implicit in the experiments on LGI in terms of weak measurements. 

Therefore, the upshot of the above considerations is that for avoiding the use of NIM in testing macrorealism, some other suitable assumption has to be necessarily invoked. In this context, we note that recently another approach has been suggested to avoid the use of NIM using a single Mach-Zehnder interferometric setup by putting forth an interesting plausible argument for inferring the violation of macrorealism directly from the occurrence of destructive interference per se \cite{PhysRevA.102.032206}. To what extent such an argument is general is not yet clear. Similarly, the implications of another recent interesting finding \cite{halliwell2020leggettgarg} that the quantum mechanical violation of LGI in the double-slit interferometric setup is necessarily accompanied by destructive interference need to be further analysed. On the other hand, for analysing the experiment testing LGI using neutrino flavour oscillations \cite{PhysRevLett.117.050402}, the NIM condition has been replaced by what has been called the `stationarity' assumption.

Against the above backdrop, our photon-based experiment presented in this paper focuses on testing macrorealism based on NIM, using LGI and its variant WLGI in terms of projective and negative result measurements.  This has been achieved \blk by rigorously addressing all the relevant loopholes. The reasons for showing violation of both the LGI and the WLGI are the following. First, both the LGI and WLGI are only necessary and not sufficient conditions of macrorealism.  Hence, \blk 
simultaneous violation of both the inequalities in a loophole free way provides a more robust violation of the notion of macrorealism.
Secondly, there is an advantage in working with WLGI in our experimental setup.  This is because here \blk the WLGI expression involves significantly lower number of measurable joint probabilities, compared to the number of measurable joint probabilities occurring in the LGI expression. Hence, it is expected to generate a lower error value, thus leading to better agreement with the corresponding quantum mechanical prediction. A notable feature of our treatment is the way we have analysed the comparison between the measured values of LGI, WLGI, NSIT and the corresponding quantum mechanical predicted values, which have been carefully evaluated by considering the relevant experimental imperfections in the actual setup we have used.

Our experimental proposal consists of two Mach-Zehnder interferometers in tandem with standard optical elements (Fig. \ref{fig:schematic1}). This is amenable to be adapted for atoms, molecules and other architectures towards studies in the macroscopic regime. While the use of heralded single-photons from the process of spontaneous parametric down-conversion (SPDC) in this experiment aids in closing various loopholes due to the lower multi-photon generation probability and higher noise tolerance, a significant feature of our experiment is the way we use the NSIT condition for closing the clumsiness loophole. 

 In order to explain our procedure for this purpose, we first recall the proposal in \cite{Wilde2012} to invoke a testable statistical relation to detect the amount of disturbance resulting from the measurement procedure, that is distinct from the quantum mechanical effect due to measurement. For this purpose, \cite{Wilde2012} suggests the use of a control experiment which can detect the amount of classical or macrorealist invasiveness of an intermediate measurement affecting the joint probability distribution of the outcomes of the preceding and subsequent measurements. The authors have called the relation used to test this as the “adroitness” condition, the satisfaction of which would certify the measurement as macro-realistically non-invasive or “adroit” in the words of the authors of \cite{Wilde2012}. The accuracy with which this relation is tested to be valid is argued to give a quantitative measure of macrorealist invasiveness or “adroitness”. This procedure has been illustrated for a two-state oscillating system by choosing the measurement times such that the “adroitness” condition is satisfied by the quantum mechanical treatment, while also showing the quantum mechanical violation of macrorealism by using LGI.

Taking clue from the treatment of \cite{Wilde2012}, we adopt the following procedure towards closing the clumsiness loophole. In our experiment, the testing of all the three independent NSIT conditions (Eqs. \eqref{eq:nsit12main}-\eqref{eq:nsit13main} given in the Section \ref{sec:exp}) is invoked as the control experiment. Here a key point is that the configuration of our experimental setup (Fig. 1) and the relevant parameters are chosen such that all these three NSIT conditions are quantum mechanically predicted to be satisfied, while also showing the violations of LGI and WLGI (as explained in the Section \ref{sec:exp}). Thus, if there is an empirically significant observed deviation from any of these NSIT relations, this would be a signature of the presence of macrorealist invasiveness (clumsiness) arising from the non-ideal implementation of the relevant negative result measurement. This is because the macrorealist non-invasiveness requires that the statistical result of any measurement should not be affected by any measurement performed at an earlier instant.

Now, given the way the negative result measurements have been performed in our experiment (as explained in the Section \ref{sec:loophole}), the measurement statistics of the obtained relevant outcomes are found to satisfy all these three NSIT relations with an accuracy of the order of $10^{-2}$. This means that any possible deviation from any of these NSIT relations is ensured to be within the corresponding measurement error range (as can be seen from the relevant experimental results given in the Section \ref{sec:result}, Fig. \ref{tab:resultTab2}). Thus, this procedure enables quantitative determination of the extent to which the macrorealist or classical non-invasiveness is guaranteed for the negative result measurements employed in our experiment. Concurrently, the violations of LGI and WLGI are demonstrated which are significantly larger than their respective measurement error bounds (see Fig. \ref{tab:resultTab1} given in the Section \ref{sec:result}).

Further discussion of the way this approach is realised in our experiment is given in Section \ref{sec:loophole}.  There \blk we also elaborately explain the procedures we have adopted for tackling a range of other loopholes like the detection efficiency loophole, multiphoton emission loophole, coincidence loophole, and preparation state loophole. Such detailed analyses enabling all these loopholes to be either circumvented or closed in a given optical experiment testing macrorealism are yet missing in existing literature and the present work fills this important gap.

The manuscript is structured as follows. In Section \ref{sec:basics}, we provide a brief discussion of the basics concerning the different forms of the macrorealist conditions and the experimental setup used in our work. Subsequently, after explaining in Section \ref{sec:loophole} the way we address the various loopholes, details of the experiment performed are discussed in Section \ref{sec:exp}. This is followed by the presentation of experimental results with the relevant error analyses in Section \ref{sec:result}. We show that the LGI and WLGI measured range of values show a decisive violation of macrorealism and are perfectly compatible with the quantum mechanical predicted ranges, which we estimate taking into account different forms of experimental non-idealities.  Finally, in Section \ref{sec:conclusion}, we summarize the work presented in this paper and also indicate some future directions of study.

\section{Basics}
\label{sec:basics}

The notion of macrorealism, which is a basic tenet of any classical theory can be regarded as consisting the following assumptions:\\
(1) \textit{Realism per se:} A system with two or more macroscopically distinct states available to it, is at any instant in one or the other of these states.\\
(2) \textit{Non-invasive measurability:} It is possible, in principle, to determine the state of a system
with an arbitrarily small perturbation on its subsequent dynamics.\\
(3) \textit{Induction or Arrow of time:} The outcome of a measurement on a system is not affected by what will or will not be measured on it later.\\
Based on these assumptions, Leggett and Garg first proposed a set of inequalities (LGI) \cite{PhysRevLett.54.857}. One of such inequalities is the following
\begin{equation}
\label{eq:LGI}
 \langle Q_{t_{1}}Q_{t_{2}}\rangle + \langle Q_{t_{2}}Q_{t_{3}}\rangle - \langle Q_{t_{1}}Q_{t_{3}}\rangle\leq 1   .
\end{equation}
$Q_{t_{i}}$ is a dichotomic observable measured at time $t_{i}$, which has two possible outcomes with eigenvalues +1 or -1. Here, $t_1<t_2<t_3$ represents the flow of time. $\langle Q_{t_{i}}Q_{t_{j}}\rangle$ is the correlation function of the measurement outcomes at $t_{i}$ and $t_{j}$.

This inequality has the upper bound of 1.5 given by the relevant quantum mechanical treatment \cite{Emary_2013}. Further, as a consequence of macrorealism, 
another set of inequalities called Wigner form of LG inequalities (WLGI)  have been derived \cite{PhysRevA.91.032117}. These inequalities consist of two-time joint probabilities of the form $P_{t_{i},t_{j}}(q_{t_{i}},q_{t_{j}})$ which is the joint probability of obtaining the outcomes denoted by $q_{t_{i}}$ and $q_{t_{j}}$ when the observable $Q_{t_{i}}$ is measured at $t_i$ and $Q_{t_j }$ is measured at $t_j$. One of such WLG inequalities is given by 
\begin{equation}
\label{eq:WLGI}
  P_{t_{1},t_{3}}(-,+)-P_{t_{1},t_{2}}(-,+)-P_{t_{2},t_{3}}(-,+)\leq 0 .
\end{equation}
This form of WLGI has the maximum quantum mechanical violation of approximately 0.4034 (see \eqref{wlgi} given later). Since both LGI and WLGI are the necessary conditions for macrorealism, the violations of both LGI and WLGI have been demonstrated in our present experiment for invalidating the notion of macrorealism. The interferometric setup (Fig. \ref{fig:schematic1}) used in our experiment has two key features:

\begin{figure}[!ht]
\includegraphics[width=\linewidth]{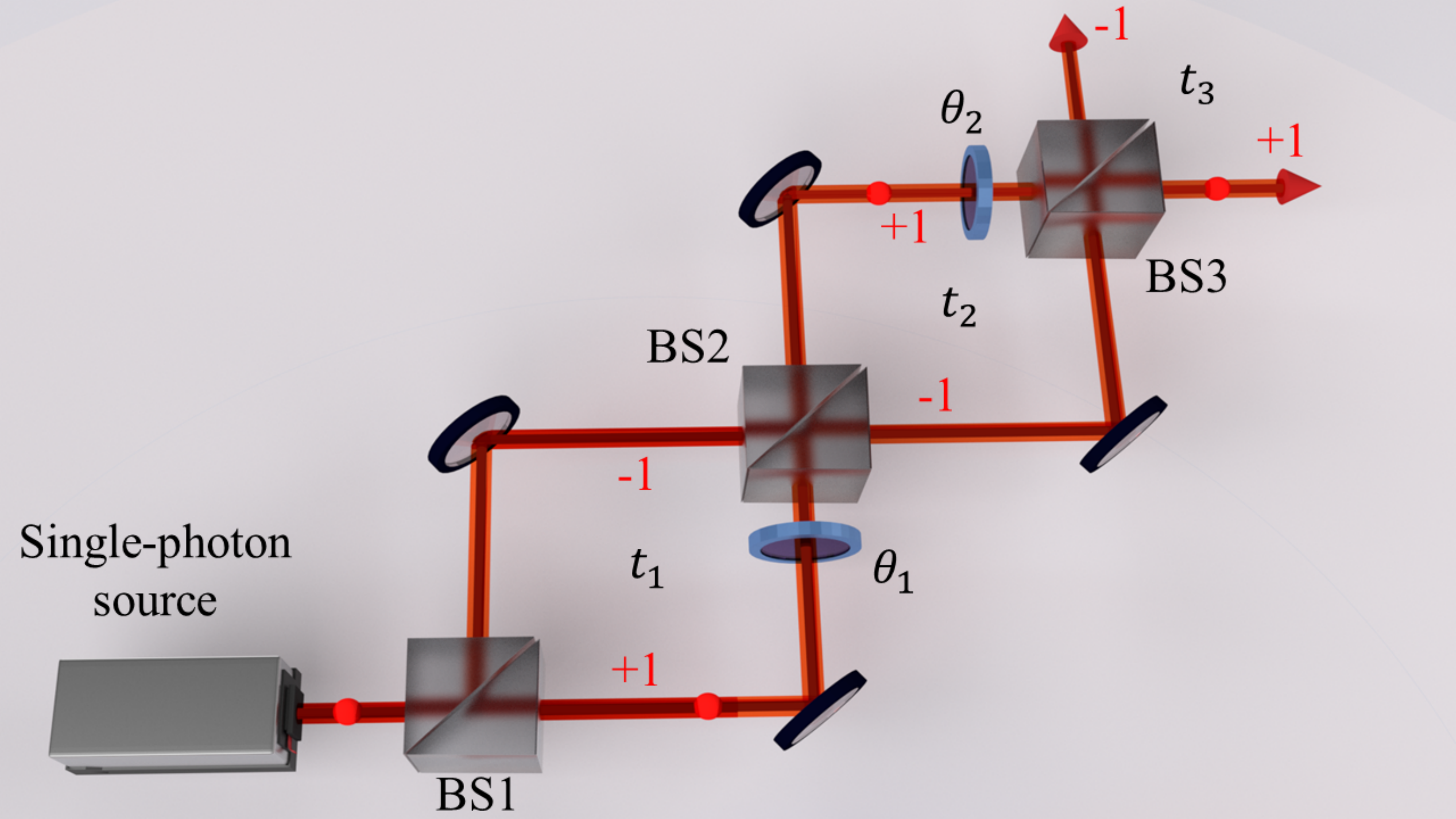}
\caption{Basic schematic of the experimental setup. \textbf{BS1, BS2, BS3}: beamsplitter; \textbf{$\theta_{1}$, $\theta_{2}$}: phase modulator.}
\label{fig:schematic1}
\end{figure}

(a) Operationally, the measurement time $t_1$ in the inequalities given by Eqs. \eqref{eq:LGI} and \eqref{eq:WLGI} corresponds to any time the photon is within the first interferometer between the first and second beamsplitters BS1 and BS2. The measurement time $t_2$ pertains to any time the photon is within the second interferometer between BS2 and BS3, while $t_3$ corresponds to the photon emerging from BS3.

(b) The dichotomic observable $Q_{t_i}$ measured in our setup has the eigenvalues +1(-1) corresponding to the states of the photon being in one or the other arm of the interferometer. It is the interference effect arising from the superposition of these states that leads to the quantum violation of LGI/WLGI. Thus, here the ‘macroscopic distinctness’ between the two superposing states is characterised by the spatial separation between the two arms of the interferometer which in the present setup is around 1 cm. This is much larger than the photonic wavelength of 810 nm used here, implying the parameter that can be taken to denote ‘macroscopicity’ in this context to be of the order of $1.2\times 10^4$. 

It has been comprehensively argued \cite{Leggett_2002,Knee2016} that the figure of merit used to characterize `macroscopicity', or, equivalently, the notion of macroscopic distinctness of the superposed states involved in any test of macrorealism is, in general, context dependent.  This \blk need not be necessarily linked to mass. It is essentially depending on the experimental specifics that an appropriate measure has to be defined for the relevant macroscopicity scale. For example, in the experiment involving a superconducting flux qubit \cite{Knee2016}, one plausible choice of the `macroscopicity' measure is taken to be the difference in some extensive physical quantity like the magnetic moment in the two superconducting current states whose superposition is being tested, normalized to some natural  atomic-scale unit which, in this case, is the Bohr magneton. Another plausible choice of such  `macroscopicity' measure for the same experiment that has also been discussed \cite{Knee2016} is in terms of what has been called the notion of `disconnectivity'. This is related to the number of electrons that behave differently in the two branches of the superposition of the clockwise and counterclockwise circulating current states of the flux qubit. 


Among the other diverse measures of `macroscopicity'  that have been proposed in different contexts, the following representative examples are worth noting, all of which considered together emphatically illustrate the inherent context dependence of the criterion of `macroscopicity' :
\begin{itemize}

\item{In the IBM QE study of macrorealism  \cite{Ku2020}, the 'macroscopicity' is characterised by the number of qubits which can be considerably enhanced by making use of the capabilities of IBM QE. }

\item{The energy separation between the superposed states has been argued to be a plausible measure of 'macroscopicity' \cite{PhysRevA.44.5401} in the context of the experiments seeking to provide bounds on proposed modifications of standard quantum mechanics based on the idea of wave function collapse/intrinsic decoherence. }

\item{In the experiment using neutrino flavour oscillation \cite{PhysRevLett.117.050402} for showing the violation of LGI, the relevant 'macroscopicity' measure is argued to be related to the length scale over which the effect of neutrino oscillation is observed. }
\end{itemize}

Now, focusing on the specific context of our experiment, let us recall the way we have specified the 'macroscopicity' figure of merit. The large value of this measure as defined in our setup means that the $Q_{t_i } = +1$ and $Q_{t_i }= -1$ states are so well separated in space that the possibility of, say, the $Q_{t_i } = +1$ state being classically affected by a measuring device triggered essentially by the $Q_{t_i} = -1$ state is considerably small. This therefore, aids in closing the clumsiness loophole. 
Further explanation of the way this loophole is tackled will be provided next, together with the discussions concerning other loopholes.


\section{Addressing the various loopholes}
\label{sec:loophole}


\subsection{Clumsiness loophole} 
In our present experiment, for testing the LGI (\ref{eq:LGI}) and WLGI inequalities (\ref{eq:WLGI}) by measuring each of the correlation functions $\langle Q_{t_i } Q_{t_j } \rangle$ and each of the joint probabilities $P_{t_i,t_j} (q_{t_i },q_{t_j })$, the first measurement of each such pair is required to satisfy NIM. This is ensured through negative result measurement by placing the measuring device (detector or blocker) in one of the arms of the interferometer, corresponding to, say, $q_{t_i }= +1$ so that the device being untriggered constitutes the measurement of $q_{t_i } = -1$. Then the results of only these runs are used for determining the joint probabilities like $P_{t_i,t_j } (-,+)$ and $P_{t_i,t_j } (-,-)$. Similarly, the other two joint probabilities $P_{t_i,t_j } (+,-)$ and $P_{t_i,t_j } (+,+)$ are determined by shifting the measuring device to the other arm. Thus, the LGI experiment takes place as three piece-wise separate sets of experimental runs measuring the joint correlations, $\langle Q_{t_1 } Q_{t_2 } \rangle, \langle Q_{t_2 } Q_{t_3} \rangle$ and $\langle Q_{t_1 } Q_{t_3 } \rangle$, with the negative result measurement performed at $t_1$ or $t_2$ depending upon the set considered. Now, to determine the accuracy with which such negative result measurements satisfy NIM, the relevant experimental parameters need to be chosen such that all the pertinent two-time NSITs are quantum mechanically predicted to be satisfied. Consequently, the testing of the validity of these relations would provide a quantitative measure of the extent to which the classical disturbance induced by the negative result measurements is minimised in the setup we use. These NSIT relations are given by,
\begin{subequations}
\begin{eqnarray}
\label{eq:NSIT12}
P_{t_2} (q_{t_2})= P_{t_1,t_2 } (+,q_{t_2})+ P_{t_1,t_2} (-,q_{t_2}) 	\\
\label{eq:NSIT13}			
P_{t_3 } (q_{t_3})= P_{t_1,t_3} (+,q_{t_3})+ P_{t_1,t_3} (-,q_{t_3}) \\
\label{eq:NSIT23}
P_{t_3 } (q_{t_3})= P_{t_2,t_3} (+,q_{t_3})+ P_{t_2,t_3} (-,q_{t_3})	.	
\end{eqnarray}
\end{subequations}
\eqref{eq:NSIT12}-\eqref{eq:NSIT13} are valid for any $q_{t_2}=\{+1,-1\}$ and $q_{t_3}=\{+1,-1\}$. Specifics concerning the choice of the experimental parameters for this purpose and the results of our measurements testing the relations (\ref{eq:NSIT12}-\ref{eq:NSIT13}) will be discussed in Section \ref{sec:result}. At this stage, we need to highlight an important additional feature of our setup that arises particularly because of the way we strategise our setup to simultaneously sidestep the detection efficiency loophole. As will be elaborated in the next subsection, apart from using the perfect blockers (metallic blocking device that perfectly absorbs all incident photons, so that one can model it as a detector with 0\% efficiency) instead of detectors having finite efficiency, a key strategy we employ is as follows. We measure the joint probability $P_{t_1,t_2 } (q_{t_1},q_{t_2} )$  and hence the correlation function $\langle Q_{t_1} Q_{t_2} \rangle$ through implementing negative result measurement by placing the two blockers jointly in the respective arms of the two interferometers and determining all the three-time joint probabilities $P_{t_1,t_2,t_3} (q_{t_1},q_{t_2},q_{t_3})$,  $\forall q_{t_i}= \pm 1$. We then obtain the marginal joint probabilities $P_{t_1,t_2} (q_{t_1},q_{t_2})$ by using the following relation at the level of the observable probabilities, implied by the assumption of induction (entailing the absence of retro-causality) stated earlier as an ingredient of the notion of macrorealism:

\begin{equation}
\label{eqn:AoT}
P_{t_1,t_2} (q_{t_1},q_{t_2})= \sum_{q_{t_3}=\pm 1}P_{t_1,t_2,t_3} (q_{t_1},q_{t_2},q_{t_3})	.
\end{equation}
As will be argued in the next subsection, the measurement of $P_{t_1,t_2 }(q_{t_1},q_{t_2})$ in the above way enables us to show the violation of both LGI and WLGI for any non-zero detection efficiency. Now, since for this procedure, the negative result measurements are required to be performed at both $t_1$ and $t_2$, in order to witness the precision up to which NIM is satisfied in this case, ideally one would have to test here the quantum mechanically predicted validity of the relevant three-time NSIT conditions involving the placing of blockers at both $t_1$ and $t_2$. However, in our present setup, the experimental parameters cannot be chosen to ensure that all such three-time NSIT conditions are quantum mechanically predicted to be satisfied, along with the quantum mechanically predicted violation of both LGI and WLGI given by the inequalities (\ref{eq:LGI}) and (\ref{eq:WLGI}).  The essential reason for this is that the two-time NSIT relations given by Eqs. (\ref{eq:NSIT12}-\ref{eq:NSIT13}) in our paper are not sufficient conditions for the validity of macrorealism. These relations, together with all the three-time NSIT relations as applied to our setup, constitute the necessary and sufficient conditions for ensuring macrorealism, as has been shown by \cite{PhysRevA.91.062103}. Hence, if the quantum mechanical results satisfy the two-time NSIT conditions given by Eqs. (\ref{eq:NSIT12}-\ref{eq:NSIT13}), but LGI is violated by quantum mechanical, then one or more three-time NSIT conditions would also have to be necessarily violated by quantum mechanical. Thus, in our experimental architecture, it is not possible to choose the relevant parameters such that all the two-time and three-time NSIT conditions are satisfied by quantum mechanical, while also showing the violation of LGI. The above point is essentially the reason why the validity of the two-time NSIT conditions is a natural choice for our experimental design in order to quantitatively determine the effect of classical or macrorealist invasiveness of the performed measurements, which violate LGI.

Nevertheless, a crucial pertinent feature is that the blockers placed in the different interferometers are spatially well separated. Hence it is reasonable to argue from the classical or macrorealist point of view that even if these blockers would have been placed at both $t_1$ {\it and} $t_2$, there would have been a negligibly small possibility of their respective local actions influencing each other (say, through some environmental degree of freedom) and producing a significant cumulative effect on the measured photon in the other spatially separated arms of the respective interferometers. Thus, the use of the two-time NSIT conditions given by Eqs. (\ref{eq:NSIT12}-\ref{eq:NSIT13}) for verifying the precision of the smallness of the classical disturbance induced by the blockers when they are placed separately at $t_1$ or $t_2$ suffices for plugging the clumsiness loophole. This also enables getting around the detection efficiency loophole in the way further explained in the next subsection. Further, it is worth noting here that the "Collusion loophole" proposed by Wilde \textit{et al.} \cite{Wilde2012} implies that an intermediate measurement may have an effect at the macrorealist level that gets washed out and is not revealed at the statistical level through the testing of the “adroitness” condition, or, in our case, using the NSIT condition. The authors of \cite{Wilde2012} have just briefly remarked suggesting the “unnaturalness” of this loophole.  We now provide an argument corroborating the unnaturalness of this loophole in the context of our experiments. \\
In our setup, we have tested the three independent NSIT conditions Eqs. (\ref{eq:NSIT12}-\ref{eq:NSIT13}), or equivalently, Eqs. (\ref{eq:nsit12main} - \ref{eq:nsit13main}). Let us first consider the ensembles of photons in the two scenarios corresponding to Eqs. \eqref{eq:nsit12main} and \eqref{eq:nsit13main} respectively. While in both these cases, negative result measurement is implemented at the same instant $t_1$, the photons are subsequently detected at two different instants $t_2$ and $t_3$ respectively undergoing different evolutions - in the former case, passing through a single beam splitter, while in the latter case, passing through two beam splitters. Thus, for the given negative result measurement at $t_1$, the macrorealist invasiveness has to occur in a conspiratorial way for these two differently evolving sets of photons such that, in both these cases, the ‘averaging washing out process’ of the invasiveness over different runs still results in the validity of the statistical non-invasive conditions (\ref{eq:nsit12main}) and (\ref{eq:nsit13main}) respectively. Similarly, considering the ensemble of photons corresponding to Eq. (\ref{eq:nsit23main}), these pass through a single beam splitter like that of Eq. (\ref{eq:nsit12main}), but after being subjected to negative result measurement at an instant $t_2$ different from that corresponding to Eq.(\ref{eq:nsit12main}). Then for this ensemble, too, the ‘averaging washing out process’ of the macrorealist invasiveness over different runs has to occur in such a way that the statistical non-invasiveness condition (\ref{eq:nsit23main}) holds well. 

Here the key point is that although the three different ensembles corresponding respectively to the NSIT conditions Eqs. \eqref{eq:nsit12main}-\eqref{eq:nsit13main} evolve in different ways, nevertheless, these three mutually independent NSIT conditions of non-invasiveness are found to be empirically satisfied, consistent with the relevant quantum mechanical predictions. Thus, the only possibility that remains is that there could have been classical or macrorealist invasiveness which may have occurred in individual runs due to the negative result measurements, but have not evidenced at the statistical level beyond the measurement errors involved in the checking of these relations. However, for this possibility to be valid, one would require the statistical washing out of such individual effects of invasiveness separately for all these three ensembles - this is the so-called 'collusion loophole'. Therefore, what such 'collusion loophole' implies is a number of independent constraints to be satisfied by an allowed macrorealist model so that the invasiveness occurs in a way ensuring that the behaviors of all these different ensembles conform to the statistical non-invasiveness tested by the relevant NSIT conditions.

It is then evident that for satisfying the above requirement, one would require considerable fine tuning in the formulation of the desired macrorealist model in order to be consistent with our experimental results. To what extent such fine tuning can be considered to be plausible, or can be physically well motivated within the framework of a macrorealist model remains an open question. Hence, modulo the possibility of the formulation of such a required macrorealist model, we contend that the collusion loophole has been significantly addressed and closed. Of course, independent of whether such a macrorealist model is explicitly formulated, it would be interesting to explore whether our present experimental architecture can possibly be extended for imposing tighter constraints. For instance, by introducing more beam splitters and blockers to test more than three NSIT conditions as the control experiment which would impose more tighter constraints on such a macrorealist model. \\

For the relevant perspective on the experimental studies of this issue, see Table \ref{tab:list} where different experiments testing LGI using projective measurements which have addressed the clumsiness loophole have been listed.

\begin{widetext}

\begin{table}[h!]
\centering
\caption{Experiments testing LGI based on projective measurements and by addressing the clumsiness loophole in different ways, with the relevant references cited in the text.}
\label{tab:list}
\begin{tabular}{|| c|| c || c ||} 
\hline \hline
 Systems & Macroscopicity parameter & Procedure \\
 \hline \hline 
  \begin{tabular}{@{}c@{}}
Nitrogen vacancy center in \\ diamond
hosting a three \\ level quantum system \cite{George3777}
  \end{tabular}
  & Atomic mass & 
  \begin{tabular}{@{}c@{}}
Testing the macrorealist bound on the \\ success rate in a `three-box' game \\ while satisfying  the NSIT conditions.
  \end{tabular}  \\ 
   \hline
    \begin{tabular}{@{}c@{}} 
    A single cesium atom performing \\
    `quantum walk' on a discrete lattice \cite{PhysRevX.5.011003}
    \end{tabular}
   &   Atomic mass &
    \begin{tabular}{@{}c@{}} 
    Testing the violation of LGI by\\ determining classical invasiveness \\
    using control experiments.
    \end{tabular} \\
  \hline 
   \begin{tabular}{@{}c@{}} 
   Spin-bearing phosphorus \\ impurity atom in silicon \cite{Knee2012} 
   \end{tabular}
   & Atomic mass & 
    \begin{tabular}{@{}c@{}} 
   Testing LGI by suitable implementation \\ of CNOT and anti-CNOT gates.
   \end{tabular} \\
 \hline 
 Neutrino flavour oscillation \cite{PhysRevLett.117.050402} & Length scale of the neutrino oscillation & 
 \begin{tabular}{@{}c@{}}
 Testing the violation of LGI using \\
 the `stationarity' condition. 
 \end{tabular} \\
 \hline 
 Three-level single photons \cite{PhysRevA.97.020101} & Spatial separation between the paths & 
 \begin{tabular}{@{}c@{}} 
 Testing the violation of modified LGI \cite{PhysRevA.96.042102} \\
 using ambiguous measurements. 
 \end{tabular}\\ 
 \hline
 Superconducting flux qubit \cite{Knee2016} &  \begin{tabular}{@{}c@{}} 
 Difference in the magnetic moments \\
 of the two oppositely circulating \\
 superconducting superposed \\
 current states of the flux qubit
 \end{tabular}
 & 
 \begin{tabular}{@{}c@{}} 
 Testing the violation of NSIT as a \\ consequence of macrorealism by \\ determining classical invasiveness \\
 using control experiment. 
 \end{tabular} \\ 
 \hline 
  \begin{tabular}{@{}c@{}} 
  Two and four qubit `cat-states' \\ 
  studied in cloud-based quantum \\  computing device \\
  `IBM Quantum experience' \cite{Ku2020} 
  \end{tabular}
  & Number of constituent qubits & 
 \begin{tabular}{@{}c@{}} 
  Testing the violation of `clumsy- \\macrorealist' bound of LGI.
  \end{tabular} \\  
 \hline \hline 
\end{tabular}
\end{table}

\end{widetext}


\subsection{Detection efficiency loophole} 

In order to test the quantum mechanically predicted bounds of LGI and WLGI given by the inequalities (\ref{eq:LGI}) and (\ref{eq:WLGI}), one has to assume that the statistical properties of the sub-ensembles of photon pairs actually detected in the relevant experimental runs are identical to those of the entire ensemble of all the pairs. This is known as the fair-sampling assumption which is strictly true if all the detectors used are ideal, i.e., $100\%$ efficient. However, in practice, single photon detectors have low efficiency of the order of $50-60\%$. Hence, by exploiting this detection efficiency loophole, it may be possible to reproduce the quantum mechanically predicted measured statistics of the detected sample based on the macrorealist considerations. A similar situation exists in the case of testing local realism for which considerable studies have been done and different approaches have been explored to take into account the effect of detector efficiency in the testing of the relevant inequalities. For example, one of the approaches in terms of the probabilities has been to assign the measurement result 0 to an undetected particle and generalize the Bell-Clauser-Horne-Shimony-Holt (Bell-CHSH) inequality to include detector inefficiency \cite{PhysRevD.35.3831}. Another approach has been to use the correlation functions, but without requiring to assign any measurement result to the undetected particles while incorporating the detection efficiency parameter in the original form of Bell inequality \cite{PhysRevA.57.3304}. It is the argument used in the latter approach that we adapt for the LGI/WLGI case by coming up with an appropriate measurement strategy. 

Here, a key idea we consider in the context of a typical macrorealist hidden variable based model is to regard the behaviour of a photon, say whether it is detected or not, to be determined by the hidden variable $\lambda$ characterising its state. Further, an important relevant point is that, in our setup, the subspace of hidden variables determining the detection of photon at time $t_i$ is, in general, different from the hidden variable subspace corresponding to the detection of photons at time $t_j$. Consequently, it is with respect to various such subspaces of hidden variable distributions corresponding to the detection of photons at $t_1, t_2, t_3$, and by considering the negative result measurements for the first of each pair of such time-separated measurements, a detailed treatment is required which is provided in 
Appendix \ref{appendix:a}. 
This is based on the appropriately written forms for the measured joint probabilities and correlations, leading to expressions for the left-hand sides of LGI and WLGI given by Eqs. (\ref{eq:LGI}) and (\ref{eq:WLGI}) respectively, recast in terms of the detector efficiency, $\eta \in (0,1]$, which is assumed to be the same for all the detectors that have been used.
From the modified upper bounds, i. e., $2/\eta-\eta$ for LGI and $(1-\eta)/(2\eta-1)$ for WLGI whenever $\eta\ge 2/3$, it is seen that the minimum detection efficiency required is greater than 85\% for which the quantum mechanical violation of LGI cannot be reproduced by the type of hidden variable model we have considered. Similarly, the requirement for WLGI is greater than 78\%. Detailed calculations for the modified upper bound of LGI/WLGI and their respective efficiency requirements have been provided in Appendix \ref{appendix:a1}. 

Now, comes the crunch. The above argued detection efficiency requirements get remarkably relaxed by taking advantage of the following two key modifications in the measurement procedure, already alluded to in the preceding subsection. First, for determining the quantities $P_{t_1,t_3} (-,+)$, $P_{t_2,t_3} (-,+)$ and $\langle Q_{t_1} Q_{t_3}\rangle$, $\langle Q_{t_2} Q_{t_3}\rangle$ in the WLGI and LGI expressions (\ref{eq:WLGI}) and (\ref{eq:LGI}) respectively, we replace the detectors used for the negative result measurements at $t_1$ or $t_2$ by metallic blocking devices which act as perfect absorbers of incident photons. Then this procedure ensures that such measured joint probability distributions $P_{t_1,t_3 } ( q_{t_1 },q_{t_3 })$ and $P_{t_2,t_3} (q_{t_2},q_{t_3})$ expressed in terms of hidden variables involve only the hidden variable subspace corresponding to photon detection at $t_3$, as explicitly discussed in Appendix \ref{appendix:a2}. 
Secondly, the two-time joint probabilities $P_{t_1,t_2} (q_{t_1}, q_{t_2})$ are evaluated from the measured three-time joint probabilities $P_{t_1,t_2,t_3} (q_{t_1},q_{t_2},q_{t_3})$ using Eq. \ref{eqn:AoT}, whence the hidden variable expressions for such evaluated two-time joint probabilities and correlation functions contain integration over only the particular hidden variable subspace determining photon detection at $t_3$ (see Appendix \ref{appendix:a2}). 
Subsequently, writing such hidden variable expressions explicitly for the left hand sides of the inequalities (\ref{eq:LGI}) and (\ref{eq:WLGI}), it is seen that the forms of LGI and WLGI reduce respectively to \eqref{eq:hvmLGI} and \eqref{eq:hvmWLGI} in Appendix \ref{appendix:a2}. 
This ensures that the dependence on the detector efficiency parameter $\eta$ is such that for the measurement strategy we have employed, both the LGI and the WLGI are satisfied by the relevant hidden variable model, importantly, for any value of $\eta$. Thus, the violations of both the LGI and the WLGI measured in this way cannot be reproduced by hidden variable model, whatever be the detector efficiency, thereby rendering \textit{the detection efficiency loophole irrelevant in this context}.

\subsection{Multiphoton emission loophole} 
For testing the macrorealist inequalities in the context of our setup, a crucial requirement is to have a single photon at a time within our setup comprising the interferometers. Thus, it is important to rigorously take into account the effect of any deviation from this condition that may occur in our experiment stemming from the non-idealness of the single photon source that is used. We call this the multiphoton emission loophole, meaning that any non-vanishing probability of the multiphoton generation can give rise to a pseudo violation of the macrorealist inequalities. Hence, by taking into account the possibility of such multiphoton occurrences, it is necessary to obtain the appropriately modified respective upper bounds of LGI and WLGI. 

Now, to estimate this, for example, in the case of LGI, the strategy we employ is to first show that one can formulate, in principle, an internally consistent model of two photons, each of which is present in either of the two different paths, that would give rise to the violation of LGI corresponding to the maximum algebraic bound of the expression given by the left hand side of the inequality \eqref{eq:LGI}. Such a model is outlined in Appendix \ref{appendix:b1}. 
The existence of such a model implies that in terms of a suitable parameter, say $\gamma$, which characterises the fraction of the total set of runs that corresponds to the occurrence of two photons within our experimental setup, we can write the modified upper bound of LGI in the following form (see Appendix \ref{appendix:b1})
\begin{equation}\label{eq:modboundlgi}
\langle Q_{t_{1}}Q_{t_{2}}\rangle + \langle Q_{t_{2}}Q_{t_{3}}\rangle - \langle Q_{t_{1}}Q_{t_{3}}\rangle\leq 1 + 2\gamma . 
\end{equation}

Based on similar reasoning (see Appendix \ref{appendix:b1}
), it can be shown that the modified form of WLGI in the presence of multiple photons is given by, %
\begin{equation}\label{eq:modboundwlgi} P_{t_{1},t_{3}}(-,+)-P_{t_{1},t_{2}}(-,+)-P_{t_{2},t_{3}}(-,+)\leq \frac{\gamma}{2} .
\end{equation}

For the SPDC based heralded single photon source used in our experiment with low pump power and suitable filtering, the probability of producing multiple photons is very small compared to that using other sources like a weak coherent pulse. 
Now, in order to estimate the precise value of the modified upper bound of LGI/WLGI relevant to our experiment, we follow the procedure for measuring $\gamma$ whose specifics are discussed in Appendix \ref{appendix:b2}.  
This is based on our experimental data pertaining to a set of measured relevant single counts and coincidence counts, along with the values of the key parameters of our setup like the beamsplitter coefficients and detector efficiencies. As shown in Appendix \ref{appendix:b2}, 
the value of the parameter $\gamma$ evaluated in this way turns out to be 0.0023. Putting this value of $\gamma$ in Eqs. \eqref{eq:modboundlgi} and \eqref{eq:modboundwlgi}, the modified upper bounds of LGI and WLGI are found to be 1.0046 and 0.0012 respectively. Thus, in our experiment, the changes in the upper bounds of the macrorealist inequalities due to the contribution from the possible presence of multiple photons are ensured to be indeed negligibly small in comparison with the observed magnitudes of the violations of these inequalities.

\subsection{Coincidence loophole} For the case of non-heralded single photons used in an experiment and for a specified detection time window, a loophole (similar to the coincidence loophole \cite{Larsson_2004} in case of Bell's inequality) can occur because of fluctuations in the passage time of the single photons within the setup, particularly as the measurement settings are varied. In our experiment, this loophole is closed by using heralded single photons having timing reference that is used in the post-processing stage for appropriately adjusting the corresponding coincidence time windows between the heralding and heralded photons of the same pair, for different measurement settings. More details on this can be found in Appendix \ref{appendix:d}. 

\subsection{Preparation state loophole} 
In deriving the macrorealist inequalities, an inherent assumption is that in different runs of the experiment, the photon considered is always prepared in the same initial state. While the heralded single photons originating from a stable SPDC source are expected to be prepared in the same initial state, the possibility remains that detectors can register photons from other sources, such as background stray noises (we call this loophole as the preparation state loophole). Nevertheless, in the present experiment, by choosing for different measurement settings the corresponding coincidence time windows having high signal to noise ratios, and by subtracting the accidental coincidences, we post-select only those detected photons that come from the SPDC source, and not from the background (see Appendix \ref{appendix:d}). 
Thus, in this way, the ‘preparation state loophole’ is also closed in our experiment.

 Finally, we note that a number of experiments testing macrorealism have attempted to address the clumsiness loophole  (see Table \ref{tab:list}). However, the other loopholes that have been analysed in this work remain largely unaddressed in the relevant literature. For example, the issue of detection efficiency loophole has not been substantially analysed in any other experiment testing macrorealism. \blk


\section{Details of the experiment}
\label{sec:exp}
An experimental scheme to investigate the violation of macrorealism consists of three consecutive stages. These may be denoted as  preparation stage ($P$), unitary transformation stage ($U_{t_{i}, t_{j}}$) and measurement stage ($M_{t_k}$).
For the simplest case of three-time inequalities, measurements are carried out at three different times, $t_1,t_2,t_3$ (denoted by $M_{t_{1}}$, $M_{t_{2}}$, and $M_{t_{3}}$), with the unitary transformations (denoted by $U_{t_{1}, t_{2}}$, and $U_{t_{2}, t_{3}}$) in between.

As illustrated in Fig. \ref{fig:schematic1}, the preparation stage consists of a single photon source and a beamsplitter (BS1) with splitting ratio of $T_1:R_1$. A unitary transformation $U_{t_{1}, t_{2}}$ is simulated as a combination of a phase-modulator ($\theta_{1}$) and a beamsplitter ($BS2$). $BS2$ has a splitting ratio of $T_{2}:R_{2}$. Similarly, $U_{t_{2}, t_{3}}$ is simulated as a combination of phase-modulator ($\theta_{2}$) and a beamsplitter ($BS3$) with a splitting ratio $T_{3}:R_{3}$. Here, $T_i+R_i=1$, $\forall~i={1,2,3}$. We 
perform three separate experiments to measure $\langle Q_{t_{1}}Q_{t_{2}}\rangle$, $\langle Q_{t_{2}}Q_{t_{3}}\rangle$ and $\langle Q_{t_{1}}Q_{t_{3}}\rangle$, wherein for each experiment, negative result measurement is implemented by 
inserting one blocker in any one arm at any time and considering only those photons that have not interacted with that blocker. 

To illustrate this point, note that if a blocker is placed at $-1$ arm at time $t_{2}$, and a photon is detected at time $t_{3}$, this 
implies that the photon must have traversed the path $+1$ at time $t_{2}$, without directly interacting with that blocker. Here it may be noted that we have explicitly verified that the number of photons detected after passing through a blocker is comparable to the detector dark count, which implies that the blocker is behaving ideally. For this setup, the quantum mechanically predicted violation of the LGI given in \eqref{eq:LGI}  is then obtained as follows,
\begin{eqnarray} \label{lgi}
&& \langle Q_{t_{1}} Q_{t_{2}}\rangle + \langle Q_{t_{2}} Q_{t_{3}}\rangle - \langle Q_{t_{1}} Q_{t_{3}}\rangle \nonumber \\
& = & 1 - 4R_2R_3 + 4\cos{\theta_{2}}\sqrt{T_2T_3R_2R_3} \nonumber \\
& \leq & 1.5 , 
\end{eqnarray}
in which the maximum value 1.5 is found for
\begin{equation*}
    \theta_{2} = 0, T_2=0.75, T_3=0.75 .
\end{equation*}
Similarly, the quantum mechanically predicted violation of the WLGI given in \eqref{eq:WLGI} is obtained as follows
\begin{eqnarray}\label{wlgi}
&& P_{t_{1},t_{3}}(-,+)-P_{t_{1},t_{2}}(-,+)-P_{t_{2},t_{3}}(-,+) \nonumber \\
&=& 2\cos{\theta_{2}}R_1\sqrt{T_2T_3R_2R_3} - 2\cos{\theta_{1}}R_3\sqrt{T_1T_2R_1R_2} - R_2R_3  \nonumber \\
&\leq & 0.4034,
\end{eqnarray}
in which the maximum value 0.4034 is found for
\begin{equation*}
 \theta_{1} = \pi,\theta_{2} = 0, T_1=0.1524, T_2=0.6952, T_3=0.4833 .
\end{equation*}
 
Now, in order to enable a loophole free version of the experiment, we apply a few modifications to the setup. For satisfying all the two-time NSITs, 
we make the two arms of the first Mach-Zehnder interferometer (MZI) non-interfering, by adding a path difference between the +1 and -1 arms at the instant $t_{1}$. This modification makes the phase term associated to the first MZI ($\theta_1$) irrelevant in our setup. Also, the second phase term ($\theta_2$) is required to be zero for the maximum violation of both LGI \eqref{lgi} and WLGI \eqref{wlgi}. So, we do not consider both the phase-modulators ($\theta_1$, $\theta_2$) in the setup. Further, in the experimental implementation, we replace the second MZI with a displaced Sagnac interferometer, as it provides the desired interferometric stability against any external vibrations, due to its geometric configuration \cite{mansuripur_2009,PhysRevLett.125.123601}.

\begin{figure*}[!ht]
\centering
\includegraphics[width=0.9\textwidth]{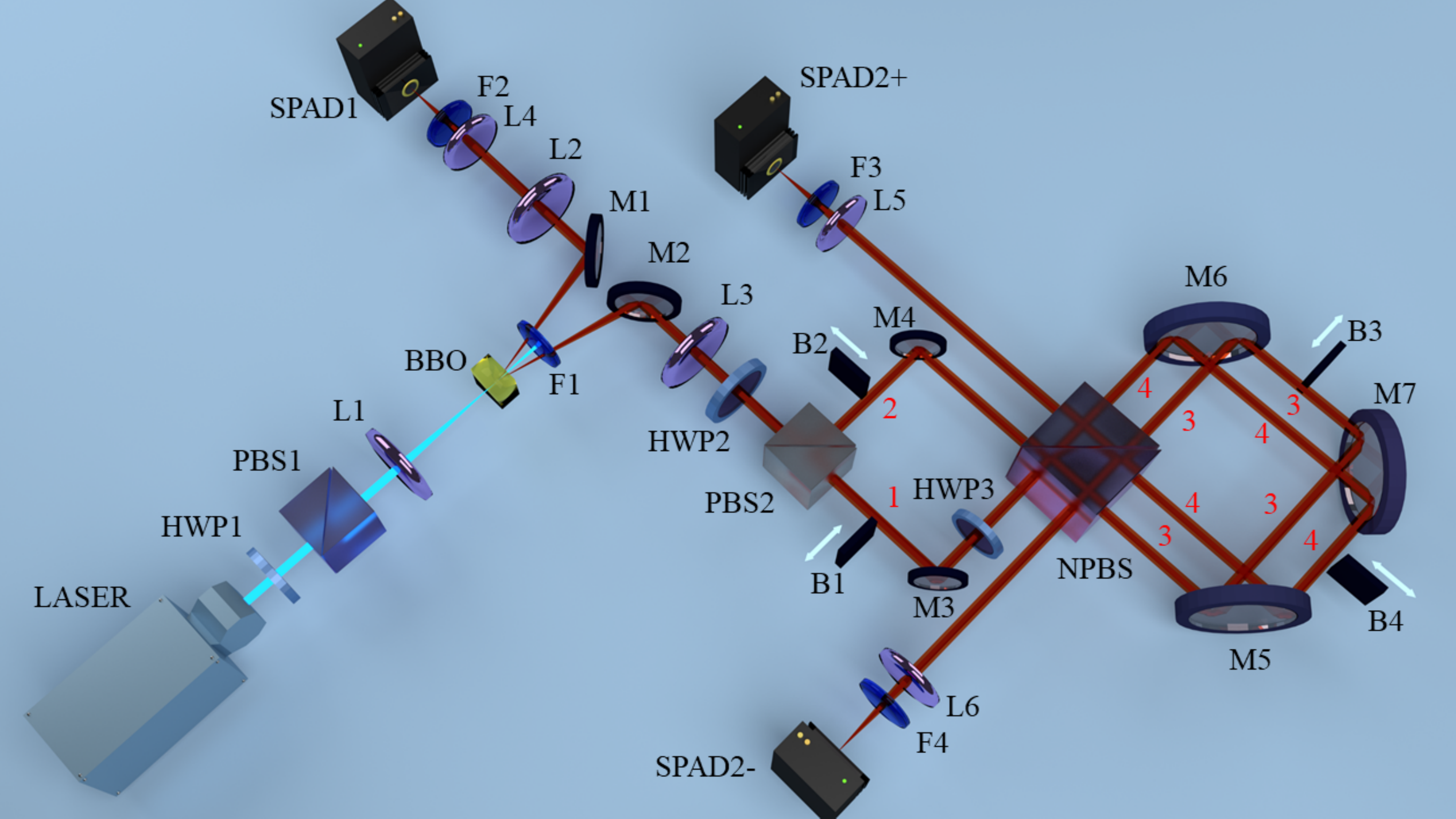}
\caption{Schematic of the modified experimental setup. \textbf{HWP1}, \textbf{HWP2}, \textbf{HWP3}: Half waveplate; \textbf{PBS1}, \textbf{PBS2}: Polarizing beam-splitter; \textbf{L1}, \textbf{L4}, \textbf{L5}, \textbf{L6}: Focusing lens; \textbf{F1}: Long-pass filter; \textbf{M}: Dielectric mirror; \textbf{L2}, \textbf{L3}: Collimating lens; \textbf{F2}, \textbf{F3}, \textbf{F4}: Band-pass filter; \textbf{B1}, \textbf{B2}, \textbf{B3}, \textbf{B4}: Blocker; \textbf{NPBS}: Non-polarizing beam-splitter; \textbf{SPAD1}, \textbf{SPAD2+}, \textbf{SPAD2-}: Single-photon avalanche detector. Two arms of the aMZI are marked as 1 and 2, which represent the $+1$ and $-1$ arm, respectively. Similarly, two arms of the dSI are marked as 3 and 4, representing $-1$ and $+1$. SPAD2+ and SPAD2- are placed in $+1$ and $-1$ arms, respectively.}
\label{fig:schematic2}
\end{figure*}

Our modified setup (see Fig. \ref{fig:schematic2}) starts with a heralded single photon source, where a diode laser (Cobolt) pumps a BBO (Beta Barium Borate) crystal with a continuous wave of light at a central wavelength of 405 nm and pump power of 10 mW. The BBO crystal is oriented in such a way that it is phase-matched for degenerate, non-collinear, type I spontaneous parametric down-conversion (SPDC) while being pumped with horizontally polarized light. To make the input pump beam Gaussian, we use an apparatus for spatial filtering, including a focusing lens, a pinhole and a collimating lens \cite{Cox_1979}. To make the pump beam horizontally polarized, a combination of a half-wave plate (HWP1) and a polarizing beamsplitter (PBS1) is used. Parametric down-conversion creates pairs of single photons, both with vertical polarization and 810 nm central wavelength. In order to increase pair generation, we also place a focusing lens (L1), to focus the pump beam into the central spot of the BBO crystal.
A long-pass filter (F1) is placed after the crystal, to block the pump beam and pass only the down-converted single photon pairs. Two dielectric mirrors (M1, M2) are placed to send one photon from each pair to the experiment, and another photon towards a single photon avalanche detector (SPAD1), for heralding. One collimating lens (L2, L3) is placed in each arm, just after the mirrors. The SPADs (ID Quantique, ID120) are all free-space detectors with a maximum quantum efficiency of 80$\%$ at 800 nm wavelength. In front of each SPAD (SPAD1, SPAD2+, SPAD2-), we place a focusing lens (L4, L5, L6) and a band-pass filter (F2, F3, F4) centered at 810 nm and filtering bandwidth of 3 nm. Focusing lenses focus single photons onto the sensors and the filters reduce background noises.

Once the single-photon pairs are generated from the SPDC source, one photon (heralded photon) from each pair is sent to the main experimental setup comprising the two interferometers that are suitably devised. The first interferometer is an asymmetric Mach-Zehnder interferometer (aMZI), whereas the second one is a displaced Sagnac interferometer (dSI). The aMZI is made asymmetric by introducing a path difference of few millimeters between the two arms of the aMZI, by slightly shifting the position of the mirror (M4) in arm 2. This modification makes the two arms of the aMZI non-interfering as the coherence length of the single photons (typically hundreds of micrometers) is well below the path difference. A combination of a HWP (HWP2) and a PBS (PBS2) controls the beamsplitting ratio in the two arms of the aMZI. Another HWP (HWP3) in arm 1 converts the horizontally (H) polarized photon back to vertically (V) polarized. In the dSI, a single non-polarizing beamsplitter (NPBS) with a measured splitting ratio of $T:R=80:20$ (w.r.t. vertically polarized light at 810 nm wavelength) is used. All alignments in the dSI are catered to achieve optimal interference visibility and stability for both input arms, simultaneously. However, as will be seen later, small inadequacies in the same lead to certain modifications in the experimentally achievable quantum mechanical predicted values for the various bounds. Two detectors (SPAD2+, SPAD2-) placed in the two output arms of the dSI detect single photons.

Any time between the photon traveling from PBS2 to the first impact on NPBS is considered as $t_{1}$. From the first impact to the second impact on NPBS is time $t_{2}$. Any time after the NPBS is $t_{3}$. Two motorized blockers (B1, B2) are placed in arm 1 and 2, and another two blockers (B3, B4) are placed in arm 3 and 4, to perform negative result measurements at time $t_{1}$ and $t_{2}$, respectively. 

Quantum mechanically predicted violations of LGI and WLGI for the modified setup are evaluated in \eqref{eq:lgi2} and \eqref{eq:wlgi2}. The splitting ratio due to the HWP2 and PBS2 combination is denoted as ${|\alpha|}^{2}:{|\beta|}^{2}$, where ${|\alpha|}^{2}+{|\beta|}^{2}=1$. The splitting ratio of the NPBS is denoted as $T:R$. All the three NSIT conditions (using Eqs. \ref{eq:NSIT12}-\ref{eq:NSIT13}) are expressed as
\begin{equation}
\label{eq:nsit12main}
 NSIT_{(t_1)t_2}=|P_{t_2}(+)-P_{t_1,t_2}(+,+)-P_{t_1,t_2}(-,+)|=0
\end{equation}
\begin{equation}\label{eq:nsit13main}
 NSIT_{(t_1)t_3} = |P_{t_3}(+)-P_{t_1,t_3}(+,+)-P_{t_1,t_3}(-,+)|=0 . 
\end{equation}
\begin{equation}
\label{eq:nsit23main}
\begin{split}
 NSIT_{(t_2)t_3}& =|P_{t_3}(+)-P_{t_2,t_3}(+,+)-P_{t_2,t_3}(-,+)| \\
 &=2TR~cos(\theta_2)|\{|\alpha|^2-|\beta|^2\}|
\end{split}
\end{equation}
Due to the introduced path difference, there is no interference as the photon evolves independently with the amplitudes $|\alpha|^2$ and $|\beta|^2$ corresponding to the +1 and -1 arms respectively. Hence, the two NSIT conditions \eqref{eq:nsit12main} and \eqref{eq:nsit13main} are valid for any value of $T:R$ and ${|\alpha|}^{2}:{|\beta|}^{2}$. On the other hand, a simple calculation leads to the expression of $NSIT_{(t_2)t_3}$ in \eqref{eq:nsit23main} (see Eq. (A4) in \cite{PhysRevA.87.052115}), which vanishes if we fix ${|\alpha|}^{2}={|\beta|}^{2}=0.5$.

Another point to be noted is that we have used $q_{t_2}=+1$ and $q_{t_3}=+1$ in Eqs. \eqref{eq:NSIT12}-\eqref{eq:NSIT13}, while deriving the three NSIT expressions. Here, one may also use $q_{t_2}=-1$ and $q_{t_3}=-1$, which will give the same result.  We evaluate the maximum values of the LGI expression as
\begin{equation}
\label{eq:lgi2}
\begin{split}
& \quad \langle Q_{t_{1}} Q_{t_{2}}\rangle + \langle Q_{t_{2}} Q_{t_{3}}\rangle - \langle Q_{t_{1}} Q_{t_{3}}\rangle\\ & = 1-4R^{2}+4TR \\&
= 1.50 ~~~[\text{for}~ T=0.75],
\end{split}
\end{equation}
 and the WLGI expression as,
\begin{equation}
\label{eq:wlgi2}
\begin{split}
& \quad P_{t_{1},t_{3}}(-,+)-P_{t_{1},t_{2}}(-,+)-P_{t_{2},t_{3}}(-,+) \\
& = 2{|\beta|}^{2}TR - R^{2} \\
& = 0.125 ~~[\text{for}~T=0.75,~ \text{while}~ {|\alpha|}^{2}={|\beta|}^{2}=0.5] .
\end{split}
\end{equation}
In order to measure $\langle Q_{t_{2}}Q_{t_{3}}\rangle$, we perform two runs of the experiment. In the first run, a blocker is placed at $-1$ arm at time $t_{2}$, and coincidence counts are measured between SPAD1, SPAD2+ and between SPAD1 and SPAD2-. We denote the first coincidence as $C_{t_{2},t_{3}}(+,+)$, because any photon that gets detected in SPAD2+ must be in $+1$ arm at $t_{2}$ and $+1$ arm at $t_{3}$. Similarly, the second coincidence count is denoted as $C_{t_{2},t_{3}}(+,-)$. In the second run of the experiment, a blocker is placed at $+1$ arm at $t_{2}$. So, coincidence count between SPAD1 and SPAD2+ is $C_{t_{2},t_{3}}(-,+)$, and coincidence count between SPAD1 and SPAD2- is $C_{t_{2},t_{3}}(-,-)$. To make sure that both runs of the experiment are consistent with each other, we measure all the coincidences for the same duration of time. However, the coincidence counts may also experience some systematic as well as random fluctuations with time which have been taken into consideration in the error-analysis (see 
Appendix \ref{appendix:c2}). Now, the total coincidence count becomes $C^{T}_{t_{2},t_{3}}=C_{t_{2},t_{3}}(+,+)+C_{t_{2},t_{3}}(+,-)+C_{t_{2},t_{3}}(-,+)+C_{t_{2},t_{3}}(-,-)$. All four joint probabilities, $P_{t_{2},t_{3}}(q_{t_{2}},q_{t_{3}})$, where $q_{t_{2}}=\pm 1$, $q_{t_{3}}=\pm 1$, are measured from these four coincidence counts.
\begin{equation}
\label{eq:prob13}
P_{t_{2},t_{3}}(q_{t_{2}},q_{t_{3}})=\dfrac{C_{t_{2},t_{3}}(q_{t_{2}},q_{t_{3}})}{C^{T}_{t_{2},t_{3}}}
\end{equation}
Similar strategy is used to measure $\langle Q_{t_{1}}Q_{t_{3}}\rangle$ as well, where in the first run, blocker is placed in $-1$ arm at time $t_{1}$, and in $+1$ arm at time $t_{1}$ for the second run.

\begin{equation}
\label{eq:prob23}
P_{t_{1},t_{3}}(q_{t_{1}},q_{t_{3}})=\dfrac{C_{t_{1},t_{3}}(q_{t_{1}},q_{t_{3}})}{C^{T}_{t_{1},t_{3}}}
\end{equation}

Measurement of $\langle Q_{t_{1}}Q_{t_{2}}\rangle$ is slightly different, as we do not place the detectors (SPAD2+ and SPAD2-) at time $t_{2}$, in order to close the detection efficiency loophole. Here we use two blockers at both times $t_{1}$ and $t_{2}$ and measure all eight, three-time joint probabilities. We perform four runs of the experiment to get all eight probabilities.
\begin{equation}
\label{eq:prob12}
P_{t_{1},t_{2},t_{3}}(q_{t_{2}},q_{t_{3}},q_{t_{3}})=\dfrac{C_{t_{1},t_{2},t_{3}}(q_{t_{1}},q_{t_{2}},q_{t_{3}})}{C^{T}_{t_{1},t_{2},t_{3}}}
\end{equation}
The four marginal joint probabilities, $P_{t_{1},t_{2}}(q_{t_{1}},q_{t_{2}})$ are calculated from these eight measured probabilities, $P_{t_{1},t_{2},t_{3}}(q_{t_{2}},q_{t_{3}},q_{t_{3}})$, by using \eqref{eqn:AoT}. More details on the experimental methods have been provided in 
Appendix \ref{appendix:c1}.

\section{Results and the error analysis}
\label{sec:result} 

We obtain all the joint probabilities of the form $P_{t_{i},t_{j}}(q_{t_{i}},q_{t_{j}})$ from the experimental data, and evaluate the correlation functions $\langle Q_{t_{i}}Q_{t_{j}}\rangle$ from these joint probabilities. A representative violation of LGI and WLGI from the experiment is provided in the table of Figure \ref{tab:resultTab1}. We also verify that all the two-time NSIT conditions (\ref{eq:NSIT12}-\ref{eq:NSIT13}) are satisfied in the experiment, as shown in the table of Figure \ref{tab:resultTab2}. We show that all the measured NSIT values are within the statistical fluctuations and hence can be considered as zero with a bound of the order of $10^{-2}$; thus ensuring the validity of the maintenance of non invasive measurement (NIM) in our experiment.

\begin{widetext}

\begin{figure}[!ht]
\centering
\includegraphics[width=10cm,height=3.5cm]{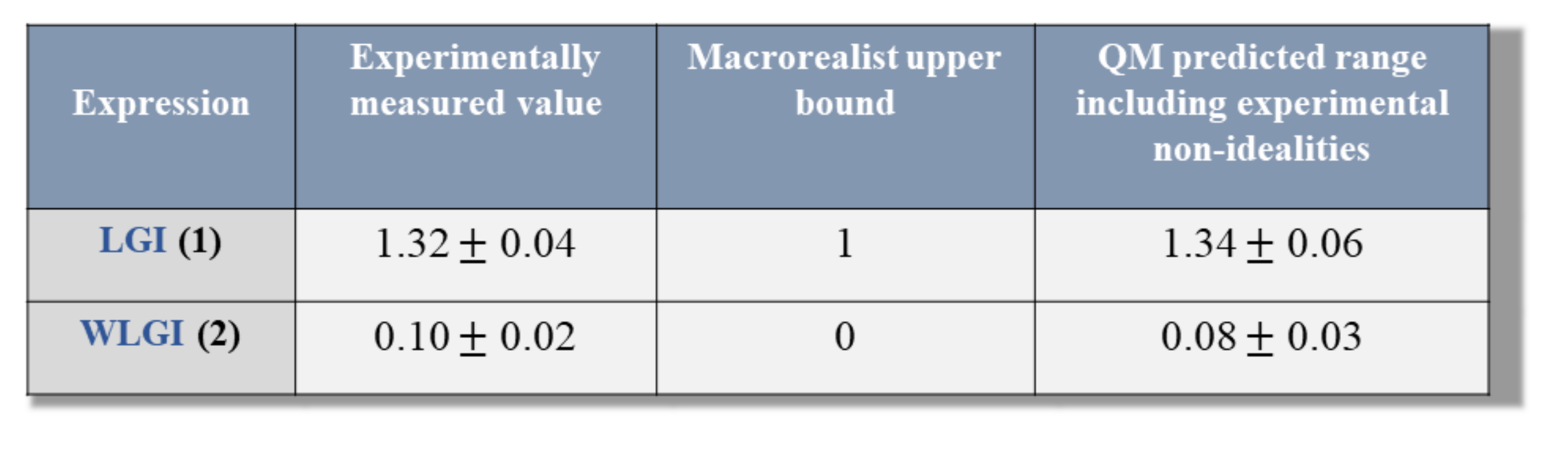}
\caption{Measured mean values of the LGI (given by \eqref{eq:LGI}) and WLGI (given by \eqref{eq:WLGI}) expressions are given with their respective error ranges in order to compare with their respective macrorealist upper bounds as well as the maximum quantum mechanical predicted ranges, that have been estimated by taking into account non idealities in our set up arising from least count related systematic limitations in our experimental components as well as the range of interferometric visibility values, as observed in our experiment. In the case of LGI, the magnitude of violation or the difference between the experimentally measured value and the upper bound from macrorealism is eight times the error value. For WLGI, the violation is five times the error value. Both the measured values show perfect compatibility with the respective quantum mechanical predicted ranges.}\label{tab:resultTab1}
\end{figure}
\begin{figure}[!ht]
\centering
\includegraphics[width=10.2cm,height=3.5cm]{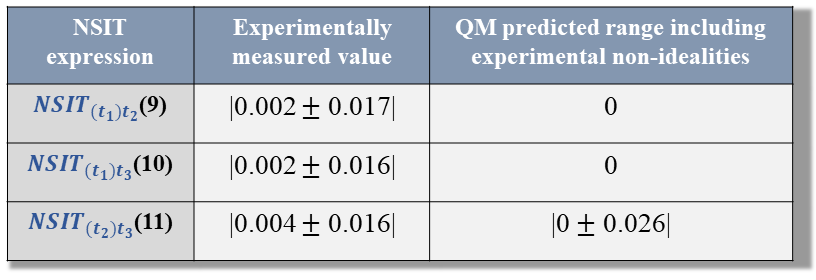}
\caption{Measured mean values of all the three NSIT expressions given by Eqs. (\ref{eq:nsit12main}-\ref{eq:nsit13main}) respectively are of the order of $10^{-3}$, smaller than their respective error ranges ($\sim 10^{-2}$). Hence all the three measured NSIT values can be considered to be zero, with a bound of $10^{-2}$, in accordance with the quantum mechanical predictions, which have been estimated taking into account experimental non idealities as for LGI/WLGI reported in the table of Figure \ref{tab:resultTab1}. This implies the maintenance of the NIM condition in the present experiment, up to this accuracy.}\label{tab:resultTab2}
\end{figure}

\end{widetext}

To show violation of LGI, we calculate the left hand side (LHS) of \eqref{eq:LGI} using experimentally measured correlation values $\langle Q_{t_{1}}Q_{t_{2}}\rangle=0.56$, $\langle Q_{t_{2}}Q_{t_{3}}\rangle=0.54$, $\langle Q_{t_{1}}Q_{t_{3}}\rangle=-0.22$. Similarly for WLGI, we use measured joint probabilities, $P_{t_{1},t_{2}}(-,+)=0.11$, $P_{t_{2},t_{3}}(-,+)=0.12$, $P_{t_{1},t_{3}}(-,+)=0.33$ to calculate the LHS of Eq. \ref{eq:WLGI}. Details on the experimental methods to measure correlation values and joint probabilities, similar to those mentioned above, can be found in 
Appendix \ref{appendix:c1}. 


Here, we make a pertinent observation. While the relative  error  in  the measured WLGI value is  of  course  larger  than  the  LGI one given the smaller value of the mean WLGI, the experimentally measured values for LGI have a higher absolute error than  the  WLGI  ones.   The  reason  for  this  can  be  easily  understood  as  follows.   The  LGI  expression (\ref{eq:LGI})  includes twelve joint probabilities, which is much larger in number than the three joint probabilities required for evaluating the WLGI expression (\ref{eq:WLGI}). The higher number of terms involves more imperfections stemming from various experimental parameters,  which in turn increases their contribution to the absolute error in the experimentally measured value for  the  LGI  expression  as  compared  to  the  WLGI  and NSIT expressions. This is thus an instructive feature of showing the violation of WLGI in our setup, along with the LGI violation.

Next, the comparison between the measured values and the quantum mechanical estimated ranges of the LGI, WLGI, NSIT values, is discussed. While calculating the quantum mechanical predicted ranges for LGI, WLGI and NSITs in the tables of Figure \ref{tab:resultTab1} and \ref{tab:resultTab2}, we consider 
 two different sources of non idealities in our set up.

\begin{itemize}
    \item One arises from least count related systematic limitations in many of our experimental components e.g., the non-polarizing beamsplitter (NPBS in Figure \ref{fig:schematic2}) having $80\%$ transmission probability instead of ideal requirement of $75\%$, the $T:R$ ratio of the NPBS being dependant on the input ports and having $\pm 2\%$ error, half-wave plate (HWP2) angle having precision error of $\pm 1^\circ$, etc. This gives rise to a quantum mechanical predicted range of $1.47\pm 0.02$ for LGI, $0.11\pm 0.02$ for WLGI, and $\left|0\pm 0.03\right|$ for $NSIT_{(t_1)t_3}$. 
    
    \item The second source of experimental non-idealities stems from our inteferometric architecture itself. We recall from the discussion in Section  \ref{sec:exp}, that in order to achieve the maximum quantum mechanical predicted value of the LGI and the WLGI expression, we need the second phase term $\theta_2$ in Eqs. \eqref{lgi} and \eqref{wlgi} to be zero. This would require achievement of optimal interferometric visibility for all experiment runs. However, although we aim for this optimal condition through painstaking alignment procedures, our relative visibility ranges between 70$\%$ to 85$\%$ during the course of all the experimental runs. This in turn is related to some unavoidable contributions that persist irrespective of the near perfect alignment. These include small temperature fluctuations in the lab, surface imperfections from mirrors that could be in the micrometer range and thus comparable to our single photon wavelength (810 nm), small polarisation dependence that is observed even in a NPBS as well as polarisation dependence of mirrors along with slight non colinearity that may still persist in the interferometer outputs due to minute angular mismatch in the beam splitter arrangements. We thus go on to estimate the modified quantum mechanical predicted range for our experiment by relaxing the $\theta_2 = 0$ condition and including the variation in relative visibility from 70$\%$ to 85$\%$ also in the calculation. This leads to a modified quantum mechanical predicted range for our experiment, of $1.34\pm0.06$ for LGI, $0.08\pm0.03$ for WLGI and $\left|0\pm0.026\right|$ for $NSIT_{(t_1)t_3}$. Detailed discussion on the calculation of the quantum mechanical predicted range for various inequalities/equalities and their dependence on different experimental parameters has been provided in Appendix \ref{appendix:c3}. 
    In order to gain further understanding for such models in general, one can refer to \cite{PhysRevA.100.013839} as well as \cite{PhysRevApplied.14.024036} wherein extensive calculations have been performed to capture such non-idealities of various experimental parameters in the context of a Hong-Ou-Mandel (HOM) experiment and a quantum key distribution experiment respectively.
\end{itemize}

An important point to note here is that the spread in the quantum mechanical predicted range without the $\theta_2 = 0$ relaxation is taking into account imperfections in optical components including least count errors. The spread in the range after relaxing the  $\theta_2 = 0$ condition is reflective of the spread in measured visibility of the interferometer. This spread clearly subsumes the optical component related non-idealities and also includes other unavoidable non idealities as discussed above, thus providing an expectation closer to actuality of our experiment.

In the table of Figure \ref{tab:resultTab2}, we give theoretical estimates of the quantum mechanical predicted values for the three NSIT expressions. It can be seen that while both the $NSIT_{(t_1)t_2}$ and $NSIT_{(t_2)t_3}$ expressions  are predicted to have the fixed value of zero, the estimated $NSIT_{(t_1)t_3}$ value ranges from 0 to 0.026. 
The reason for this is the fact that the first interferometer we use in the setup is an asymmetric Mach-Zehnder interferometer (aMZI) which does not allow any interference while the second interferometer (dSI) shows interference. So, $NSIT_{(t_1)t_2}$ and $NSIT_{(t_2)t_3}$ do not depend on any experimental parameters (see \eqref{eq:nsit12main} and \eqref{eq:nsit23main}), while $NSIT_{(t_1)t_3}$ depends on beamsplitting ratios, half-wave plate angle, etc. (see \eqref{eq:nsit13main}), hence due to the uncertainties in the values of these parameters, there is a range of the quantum mechanical predicted value for this NSIT expression. As for LGI/WLGI above, we also note the quantum mechanical predicted values for the NSIT expressions in the $\theta_2 \neq 0$ condition (as shown in Appendix \ref{appendix:c3}).\\
In order to quote the measured values for the LGI, WLGI, and NSITs, in the tables of Figure \ref{tab:resultTab1} and \ref{tab:resultTab2}, we estimate the average values and the errors from all experimental datasets. For this purpose, we repeat all the experimental runs for a significant number of times, recording the coincidence counts for 10 seconds in the different runs. In order to find the number of iterations sufficient for the experiment, we record 1000 coincidence datasets for each of the two scenarios. First, for the non-interfering setup, wherein the two blockers are placed, one in arm 1 and another in arm 3, while the coincidences between SPAD1 and SPAD2+ are measured (see Figure \ref{fig:schematic2}). For the second case, which we call the interference case, the blocker is placed only in arm 1, and the coincidences between SPAD1 and SPAD2+ are measured. We apply a bootstrapping algorithm to find the standard deviation by mean (SD/M) value of the mean coincidence for the different iterations of the experiment. We find an SD/M value lower than 0.05$\%$ for 150 iterations in the non-interference case and 300 iterations for the interference case. Hence, we take 300 iterations while measuring the correlation function $\langle Q_{t_{1}}Q_{t_{3}}\rangle$ as it includes all interference terms, and for 150 iterations, the correlation functions $\langle Q_{t_{1}}Q_{t_{2}}\rangle$ and $\langle Q_{t_{2}}Q_{t_{3}}\rangle$ are measured. For testing the LGI, the three mean values ${\langle Q_{t_{1}}Q_{t_{2}}\rangle}_{\mu}, {\langle Q_{t_{2}}Q_{t_{3}}\rangle}_{\mu}, {\langle Q_{t_{1}}Q_{t_{3}}\rangle}_{\mu}$ are measured, along with their respective standard deviations denoted as $\sigma_{1,2}, \sigma_{2,3}, \sigma_{1,3}$. Here, the subscript $\mu$ is used to identify these correlation values as experimentally measured averages of the large iterations. The LGI expression ${\langle Q_{t_{1}}Q_{t_{2}}\rangle}_{\mu}+{\langle Q_{t_{2}}Q_{t_{3}}\rangle}_{\mu}-{\langle Q_{t_{1}}Q_{t_{3}}\rangle}_{\mu}$ in this way has a maximum error of $\pm\Delta$. We consider the worst-case scenario where we assume that all the three experimental runs are independent of each other, and hence the error from each run adds up to form the maximum error, i. e., $\Delta=\sigma_{1,2}+\sigma_{2,3}+\sigma_{1,3}$. 
A similar strategy is also applied for testing the WLGI and the NSITs. Detailed discussion on the bootstrapping strategy as well as the standard deviation can be found in Appendix \ref{appendix:c2}. 


\section{Summary and Outlook}
\label{sec:conclusion}

In this work, we have demonstrated experimental violations of both the LGI (\ref{eq:LGI}) and WLGI (\ref{eq:WLGI}) using an interferometric setup where a two-level system is realized by measuring the position of the heralded single photon in either of the two arms of the interferometer at three different instants ($t_1, t_2, t_3$). All the measurements are unambiguous and projective, and negative result measurements have been performed at times $t_1$ and $t_2$, in order to satisfy the non-invasiveness condition. We report LGI violation of $1.32 \pm 0.04$ and WLGI violation of $0.10 \pm 0.02$ from the experimental data whereas the macrorealist upper bound for these inequalities are $1$ and $0$ respectively. We have also estimated the LGI and WLGI ranges from quantum mechanical predictions for our experimental conditions to be $1.34\pm 0.06$ and $0.08 \pm 0.03$ respectively and we find both the WLGI value and the LGI measured values are perfectly consistent with quantum mechanical predictions. Thus our experiment not only provides a decisive violation of macrorealism but also a convincing compatibility with quantum mechanical. All the measured NSIT values are zero with a bound of the order of $10^{-2}$; thus ensuring the required non-invasiveness of the measurements in our experiment. 

In a nutshell, the experiment we have reported in the present paper provides for the first time a rigorous test of the notion of macrorealism as applied to single photons by closing the relevant loopholes. The conclusive violation of the macrorealist inequalities (LGI/WLGI) observed in our experiment establishes that, independent of the specifics of any theoretical framework, the behaviour of single photons in the interferometric setup we have used is fundamentally incompatible with the tenet of macrorealism which is a generic trait characterising any classical description of the physical world. This deep-seated implication of such an experimental demonstration, thus, enriches its significance beyond showing the quantumness of single photons per se. The present experimental study, therefore, complements the loophole-free empirical repudiation of another fundamental notion, viz. local realism that has been convincingly demonstrated in the recent years using the entangled photons \cite{PhysRevLett.115.250401,PhysRevLett.115.250402}. While the locality loophole has been significantly closed in these latter experiments, our present experiment addresses the clumsiness loophole by suitably employing negative result measurement and using the NSIT condition as a tool for confirming the extent to which this loophole has been closed. For getting around the detection efficiency loophole, the scheme adopted by the tests of local realism using the entangled photons rely on achieving the required critical detection efficiency. On the other hand, a striking feature of our experiment is that the procedure based on the experimental configuration we have used facilitates avoiding the detection efficiency loophole by showing the violation of the macrorealist inequalities for any value of the detection efficiency. Surprisingly, the detection efficiency loophole has remained unaddressed in any of the experiments performed so far towards testing macrorealism using photons. Another salient feature is the way we have closed the multiphoton emission loophole which, too, has not been tackled in any of the photonic experiments that have been performed to date for testing macrorealism. 

Thus, having plugged the relevant loopholes, our present experiment with single photons not only decisively falsifies both the macrorealist inequalities LGI and WLGI in the same experiment for the first time, but also demonstrates a convincing compatibility with quantum mechanical predictions that have been estimated taking into account experimental non idealities. This, therefore, potentially provides a reliable, robust and efficient platform towards exploring the single photon based applications of such macrorealist inequalities, for which some relevant ideas have already been discussed \cite{Emary_2013,PhysRevA.73.052308,SHENOYH20172478,arXiv:quant-ph/0402127}. 
For instance, in Ref. \cite{arXiv:quant-ph/0402127}, specific examples have been given of the information theoretic tasks for which the quantum advantage stems from the violation of the relevant macrorealist bounds. Interestingly, it has been suggested that quantum communication complexity protocols that are based on exchange of qubits can be reformulated within the framework of such tasks, thereby contributing to the possibility of increasing capacities of classical computational devices. In the context of quantum crptography, in Ref. \cite{SHENOYH20172478}, the question whether the violation of LGI can contribute towards enhancing the security of the BB84 protocol has been analysed. On the other hand, relevant to quantum computation, information-theoretic temporal versions of Bell inequality have also been proposed \cite{PhysRevA.73.052308}. These directions of studies should be worth exploring for harnessing the violation of macrorealism for practical applications.

Of course, no experiment, as ideal as it is, can unequivocally be regarded as entirely free from loopholes. Similar to what underlies the memory loophole in the context of testing the Bell-type inequalities \cite{PhysRevA.66.042111}, the analysis of the experiments testing macrorealist inequalities also involves a subtle assumption that the behaviours of single photons in the different runs of the experiment are mutually independent. In particular, in the context of our experiment, it means assuming that the behaviour of a single photon in the $n$th run is independent of the measurement choices and outcomes for the preceding $(n-1)$ runs. With respect to the testing of the Bell-CHSH inequality, it has been argued \cite{PhysRevA.66.042111} that the probability of simulating a significant violation of this inequality by a local realist model exploiting this memory loophole can be made negligibly small by taking a sufficiently large number of runs. Thus, by taking cue from the way this martingale-based analysis has been formulated \cite{PhysRevA.66.042111} in the Bell-CHSH context, it should be  instructive to investigate as to  what extent one can minimise the possibility of simulating the observed violation of LGI/WLGI using a macrorealist model based on the deviation from the assumed condition of mutual independence of different runs.  

Another possible direction in which our setup may be used is by adopting the type of negative result measurement protocol that has been proposed for arbitrary choices of the measurement times, as well as for testing higher-order correlations involved in the generalization of the macrorealist framework \cite{PhysRevA.99.022119}.

Finally, we note that the macrorealist inequalities (LGI and WLGI) tested in the present experiment are essentially the necessary conditions for macrorealism. In the recent years, interestingly, the necessary and sufficient conditions for macrorealism have been formulated in terms of a suitable combination of the three-time and two-time LG inequalities \cite{PhysRevA.96.012121}. The platform provided by our present experimental study can, thus, be appropriately utilised to empirically check in a loophole-free way the full set of such necessary and sufficient macrorealist conditions. \\


\subsection*{Acknowledgements}
US acknowledges partial support provided by the Ministry of Electronics and Information Technology (MeitY), Government of India under grant for Centre for Excellence in Quantum Technologies with Ref. No. 4(7)/2020 - ITEA as well as partial support of the QuEST-DST Project Q-97 of the Govt. of India. DH thanks NASI for the Senior Scientist Platinum Jubilee Fellowship and acknowledges support of the QuEST-DST Project Q-98 of the Govt. of India.

\bibliography{ref}

\begin{thebibliography}{48}%
\makeatletter
\providecommand \@ifxundefined [1]{%
 \@ifx{#1\undefined}
}%
\providecommand \@ifnum [1]{%
 \ifnum #1\expandafter \@firstoftwo
 \else \expandafter \@secondoftwo
 \fi
}%
\providecommand \@ifx [1]{%
 \ifx #1\expandafter \@firstoftwo
 \else \expandafter \@secondoftwo
 \fi
}%
\providecommand \natexlab [1]{#1}%
\providecommand \enquote  [1]{``#1''}%
\providecommand \bibnamefont  [1]{#1}%
\providecommand \bibfnamefont [1]{#1}%
\providecommand \citenamefont [1]{#1}%
\providecommand \href@noop [0]{\@secondoftwo}%
\providecommand \href [0]{\begingroup \@sanitize@url \@href}%
\providecommand \@href[1]{\@@startlink{#1}\@@href}%
\providecommand \@@href[1]{\endgroup#1\@@endlink}%
\providecommand \@sanitize@url [0]{\catcode `\\12\catcode `\$12\catcode
  `\&12\catcode `\#12\catcode `\^12\catcode `\_12\catcode `\%12\relax}%
\providecommand \@@startlink[1]{}%
\providecommand \@@endlink[0]{}%
\providecommand \url  [0]{\begingroup\@sanitize@url \@url }%
\providecommand \@url [1]{\endgroup\@href {#1}{\urlprefix }}%
\providecommand \urlprefix  [0]{URL }%
\providecommand \Eprint [0]{\href }%
\providecommand \doibase [0]{http://dx.doi.org/}%
\providecommand \selectlanguage [0]{\@gobble}%
\providecommand \bibinfo  [0]{\@secondoftwo}%
\providecommand \bibfield  [0]{\@secondoftwo}%
\providecommand \translation [1]{[#1]}%
\providecommand \BibitemOpen [0]{}%
\providecommand \bibitemStop [0]{}%
\providecommand \bibitemNoStop [0]{.\EOS\space}%
\providecommand \EOS [0]{\spacefactor3000\relax}%
\providecommand \BibitemShut  [1]{\csname bibitem#1\endcsname}%
\let\auto@bib@innerbib\@empty
\bibitem [{\citenamefont {Bell}(1964)}]{PhysicsPhysiqueFizika.1.195}%
  \BibitemOpen
  \bibfield  {author} {\bibinfo {author} {\bibfnamefont {J.~S.}\ \bibnamefont
  {Bell}},\ }\href {\doibase 10.1103/PhysicsPhysiqueFizika.1.195} {\bibfield
  {journal} {\bibinfo  {journal} {Physics Physique Fizika}\ }\textbf {\bibinfo
  {volume} {1}},\ \bibinfo {pages} {195} (\bibinfo {year} {1964})}\BibitemShut
  {NoStop}%
\bibitem [{\citenamefont {Leggett}\ and\ \citenamefont
  {Garg}(1985)}]{PhysRevLett.54.857}%
  \BibitemOpen
  \bibfield  {author} {\bibinfo {author} {\bibfnamefont {A.~J.}\ \bibnamefont
  {Leggett}}\ and\ \bibinfo {author} {\bibfnamefont {A.}~\bibnamefont {Garg}},\
  }\href {\doibase 10.1103/PhysRevLett.54.857} {\bibfield  {journal} {\bibinfo
  {journal} {Phys. Rev. Lett.}\ }\textbf {\bibinfo {volume} {54}},\ \bibinfo
  {pages} {857} (\bibinfo {year} {1985})}\BibitemShut {NoStop}%
\bibitem [{\citenamefont {Palacios-Laloy}\ \emph {et~al.}(2010)\citenamefont
  {Palacios-Laloy}, \citenamefont {Mallet}, \citenamefont {Nguyen},
  \citenamefont {Bertet}, \citenamefont {Vion}, \citenamefont {Esteve},\ and\
  \citenamefont {Korotkov}}]{Palacios-Laloy2010}%
  \BibitemOpen
  \bibfield  {author} {\bibinfo {author} {\bibfnamefont {A.}~\bibnamefont
  {Palacios-Laloy}}, \bibinfo {author} {\bibfnamefont {F.}~\bibnamefont
  {Mallet}}, \bibinfo {author} {\bibfnamefont {F.}~\bibnamefont {Nguyen}},
  \bibinfo {author} {\bibfnamefont {P.}~\bibnamefont {Bertet}}, \bibinfo
  {author} {\bibfnamefont {D.}~\bibnamefont {Vion}}, \bibinfo {author}
  {\bibfnamefont {D.}~\bibnamefont {Esteve}}, \ and\ \bibinfo {author}
  {\bibfnamefont {A.~N.}\ \bibnamefont {Korotkov}},\ }\href {\doibase
  10.1038/nphys1641} {\bibfield  {journal} {\bibinfo  {journal} {Nature
  Physics}\ }\textbf {\bibinfo {volume} {6}},\ \bibinfo {pages} {442} (\bibinfo
  {year} {2010})}\BibitemShut {NoStop}%
\bibitem [{\citenamefont {Katiyar}\ \emph {et~al.}(2013)\citenamefont
  {Katiyar}, \citenamefont {Shukla}, \citenamefont {Rao},\ and\ \citenamefont
  {Mahesh}}]{PhysRevA.87.052102}%
  \BibitemOpen
  \bibfield  {author} {\bibinfo {author} {\bibfnamefont {H.}~\bibnamefont
  {Katiyar}}, \bibinfo {author} {\bibfnamefont {A.}~\bibnamefont {Shukla}},
  \bibinfo {author} {\bibfnamefont {K.~R.~K.}\ \bibnamefont {Rao}}, \ and\
  \bibinfo {author} {\bibfnamefont {T.~S.}\ \bibnamefont {Mahesh}},\ }\href
  {\doibase 10.1103/PhysRevA.87.052102} {\bibfield  {journal} {\bibinfo
  {journal} {Phys. Rev. A}\ }\textbf {\bibinfo {volume} {87}},\ \bibinfo
  {pages} {052102} (\bibinfo {year} {2013})}\BibitemShut {NoStop}%
\bibitem [{\citenamefont {Athalye}\ \emph {et~al.}(2011)\citenamefont
  {Athalye}, \citenamefont {Roy},\ and\ \citenamefont
  {Mahesh}}]{PhysRevLett.107.130402}%
  \BibitemOpen
  \bibfield  {author} {\bibinfo {author} {\bibfnamefont {V.}~\bibnamefont
  {Athalye}}, \bibinfo {author} {\bibfnamefont {S.~S.}\ \bibnamefont {Roy}}, \
  and\ \bibinfo {author} {\bibfnamefont {T.~S.}\ \bibnamefont {Mahesh}},\
  }\href {\doibase 10.1103/PhysRevLett.107.130402} {\bibfield  {journal}
  {\bibinfo  {journal} {Phys. Rev. Lett.}\ }\textbf {\bibinfo {volume} {107}},\
  \bibinfo {pages} {130402} (\bibinfo {year} {2011})}\BibitemShut {NoStop}%
\bibitem [{\citenamefont {Emary}\ \emph {et~al.}(2012)\citenamefont {Emary},
  \citenamefont {Lambert},\ and\ \citenamefont {Nori}}]{PhysRevB.86.235447}%
  \BibitemOpen
  \bibfield  {author} {\bibinfo {author} {\bibfnamefont {C.}~\bibnamefont
  {Emary}}, \bibinfo {author} {\bibfnamefont {N.}~\bibnamefont {Lambert}}, \
  and\ \bibinfo {author} {\bibfnamefont {F.}~\bibnamefont {Nori}},\ }\href
  {\doibase 10.1103/PhysRevB.86.235447} {\bibfield  {journal} {\bibinfo
  {journal} {Phys. Rev. B}\ }\textbf {\bibinfo {volume} {86}},\ \bibinfo
  {pages} {235447} (\bibinfo {year} {2012})}\BibitemShut {NoStop}%
\bibitem [{\citenamefont {Williams}\ and\ \citenamefont
  {Jordan}(2008)}]{PhysRevLett.100.026804}%
  \BibitemOpen
  \bibfield  {author} {\bibinfo {author} {\bibfnamefont {N.~S.}\ \bibnamefont
  {Williams}}\ and\ \bibinfo {author} {\bibfnamefont {A.~N.}\ \bibnamefont
  {Jordan}},\ }\href {\doibase 10.1103/PhysRevLett.100.026804} {\bibfield
  {journal} {\bibinfo  {journal} {Phys. Rev. Lett.}\ }\textbf {\bibinfo
  {volume} {100}},\ \bibinfo {pages} {026804} (\bibinfo {year}
  {2008})}\BibitemShut {NoStop}%
\bibitem [{\citenamefont {George}\ \emph {et~al.}(2013)\citenamefont {George},
  \citenamefont {Robledo}, \citenamefont {Maroney}, \citenamefont {Blok},
  \citenamefont {Bernien}, \citenamefont {Markham}, \citenamefont {Twitchen},
  \citenamefont {Morton}, \citenamefont {Briggs},\ and\ \citenamefont
  {Hanson}}]{George3777}%
  \BibitemOpen
  \bibfield  {author} {\bibinfo {author} {\bibfnamefont {R.~E.}\ \bibnamefont
  {George}}, \bibinfo {author} {\bibfnamefont {L.~M.}\ \bibnamefont {Robledo}},
  \bibinfo {author} {\bibfnamefont {O.~J.~E.}\ \bibnamefont {Maroney}},
  \bibinfo {author} {\bibfnamefont {M.~S.}\ \bibnamefont {Blok}}, \bibinfo
  {author} {\bibfnamefont {H.}~\bibnamefont {Bernien}}, \bibinfo {author}
  {\bibfnamefont {M.~L.}\ \bibnamefont {Markham}}, \bibinfo {author}
  {\bibfnamefont {D.~J.}\ \bibnamefont {Twitchen}}, \bibinfo {author}
  {\bibfnamefont {J.~J.~L.}\ \bibnamefont {Morton}}, \bibinfo {author}
  {\bibfnamefont {G.~A.~D.}\ \bibnamefont {Briggs}}, \ and\ \bibinfo {author}
  {\bibfnamefont {R.}~\bibnamefont {Hanson}},\ }\href {\doibase
  10.1073/pnas.1208374110} {\bibfield  {journal} {\bibinfo  {journal}
  {Proceedings of the National Academy of Sciences}\ }\textbf {\bibinfo
  {volume} {110}},\ \bibinfo {pages} {3777} (\bibinfo {year}
  {2013})}\BibitemShut {NoStop}%
\bibitem [{\citenamefont {Robens}\ \emph {et~al.}(2015)\citenamefont {Robens},
  \citenamefont {Alt}, \citenamefont {Meschede}, \citenamefont {Emary},\ and\
  \citenamefont {Alberti}}]{PhysRevX.5.011003}%
  \BibitemOpen
  \bibfield  {author} {\bibinfo {author} {\bibfnamefont {C.}~\bibnamefont
  {Robens}}, \bibinfo {author} {\bibfnamefont {W.}~\bibnamefont {Alt}},
  \bibinfo {author} {\bibfnamefont {D.}~\bibnamefont {Meschede}}, \bibinfo
  {author} {\bibfnamefont {C.}~\bibnamefont {Emary}}, \ and\ \bibinfo {author}
  {\bibfnamefont {A.}~\bibnamefont {Alberti}},\ }\href {\doibase
  10.1103/PhysRevX.5.011003} {\bibfield  {journal} {\bibinfo  {journal} {Phys.
  Rev. X}\ }\textbf {\bibinfo {volume} {5}},\ \bibinfo {pages} {011003}
  (\bibinfo {year} {2015})}\BibitemShut {NoStop}%
\bibitem [{\citenamefont {Knee}\ \emph {et~al.}(2012)\citenamefont {Knee},
  \citenamefont {Simmons}, \citenamefont {Gauger}, \citenamefont {Morton},
  \citenamefont {Riemann}, \citenamefont {Abrosimov}, \citenamefont {Becker},
  \citenamefont {Pohl}, \citenamefont {Itoh}, \citenamefont {Thewalt},
  \citenamefont {Briggs},\ and\ \citenamefont {Benjamin}}]{Knee2012}%
  \BibitemOpen
  \bibfield  {author} {\bibinfo {author} {\bibfnamefont {G.~C.}\ \bibnamefont
  {Knee}}, \bibinfo {author} {\bibfnamefont {S.}~\bibnamefont {Simmons}},
  \bibinfo {author} {\bibfnamefont {E.~M.}\ \bibnamefont {Gauger}}, \bibinfo
  {author} {\bibfnamefont {J.~J.}\ \bibnamefont {Morton}}, \bibinfo {author}
  {\bibfnamefont {H.}~\bibnamefont {Riemann}}, \bibinfo {author} {\bibfnamefont
  {N.~V.}\ \bibnamefont {Abrosimov}}, \bibinfo {author} {\bibfnamefont
  {P.}~\bibnamefont {Becker}}, \bibinfo {author} {\bibfnamefont {H.-J.}\
  \bibnamefont {Pohl}}, \bibinfo {author} {\bibfnamefont {K.~M.}\ \bibnamefont
  {Itoh}}, \bibinfo {author} {\bibfnamefont {M.~L.}\ \bibnamefont {Thewalt}},
  \bibinfo {author} {\bibfnamefont {G.~A.~D.}\ \bibnamefont {Briggs}}, \ and\
  \bibinfo {author} {\bibfnamefont {S.~C.}\ \bibnamefont {Benjamin}},\ }\href
  {\doibase 10.1038/ncomms1614} {\bibfield  {journal} {\bibinfo  {journal}
  {Nature Communications}\ }\textbf {\bibinfo {volume} {3}},\ \bibinfo {pages}
  {606} (\bibinfo {year} {2012})}\BibitemShut {NoStop}%
\bibitem [{\citenamefont {Formaggio}\ \emph {et~al.}(2016)\citenamefont
  {Formaggio}, \citenamefont {Kaiser}, \citenamefont {Murskyj},\ and\
  \citenamefont {Weiss}}]{PhysRevLett.117.050402}%
  \BibitemOpen
  \bibfield  {author} {\bibinfo {author} {\bibfnamefont {J.~A.}\ \bibnamefont
  {Formaggio}}, \bibinfo {author} {\bibfnamefont {D.~I.}\ \bibnamefont
  {Kaiser}}, \bibinfo {author} {\bibfnamefont {M.~M.}\ \bibnamefont {Murskyj}},
  \ and\ \bibinfo {author} {\bibfnamefont {T.~E.}\ \bibnamefont {Weiss}},\
  }\href {\doibase 10.1103/PhysRevLett.117.050402} {\bibfield  {journal}
  {\bibinfo  {journal} {Phys. Rev. Lett.}\ }\textbf {\bibinfo {volume} {117}},\
  \bibinfo {pages} {050402} (\bibinfo {year} {2016})}\BibitemShut {NoStop}%
\bibitem [{\citenamefont {Xu}\ \emph {et~al.}(2011)\citenamefont {Xu},
  \citenamefont {Li}, \citenamefont {Zou},\ and\ \citenamefont {Guo}}]{Xu2011}%
  \BibitemOpen
  \bibfield  {author} {\bibinfo {author} {\bibfnamefont {J.-S.}\ \bibnamefont
  {Xu}}, \bibinfo {author} {\bibfnamefont {C.-F.}\ \bibnamefont {Li}}, \bibinfo
  {author} {\bibfnamefont {X.-B.}\ \bibnamefont {Zou}}, \ and\ \bibinfo
  {author} {\bibfnamefont {G.-C.}\ \bibnamefont {Guo}},\ }\href {\doibase
  10.1038/srep00101} {\bibfield  {journal} {\bibinfo  {journal} {Scientific
  Reports}\ }\textbf {\bibinfo {volume} {1}},\ \bibinfo {pages} {101} (\bibinfo
  {year} {2011})}\BibitemShut {NoStop}%
\bibitem [{\citenamefont {Dressel}\ \emph {et~al.}(2011)\citenamefont
  {Dressel}, \citenamefont {Broadbent}, \citenamefont {Howell},\ and\
  \citenamefont {Jordan}}]{PhysRevLett.106.040402}%
  \BibitemOpen
  \bibfield  {author} {\bibinfo {author} {\bibfnamefont {J.}~\bibnamefont
  {Dressel}}, \bibinfo {author} {\bibfnamefont {C.~J.}\ \bibnamefont
  {Broadbent}}, \bibinfo {author} {\bibfnamefont {J.~C.}\ \bibnamefont
  {Howell}}, \ and\ \bibinfo {author} {\bibfnamefont {A.~N.}\ \bibnamefont
  {Jordan}},\ }\href {\doibase 10.1103/PhysRevLett.106.040402} {\bibfield
  {journal} {\bibinfo  {journal} {Phys. Rev. Lett.}\ }\textbf {\bibinfo
  {volume} {106}},\ \bibinfo {pages} {040402} (\bibinfo {year}
  {2011})}\BibitemShut {NoStop}%
\bibitem [{\citenamefont {Goggin}\ \emph {et~al.}(2011)\citenamefont {Goggin},
  \citenamefont {Almeida}, \citenamefont {Barbieri}, \citenamefont {Lanyon},
  \citenamefont {O{\textquoteright}Brien}, \citenamefont {White},\ and\
  \citenamefont {Pryde}}]{Goggin1256}%
  \BibitemOpen
  \bibfield  {author} {\bibinfo {author} {\bibfnamefont {M.~E.}\ \bibnamefont
  {Goggin}}, \bibinfo {author} {\bibfnamefont {M.~P.}\ \bibnamefont {Almeida}},
  \bibinfo {author} {\bibfnamefont {M.}~\bibnamefont {Barbieri}}, \bibinfo
  {author} {\bibfnamefont {B.~P.}\ \bibnamefont {Lanyon}}, \bibinfo {author}
  {\bibfnamefont {J.~L.}\ \bibnamefont {O{\textquoteright}Brien}}, \bibinfo
  {author} {\bibfnamefont {A.~G.}\ \bibnamefont {White}}, \ and\ \bibinfo
  {author} {\bibfnamefont {G.~J.}\ \bibnamefont {Pryde}},\ }\href {\doibase
  10.1073/pnas.1005774108} {\bibfield  {journal} {\bibinfo  {journal}
  {Proceedings of the National Academy of Sciences}\ }\textbf {\bibinfo
  {volume} {108}},\ \bibinfo {pages} {1256} (\bibinfo {year}
  {2011})}\BibitemShut {NoStop}%
\bibitem [{\citenamefont {Suzuki}\ \emph {et~al.}(2012)\citenamefont {Suzuki},
  \citenamefont {Iinuma},\ and\ \citenamefont {Hofmann}}]{Suzuki_2012}%
  \BibitemOpen
  \bibfield  {author} {\bibinfo {author} {\bibfnamefont {Y.}~\bibnamefont
  {Suzuki}}, \bibinfo {author} {\bibfnamefont {M.}~\bibnamefont {Iinuma}}, \
  and\ \bibinfo {author} {\bibfnamefont {H.~F.}\ \bibnamefont {Hofmann}},\
  }\href {\doibase 10.1088/1367-2630/14/10/103022} {\bibfield  {journal}
  {\bibinfo  {journal} {New Journal of Physics}\ }\textbf {\bibinfo {volume}
  {14}},\ \bibinfo {pages} {103022} (\bibinfo {year} {2012})}\BibitemShut
  {NoStop}%
\bibitem [{\citenamefont {Wang}\ \emph {et~al.}(2018)\citenamefont {Wang},
  \citenamefont {Emary}, \citenamefont {Xu}, \citenamefont {Zhan},
  \citenamefont {Bian}, \citenamefont {Xiao},\ and\ \citenamefont
  {Xue}}]{PhysRevA.97.020101}%
  \BibitemOpen
  \bibfield  {author} {\bibinfo {author} {\bibfnamefont {K.}~\bibnamefont
  {Wang}}, \bibinfo {author} {\bibfnamefont {C.}~\bibnamefont {Emary}},
  \bibinfo {author} {\bibfnamefont {M.}~\bibnamefont {Xu}}, \bibinfo {author}
  {\bibfnamefont {X.}~\bibnamefont {Zhan}}, \bibinfo {author} {\bibfnamefont
  {Z.}~\bibnamefont {Bian}}, \bibinfo {author} {\bibfnamefont {L.}~\bibnamefont
  {Xiao}}, \ and\ \bibinfo {author} {\bibfnamefont {P.}~\bibnamefont {Xue}},\
  }\href {\doibase 10.1103/PhysRevA.97.020101} {\bibfield  {journal} {\bibinfo
  {journal} {Phys. Rev. A}\ }\textbf {\bibinfo {volume} {97}},\ \bibinfo
  {pages} {020101} (\bibinfo {year} {2018})}\BibitemShut {NoStop}%
\bibitem [{\citenamefont {Emary}\ \emph {et~al.}(2013)\citenamefont {Emary},
  \citenamefont {Lambert},\ and\ \citenamefont {Nori}}]{Emary_2013}%
  \BibitemOpen
  \bibfield  {author} {\bibinfo {author} {\bibfnamefont {C.}~\bibnamefont
  {Emary}}, \bibinfo {author} {\bibfnamefont {N.}~\bibnamefont {Lambert}}, \
  and\ \bibinfo {author} {\bibfnamefont {F.}~\bibnamefont {Nori}},\ }\href
  {\doibase 10.1088/0034-4885/77/1/016001} {\bibfield  {journal} {\bibinfo
  {journal} {Reports on Progress in Physics}\ }\textbf {\bibinfo {volume}
  {77}},\ \bibinfo {pages} {016001} (\bibinfo {year} {2013})}\BibitemShut
  {NoStop}%
\bibitem [{\citenamefont {Li}\ \emph {et~al.}(2012)\citenamefont {Li},
  \citenamefont {Lambert}, \citenamefont {Chen}, \citenamefont {Chen},\ and\
  \citenamefont {Nori}}]{Li2012}%
  \BibitemOpen
  \bibfield  {author} {\bibinfo {author} {\bibfnamefont {C.-M.}\ \bibnamefont
  {Li}}, \bibinfo {author} {\bibfnamefont {N.}~\bibnamefont {Lambert}},
  \bibinfo {author} {\bibfnamefont {Y.-N.}\ \bibnamefont {Chen}}, \bibinfo
  {author} {\bibfnamefont {G.-Y.}\ \bibnamefont {Chen}}, \ and\ \bibinfo
  {author} {\bibfnamefont {F.}~\bibnamefont {Nori}},\ }\href {\doibase
  10.1038/srep00885} {\bibfield  {journal} {\bibinfo  {journal} {Scientific
  Reports}\ }\textbf {\bibinfo {volume} {2}},\ \bibinfo {pages} {885} (\bibinfo
  {year} {2012})}\BibitemShut {NoStop}%
\bibitem [{\citenamefont {Kofler}\ and\ \citenamefont
  {Brukner}(2013)}]{PhysRevA.87.052115}%
  \BibitemOpen
  \bibfield  {author} {\bibinfo {author} {\bibfnamefont {J.}~\bibnamefont
  {Kofler}}\ and\ \bibinfo {author} {\bibfnamefont {C.}~\bibnamefont
  {Brukner}},\ }\href {\doibase 10.1103/PhysRevA.87.052115} {\bibfield
  {journal} {\bibinfo  {journal} {Phys. Rev. A}\ }\textbf {\bibinfo {volume}
  {87}},\ \bibinfo {pages} {052115} (\bibinfo {year} {2013})}\BibitemShut
  {NoStop}%
\bibitem [{\citenamefont {Clemente}\ and\ \citenamefont
  {Kofler}(2015)}]{PhysRevA.91.062103}%
  \BibitemOpen
  \bibfield  {author} {\bibinfo {author} {\bibfnamefont {L.}~\bibnamefont
  {Clemente}}\ and\ \bibinfo {author} {\bibfnamefont {J.}~\bibnamefont
  {Kofler}},\ }\href {\doibase 10.1103/PhysRevA.91.062103} {\bibfield
  {journal} {\bibinfo  {journal} {Phys. Rev. A}\ }\textbf {\bibinfo {volume}
  {91}},\ \bibinfo {pages} {062103} (\bibinfo {year} {2015})}\BibitemShut
  {NoStop}%
\bibitem [{\citenamefont {Clemente}\ and\ \citenamefont
  {Kofler}(2016)}]{PhysRevLett.116.150401}%
  \BibitemOpen
  \bibfield  {author} {\bibinfo {author} {\bibfnamefont {L.}~\bibnamefont
  {Clemente}}\ and\ \bibinfo {author} {\bibfnamefont {J.}~\bibnamefont
  {Kofler}},\ }\href {\doibase 10.1103/PhysRevLett.116.150401} {\bibfield
  {journal} {\bibinfo  {journal} {Phys. Rev. Lett.}\ }\textbf {\bibinfo
  {volume} {116}},\ \bibinfo {pages} {150401} (\bibinfo {year}
  {2016})}\BibitemShut {NoStop}%
\bibitem [{\citenamefont {Knee}\ \emph {et~al.}(2016)\citenamefont {Knee},
  \citenamefont {Kakuyanagi}, \citenamefont {Yeh}, \citenamefont {Matsuzaki},
  \citenamefont {Toida}, \citenamefont {Yamaguchi}, \citenamefont {Saito},
  \citenamefont {Leggett},\ and\ \citenamefont {Munro}}]{Knee2016}%
  \BibitemOpen
  \bibfield  {author} {\bibinfo {author} {\bibfnamefont {G.~C.}\ \bibnamefont
  {Knee}}, \bibinfo {author} {\bibfnamefont {K.}~\bibnamefont {Kakuyanagi}},
  \bibinfo {author} {\bibfnamefont {M.-C.}\ \bibnamefont {Yeh}}, \bibinfo
  {author} {\bibfnamefont {Y.}~\bibnamefont {Matsuzaki}}, \bibinfo {author}
  {\bibfnamefont {H.}~\bibnamefont {Toida}}, \bibinfo {author} {\bibfnamefont
  {H.}~\bibnamefont {Yamaguchi}}, \bibinfo {author} {\bibfnamefont
  {S.}~\bibnamefont {Saito}}, \bibinfo {author} {\bibfnamefont {A.~J.}\
  \bibnamefont {Leggett}}, \ and\ \bibinfo {author} {\bibfnamefont {W.~J.}\
  \bibnamefont {Munro}},\ }\href {\doibase 10.1038/ncomms13253} {\bibfield
  {journal} {\bibinfo  {journal} {Nature Communications}\ }\textbf {\bibinfo
  {volume} {7}},\ \bibinfo {pages} {13253} (\bibinfo {year}
  {2016})}\BibitemShut {NoStop}%
\bibitem [{\citenamefont {Saha}\ \emph {et~al.}(2015)\citenamefont {Saha},
  \citenamefont {Mal}, \citenamefont {Panigrahi},\ and\ \citenamefont
  {Home}}]{PhysRevA.91.032117}%
  \BibitemOpen
  \bibfield  {author} {\bibinfo {author} {\bibfnamefont {D.}~\bibnamefont
  {Saha}}, \bibinfo {author} {\bibfnamefont {S.}~\bibnamefont {Mal}}, \bibinfo
  {author} {\bibfnamefont {P.~K.}\ \bibnamefont {Panigrahi}}, \ and\ \bibinfo
  {author} {\bibfnamefont {D.}~\bibnamefont {Home}},\ }\href {\doibase
  10.1103/PhysRevA.91.032117} {\bibfield  {journal} {\bibinfo  {journal} {Phys.
  Rev. A}\ }\textbf {\bibinfo {volume} {91}},\ \bibinfo {pages} {032117}
  (\bibinfo {year} {2015})}\BibitemShut {NoStop}%
\bibitem [{\citenamefont {Wigner}(1970)}]{wignerajp}%
  \BibitemOpen
  \bibfield  {author} {\bibinfo {author} {\bibfnamefont {E.~P.}\ \bibnamefont
  {Wigner}},\ }\href {\doibase 10.1119/1.1976526} {\bibfield  {journal}
  {\bibinfo  {journal} {American Journal of Physics}\ }\textbf {\bibinfo
  {volume} {38}},\ \bibinfo {pages} {1005} (\bibinfo {year}
  {1970})}\BibitemShut {NoStop}%
\bibitem [{\citenamefont {Leggett}(2008)}]{Leggett_2008}%
  \BibitemOpen
  \bibfield  {author} {\bibinfo {author} {\bibfnamefont {A.~J.}\ \bibnamefont
  {Leggett}},\ }\href {\doibase 10.1088/0034-4885/71/2/022001} {\bibfield
  {journal} {\bibinfo  {journal} {Reports on Progress in Physics}\ }\textbf
  {\bibinfo {volume} {71}},\ \bibinfo {pages} {022001} (\bibinfo {year}
  {2008})}\BibitemShut {NoStop}%
\bibitem [{\citenamefont {Wilde}\ and\ \citenamefont
  {Mizel}(2012)}]{Wilde2012}%
  \BibitemOpen
  \bibfield  {author} {\bibinfo {author} {\bibfnamefont {M.~M.}\ \bibnamefont
  {Wilde}}\ and\ \bibinfo {author} {\bibfnamefont {A.}~\bibnamefont {Mizel}},\
  }\href {\doibase 10.1007/s10701-011-9598-4} {\bibfield  {journal} {\bibinfo
  {journal} {Foundations of Physics}\ }\textbf {\bibinfo {volume} {42}},\
  \bibinfo {pages} {256} (\bibinfo {year} {2012})}\BibitemShut {NoStop}%
\bibitem [{\citenamefont {Ku}\ \emph {et~al.}(2020)\citenamefont {Ku},
  \citenamefont {Lambert}, \citenamefont {Chan}, \citenamefont {Emary},
  \citenamefont {Chen},\ and\ \citenamefont {Nori}}]{Ku2020}%
  \BibitemOpen
  \bibfield  {author} {\bibinfo {author} {\bibfnamefont {H.-Y.}\ \bibnamefont
  {Ku}}, \bibinfo {author} {\bibfnamefont {N.}~\bibnamefont {Lambert}},
  \bibinfo {author} {\bibfnamefont {F.-J.}\ \bibnamefont {Chan}}, \bibinfo
  {author} {\bibfnamefont {C.}~\bibnamefont {Emary}}, \bibinfo {author}
  {\bibfnamefont {Y.-N.}\ \bibnamefont {Chen}}, \ and\ \bibinfo {author}
  {\bibfnamefont {F.}~\bibnamefont {Nori}},\ }\href {\doibase
  10.1038/s41534-020-00321-x} {\bibfield  {journal} {\bibinfo  {journal} {npj
  Quantum Information}\ }\textbf {\bibinfo {volume} {6}},\ \bibinfo {pages}
  {98} (\bibinfo {year} {2020})}\BibitemShut {NoStop}%
\bibitem [{\citenamefont {Emary}(2017)}]{PhysRevA.96.042102}%
  \BibitemOpen
  \bibfield  {author} {\bibinfo {author} {\bibfnamefont {C.}~\bibnamefont
  {Emary}},\ }\href {\doibase 10.1103/PhysRevA.96.042102} {\bibfield  {journal}
  {\bibinfo  {journal} {Phys. Rev. A}\ }\textbf {\bibinfo {volume} {96}},\
  \bibinfo {pages} {042102} (\bibinfo {year} {2017})}\BibitemShut {NoStop}%
\bibitem [{\citenamefont {Pan}(2020)}]{PhysRevA.102.032206}%
  \BibitemOpen
  \bibfield  {author} {\bibinfo {author} {\bibfnamefont {A.~K.}\ \bibnamefont
  {Pan}},\ }\href {\doibase 10.1103/PhysRevA.102.032206} {\bibfield  {journal}
  {\bibinfo  {journal} {Phys. Rev. A}\ }\textbf {\bibinfo {volume} {102}},\
  \bibinfo {pages} {032206} (\bibinfo {year} {2020})}\BibitemShut {NoStop}%
\bibitem [{\citenamefont {Halliwell}\ \emph {et~al.}(2021)\citenamefont
  {Halliwell}, \citenamefont {Bhatnagar}, \citenamefont {Ireland},
  \citenamefont {Nadeem},\ and\ \citenamefont
  {Wimalaweera}}]{halliwell2020leggettgarg}%
  \BibitemOpen
  \bibfield  {author} {\bibinfo {author} {\bibfnamefont {J.~J.}\ \bibnamefont
  {Halliwell}}, \bibinfo {author} {\bibfnamefont {A.}~\bibnamefont
  {Bhatnagar}}, \bibinfo {author} {\bibfnamefont {E.}~\bibnamefont {Ireland}},
  \bibinfo {author} {\bibfnamefont {H.}~\bibnamefont {Nadeem}}, \ and\ \bibinfo
  {author} {\bibfnamefont {V.}~\bibnamefont {Wimalaweera}},\ }\href {\doibase
  10.1103/PhysRevA.103.032218} {\bibfield  {journal} {\bibinfo  {journal}
  {Phys. Rev. A}\ }\textbf {\bibinfo {volume} {103}},\ \bibinfo {pages}
  {032218} (\bibinfo {year} {2021})}\BibitemShut {NoStop}%
\bibitem [{\citenamefont {Leggett}(2002)}]{Leggett_2002}%
  \BibitemOpen
  \bibfield  {author} {\bibinfo {author} {\bibfnamefont {A.~J.}\ \bibnamefont
  {Leggett}},\ }\href {\doibase 10.1088/0953-8984/14/15/201} {\bibfield
  {journal} {\bibinfo  {journal} {Journal of Physics: Condensed Matter}\
  }\textbf {\bibinfo {volume} {14}},\ \bibinfo {pages} {R415} (\bibinfo {year}
  {2002})}\BibitemShut {NoStop}%
\bibitem [{\citenamefont {Milburn}(1991)}]{PhysRevA.44.5401}%
  \BibitemOpen
  \bibfield  {author} {\bibinfo {author} {\bibfnamefont {G.~J.}\ \bibnamefont
  {Milburn}},\ }\href {\doibase 10.1103/PhysRevA.44.5401} {\bibfield  {journal}
  {\bibinfo  {journal} {Phys. Rev. A}\ }\textbf {\bibinfo {volume} {44}},\
  \bibinfo {pages} {5401} (\bibinfo {year} {1991})}\BibitemShut {NoStop}%
\bibitem [{\citenamefont {Garg}\ and\ \citenamefont
  {Mermin}(1987)}]{PhysRevD.35.3831}%
  \BibitemOpen
  \bibfield  {author} {\bibinfo {author} {\bibfnamefont {A.}~\bibnamefont
  {Garg}}\ and\ \bibinfo {author} {\bibfnamefont {N.~D.}\ \bibnamefont
  {Mermin}},\ }\href {\doibase 10.1103/PhysRevD.35.3831} {\bibfield  {journal}
  {\bibinfo  {journal} {Phys. Rev. D}\ }\textbf {\bibinfo {volume} {35}},\
  \bibinfo {pages} {3831} (\bibinfo {year} {1987})}\BibitemShut {NoStop}%
\bibitem [{\citenamefont {Larsson}(1998)}]{PhysRevA.57.3304}%
  \BibitemOpen
  \bibfield  {author} {\bibinfo {author} {\bibfnamefont {J.-A.}\ \bibnamefont
  {Larsson}},\ }\href {\doibase 10.1103/PhysRevA.57.3304} {\bibfield  {journal}
  {\bibinfo  {journal} {Phys. Rev. A}\ }\textbf {\bibinfo {volume} {57}},\
  \bibinfo {pages} {3304} (\bibinfo {year} {1998})}\BibitemShut {NoStop}%
\bibitem [{\citenamefont {Larsson}\ and\ \citenamefont
  {Gill}(2004)}]{Larsson_2004}%
  \BibitemOpen
  \bibfield  {author} {\bibinfo {author} {\bibfnamefont {J.-{\AA}.}\
  \bibnamefont {Larsson}}\ and\ \bibinfo {author} {\bibfnamefont {R.~D.}\
  \bibnamefont {Gill}},\ }\href {\doibase 10.1209/epl/i2004-10124-7} {\bibfield
   {journal} {\bibinfo  {journal} {Europhysics Letters ({EPL})}\ }\textbf
  {\bibinfo {volume} {67}},\ \bibinfo {pages} {707} (\bibinfo {year}
  {2004})}\BibitemShut {NoStop}%
\bibitem [{\citenamefont {Mansuripur}(2009)}]{mansuripur_2009}%
  \BibitemOpen
  \bibfield  {author} {\bibinfo {author} {\bibfnamefont {M.}~\bibnamefont
  {Mansuripur}},\ }\enquote {\bibinfo {title} {The sagnac interferometer},}\
  in\ \href {\doibase 10.1017/CBO9780511803796.017} {\emph {\bibinfo
  {booktitle} {Classical Optics and its Applications}}}\ (\bibinfo  {publisher}
  {Cambridge University Press},\ \bibinfo {year} {2009})\ p.\ \bibinfo {pages}
  {182–196},\ \bibinfo {edition} {2nd}\ ed.\BibitemShut {Stop}%
\bibitem [{\citenamefont {Sahoo}\ \emph {et~al.}(2020)\citenamefont {Sahoo},
  \citenamefont {Chakraborti}, \citenamefont {Pati},\ and\ \citenamefont
  {Sinha}}]{PhysRevLett.125.123601}%
  \BibitemOpen
  \bibfield  {author} {\bibinfo {author} {\bibfnamefont {S.~N.}\ \bibnamefont
  {Sahoo}}, \bibinfo {author} {\bibfnamefont {S.}~\bibnamefont {Chakraborti}},
  \bibinfo {author} {\bibfnamefont {A.~K.}\ \bibnamefont {Pati}}, \ and\
  \bibinfo {author} {\bibfnamefont {U.}~\bibnamefont {Sinha}},\ }\href
  {\doibase 10.1103/PhysRevLett.125.123601} {\bibfield  {journal} {\bibinfo
  {journal} {Phys. Rev. Lett.}\ }\textbf {\bibinfo {volume} {125}},\ \bibinfo
  {pages} {123601} (\bibinfo {year} {2020})}\BibitemShut {NoStop}%
\bibitem [{\citenamefont {Cox}(1979)}]{Cox_1979}%
  \BibitemOpen
  \bibfield  {author} {\bibinfo {author} {\bibfnamefont {M.~E.}\ \bibnamefont
  {Cox}},\ }\href {\doibase 10.1088/0031-9120/14/1/009} {\bibfield  {journal}
  {\bibinfo  {journal} {Physics Education}\ }\textbf {\bibinfo {volume} {14}},\
  \bibinfo {pages} {56} (\bibinfo {year} {1979})}\BibitemShut {NoStop}%
\bibitem [{\citenamefont {Sadana}\ \emph {et~al.}(2019)\citenamefont {Sadana},
  \citenamefont {Ghosh}, \citenamefont {Joarder}, \citenamefont {Lakshmi},
  \citenamefont {Sanders},\ and\ \citenamefont {Sinha}}]{PhysRevA.100.013839}%
  \BibitemOpen
  \bibfield  {author} {\bibinfo {author} {\bibfnamefont {S.}~\bibnamefont
  {Sadana}}, \bibinfo {author} {\bibfnamefont {D.}~\bibnamefont {Ghosh}},
  \bibinfo {author} {\bibfnamefont {K.}~\bibnamefont {Joarder}}, \bibinfo
  {author} {\bibfnamefont {A.~N.}\ \bibnamefont {Lakshmi}}, \bibinfo {author}
  {\bibfnamefont {B.~C.}\ \bibnamefont {Sanders}}, \ and\ \bibinfo {author}
  {\bibfnamefont {U.}~\bibnamefont {Sinha}},\ }\href {\doibase
  10.1103/PhysRevA.100.013839} {\bibfield  {journal} {\bibinfo  {journal}
  {Phys. Rev. A}\ }\textbf {\bibinfo {volume} {100}},\ \bibinfo {pages}
  {013839} (\bibinfo {year} {2019})}\BibitemShut {NoStop}%
\bibitem [{\citenamefont {Chatterjee}\ \emph {et~al.}(2020)\citenamefont
  {Chatterjee}, \citenamefont {Joarder}, \citenamefont {Chatterjee},
  \citenamefont {Sanders},\ and\ \citenamefont
  {Sinha}}]{PhysRevApplied.14.024036}%
  \BibitemOpen
  \bibfield  {author} {\bibinfo {author} {\bibfnamefont {R.}~\bibnamefont
  {Chatterjee}}, \bibinfo {author} {\bibfnamefont {K.}~\bibnamefont {Joarder}},
  \bibinfo {author} {\bibfnamefont {S.}~\bibnamefont {Chatterjee}}, \bibinfo
  {author} {\bibfnamefont {B.~C.}\ \bibnamefont {Sanders}}, \ and\ \bibinfo
  {author} {\bibfnamefont {U.}~\bibnamefont {Sinha}},\ }\href {\doibase
  10.1103/PhysRevApplied.14.024036} {\bibfield  {journal} {\bibinfo  {journal}
  {Phys. Rev. Applied}\ }\textbf {\bibinfo {volume} {14}},\ \bibinfo {pages}
  {024036} (\bibinfo {year} {2020})}\BibitemShut {NoStop}%
\bibitem [{\citenamefont {Giustina}\ \emph {et~al.}(2015)\citenamefont
  {Giustina}, \citenamefont {Versteegh}, \citenamefont {Wengerowsky},
  \citenamefont {Handsteiner}, \citenamefont {Hochrainer}, \citenamefont
  {Phelan}, \citenamefont {Steinlechner}, \citenamefont {Kofler}, \citenamefont
  {Larsson}, \citenamefont {Abell\'an}, \citenamefont {Amaya}, \citenamefont
  {Pruneri}, \citenamefont {Mitchell}, \citenamefont {Beyer}, \citenamefont
  {Gerrits}, \citenamefont {Lita}, \citenamefont {Shalm}, \citenamefont {Nam},
  \citenamefont {Scheidl}, \citenamefont {Ursin}, \citenamefont {Wittmann},\
  and\ \citenamefont {Zeilinger}}]{PhysRevLett.115.250401}%
  \BibitemOpen
  \bibfield  {author} {\bibinfo {author} {\bibfnamefont {M.}~\bibnamefont
  {Giustina}}, \bibinfo {author} {\bibfnamefont {M.~A.~M.}\ \bibnamefont
  {Versteegh}}, \bibinfo {author} {\bibfnamefont {S.}~\bibnamefont
  {Wengerowsky}}, \bibinfo {author} {\bibfnamefont {J.}~\bibnamefont
  {Handsteiner}}, \bibinfo {author} {\bibfnamefont {A.}~\bibnamefont
  {Hochrainer}}, \bibinfo {author} {\bibfnamefont {K.}~\bibnamefont {Phelan}},
  \bibinfo {author} {\bibfnamefont {F.}~\bibnamefont {Steinlechner}}, \bibinfo
  {author} {\bibfnamefont {J.}~\bibnamefont {Kofler}}, \bibinfo {author}
  {\bibfnamefont {J.-A.}\ \bibnamefont {Larsson}}, \bibinfo {author}
  {\bibfnamefont {C.}~\bibnamefont {Abell\'an}}, \bibinfo {author}
  {\bibfnamefont {W.}~\bibnamefont {Amaya}}, \bibinfo {author} {\bibfnamefont
  {V.}~\bibnamefont {Pruneri}}, \bibinfo {author} {\bibfnamefont {M.~W.}\
  \bibnamefont {Mitchell}}, \bibinfo {author} {\bibfnamefont {J.}~\bibnamefont
  {Beyer}}, \bibinfo {author} {\bibfnamefont {T.}~\bibnamefont {Gerrits}},
  \bibinfo {author} {\bibfnamefont {A.~E.}\ \bibnamefont {Lita}}, \bibinfo
  {author} {\bibfnamefont {L.~K.}\ \bibnamefont {Shalm}}, \bibinfo {author}
  {\bibfnamefont {S.~W.}\ \bibnamefont {Nam}}, \bibinfo {author} {\bibfnamefont
  {T.}~\bibnamefont {Scheidl}}, \bibinfo {author} {\bibfnamefont
  {R.}~\bibnamefont {Ursin}}, \bibinfo {author} {\bibfnamefont
  {B.}~\bibnamefont {Wittmann}}, \ and\ \bibinfo {author} {\bibfnamefont
  {A.}~\bibnamefont {Zeilinger}},\ }\href {\doibase
  10.1103/PhysRevLett.115.250401} {\bibfield  {journal} {\bibinfo  {journal}
  {Phys. Rev. Lett.}\ }\textbf {\bibinfo {volume} {115}},\ \bibinfo {pages}
  {250401} (\bibinfo {year} {2015})}\BibitemShut {NoStop}%
\bibitem [{\citenamefont {Shalm}\ \emph {et~al.}(2015)\citenamefont {Shalm},
  \citenamefont {Meyer-Scott}, \citenamefont {Christensen}, \citenamefont
  {Bierhorst}, \citenamefont {Wayne}, \citenamefont {Stevens}, \citenamefont
  {Gerrits}, \citenamefont {Glancy}, \citenamefont {Hamel}, \citenamefont
  {Allman}, \citenamefont {Coakley}, \citenamefont {Dyer}, \citenamefont
  {Hodge}, \citenamefont {Lita}, \citenamefont {Verma}, \citenamefont
  {Lambrocco}, \citenamefont {Tortorici}, \citenamefont {Migdall},
  \citenamefont {Zhang}, \citenamefont {Kumor}, \citenamefont {Farr},
  \citenamefont {Marsili}, \citenamefont {Shaw}, \citenamefont {Stern},
  \citenamefont {Abell\'an}, \citenamefont {Amaya}, \citenamefont {Pruneri},
  \citenamefont {Jennewein}, \citenamefont {Mitchell}, \citenamefont {Kwiat},
  \citenamefont {Bienfang}, \citenamefont {Mirin}, \citenamefont {Knill},\ and\
  \citenamefont {Nam}}]{PhysRevLett.115.250402}%
  \BibitemOpen
  \bibfield  {author} {\bibinfo {author} {\bibfnamefont {L.~K.}\ \bibnamefont
  {Shalm}}, \bibinfo {author} {\bibfnamefont {E.}~\bibnamefont {Meyer-Scott}},
  \bibinfo {author} {\bibfnamefont {B.~G.}\ \bibnamefont {Christensen}},
  \bibinfo {author} {\bibfnamefont {P.}~\bibnamefont {Bierhorst}}, \bibinfo
  {author} {\bibfnamefont {M.~A.}\ \bibnamefont {Wayne}}, \bibinfo {author}
  {\bibfnamefont {M.~J.}\ \bibnamefont {Stevens}}, \bibinfo {author}
  {\bibfnamefont {T.}~\bibnamefont {Gerrits}}, \bibinfo {author} {\bibfnamefont
  {S.}~\bibnamefont {Glancy}}, \bibinfo {author} {\bibfnamefont {D.~R.}\
  \bibnamefont {Hamel}}, \bibinfo {author} {\bibfnamefont {M.~S.}\ \bibnamefont
  {Allman}}, \bibinfo {author} {\bibfnamefont {K.~J.}\ \bibnamefont {Coakley}},
  \bibinfo {author} {\bibfnamefont {S.~D.}\ \bibnamefont {Dyer}}, \bibinfo
  {author} {\bibfnamefont {C.}~\bibnamefont {Hodge}}, \bibinfo {author}
  {\bibfnamefont {A.~E.}\ \bibnamefont {Lita}}, \bibinfo {author}
  {\bibfnamefont {V.~B.}\ \bibnamefont {Verma}}, \bibinfo {author}
  {\bibfnamefont {C.}~\bibnamefont {Lambrocco}}, \bibinfo {author}
  {\bibfnamefont {E.}~\bibnamefont {Tortorici}}, \bibinfo {author}
  {\bibfnamefont {A.~L.}\ \bibnamefont {Migdall}}, \bibinfo {author}
  {\bibfnamefont {Y.}~\bibnamefont {Zhang}}, \bibinfo {author} {\bibfnamefont
  {D.~R.}\ \bibnamefont {Kumor}}, \bibinfo {author} {\bibfnamefont {W.~H.}\
  \bibnamefont {Farr}}, \bibinfo {author} {\bibfnamefont {F.}~\bibnamefont
  {Marsili}}, \bibinfo {author} {\bibfnamefont {M.~D.}\ \bibnamefont {Shaw}},
  \bibinfo {author} {\bibfnamefont {J.~A.}\ \bibnamefont {Stern}}, \bibinfo
  {author} {\bibfnamefont {C.}~\bibnamefont {Abell\'an}}, \bibinfo {author}
  {\bibfnamefont {W.}~\bibnamefont {Amaya}}, \bibinfo {author} {\bibfnamefont
  {V.}~\bibnamefont {Pruneri}}, \bibinfo {author} {\bibfnamefont
  {T.}~\bibnamefont {Jennewein}}, \bibinfo {author} {\bibfnamefont {M.~W.}\
  \bibnamefont {Mitchell}}, \bibinfo {author} {\bibfnamefont {P.~G.}\
  \bibnamefont {Kwiat}}, \bibinfo {author} {\bibfnamefont {J.~C.}\ \bibnamefont
  {Bienfang}}, \bibinfo {author} {\bibfnamefont {R.~P.}\ \bibnamefont {Mirin}},
  \bibinfo {author} {\bibfnamefont {E.}~\bibnamefont {Knill}}, \ and\ \bibinfo
  {author} {\bibfnamefont {S.~W.}\ \bibnamefont {Nam}},\ }\href {\doibase
  10.1103/PhysRevLett.115.250402} {\bibfield  {journal} {\bibinfo  {journal}
  {Phys. Rev. Lett.}\ }\textbf {\bibinfo {volume} {115}},\ \bibinfo {pages}
  {250402} (\bibinfo {year} {2015})}\BibitemShut {NoStop}%
\bibitem [{\citenamefont {Morikoshi}(2006)}]{PhysRevA.73.052308}%
  \BibitemOpen
  \bibfield  {author} {\bibinfo {author} {\bibfnamefont {F.}~\bibnamefont
  {Morikoshi}},\ }\href {\doibase 10.1103/PhysRevA.73.052308} {\bibfield
  {journal} {\bibinfo  {journal} {Phys. Rev. A}\ }\textbf {\bibinfo {volume}
  {73}},\ \bibinfo {pages} {052308} (\bibinfo {year} {2006})}\BibitemShut
  {NoStop}%
\bibitem [{\citenamefont {{Shenoy H.}}\ \emph {et~al.}(2017)\citenamefont
  {{Shenoy H.}}, \citenamefont {Aravinda}, \citenamefont {Srikanth},\ and\
  \citenamefont {Home}}]{SHENOYH20172478}%
  \BibitemOpen
  \bibfield  {author} {\bibinfo {author} {\bibfnamefont {A.}~\bibnamefont
  {{Shenoy H.}}}, \bibinfo {author} {\bibfnamefont {S.}~\bibnamefont
  {Aravinda}}, \bibinfo {author} {\bibfnamefont {R.}~\bibnamefont {Srikanth}},
  \ and\ \bibinfo {author} {\bibfnamefont {D.}~\bibnamefont {Home}},\ }\href
  {\doibase https://doi.org/10.1016/j.physleta.2017.05.053} {\bibfield
  {journal} {\bibinfo  {journal} {Physics Letters A}\ }\textbf {\bibinfo
  {volume} {381}},\ \bibinfo {pages} {2478 } (\bibinfo {year}
  {2017})}\BibitemShut {NoStop}%
\bibitem [{\citenamefont {Brukner}\ \emph {et~al.}(2004)\citenamefont
  {Brukner}, \citenamefont {Taylor}, \citenamefont {Cheung},\ and\
  \citenamefont {Vedral}}]{arXiv:quant-ph/0402127}%
  \BibitemOpen
  \bibfield  {author} {\bibinfo {author} {\bibfnamefont {C.}~\bibnamefont
  {Brukner}}, \bibinfo {author} {\bibfnamefont {S.}~\bibnamefont {Taylor}},
  \bibinfo {author} {\bibfnamefont {S.}~\bibnamefont {Cheung}}, \ and\ \bibinfo
  {author} {\bibfnamefont {V.}~\bibnamefont {Vedral}},\ }\href@noop {}
  {\enquote {\bibinfo {title} {Quantum entanglement in time},}\ } (\bibinfo
  {year} {2004}),\ \Eprint {http://arxiv.org/abs/0402127} {arXiv:0402127
  [quant-ph]} \BibitemShut {NoStop}%
\bibitem [{\citenamefont {Barrett}\ \emph {et~al.}(2002)\citenamefont
  {Barrett}, \citenamefont {Collins}, \citenamefont {Hardy}, \citenamefont
  {Kent},\ and\ \citenamefont {Popescu}}]{PhysRevA.66.042111}%
  \BibitemOpen
  \bibfield  {author} {\bibinfo {author} {\bibfnamefont {J.}~\bibnamefont
  {Barrett}}, \bibinfo {author} {\bibfnamefont {D.}~\bibnamefont {Collins}},
  \bibinfo {author} {\bibfnamefont {L.}~\bibnamefont {Hardy}}, \bibinfo
  {author} {\bibfnamefont {A.}~\bibnamefont {Kent}}, \ and\ \bibinfo {author}
  {\bibfnamefont {S.}~\bibnamefont {Popescu}},\ }\href {\doibase
  10.1103/PhysRevA.66.042111} {\bibfield  {journal} {\bibinfo  {journal} {Phys.
  Rev. A}\ }\textbf {\bibinfo {volume} {66}},\ \bibinfo {pages} {042111}
  (\bibinfo {year} {2002})}\BibitemShut {NoStop}%
\bibitem [{\citenamefont {Halliwell}(2019)}]{PhysRevA.99.022119}%
  \BibitemOpen
  \bibfield  {author} {\bibinfo {author} {\bibfnamefont {J.~J.}\ \bibnamefont
  {Halliwell}},\ }\href {\doibase 10.1103/PhysRevA.99.022119} {\bibfield
  {journal} {\bibinfo  {journal} {Phys. Rev. A}\ }\textbf {\bibinfo {volume}
  {99}},\ \bibinfo {pages} {022119} (\bibinfo {year} {2019})}\BibitemShut
  {NoStop}%
\bibitem [{\citenamefont {Halliwell}(2017)}]{PhysRevA.96.012121}%
  \BibitemOpen
  \bibfield  {author} {\bibinfo {author} {\bibfnamefont {J.~J.}\ \bibnamefont
  {Halliwell}},\ }\href {\doibase 10.1103/PhysRevA.96.012121} {\bibfield
  {journal} {\bibinfo  {journal} {Phys. Rev. A}\ }\textbf {\bibinfo {volume}
  {96}},\ \bibinfo {pages} {012121} (\bibinfo {year} {2017})}\BibitemShut
  {NoStop}%
\end{thebibliography}%

\clearpage

\onecolumngrid

\appendix

\section{Closing the detection efficiency loophole}
\label{appendix:a}

Let us first recall the LGI and WLGI expressions considered in the main text
\begin{equation}
\label{eq:lgiApx}
 LGI:\,\langle Q_{t_{1}} Q_{t_{2}}\rangle + \langle Q_{t_{2}} Q_{t_{3}}\rangle - \langle Q_{t_{1}} Q_{t_{3}}\rangle   ,
\end{equation}
\begin{equation}
\label{eq:wlgiApx}
 WLGI:\,P_{t_{1},t_{3}}(-,+)-P_{t_{1},t_{2}}(-,+)-P_{t_{2},t_{3}}(-,+)  .
\end{equation}

\begin{figure}[!ht]
\centering
\includegraphics[scale=0.3]{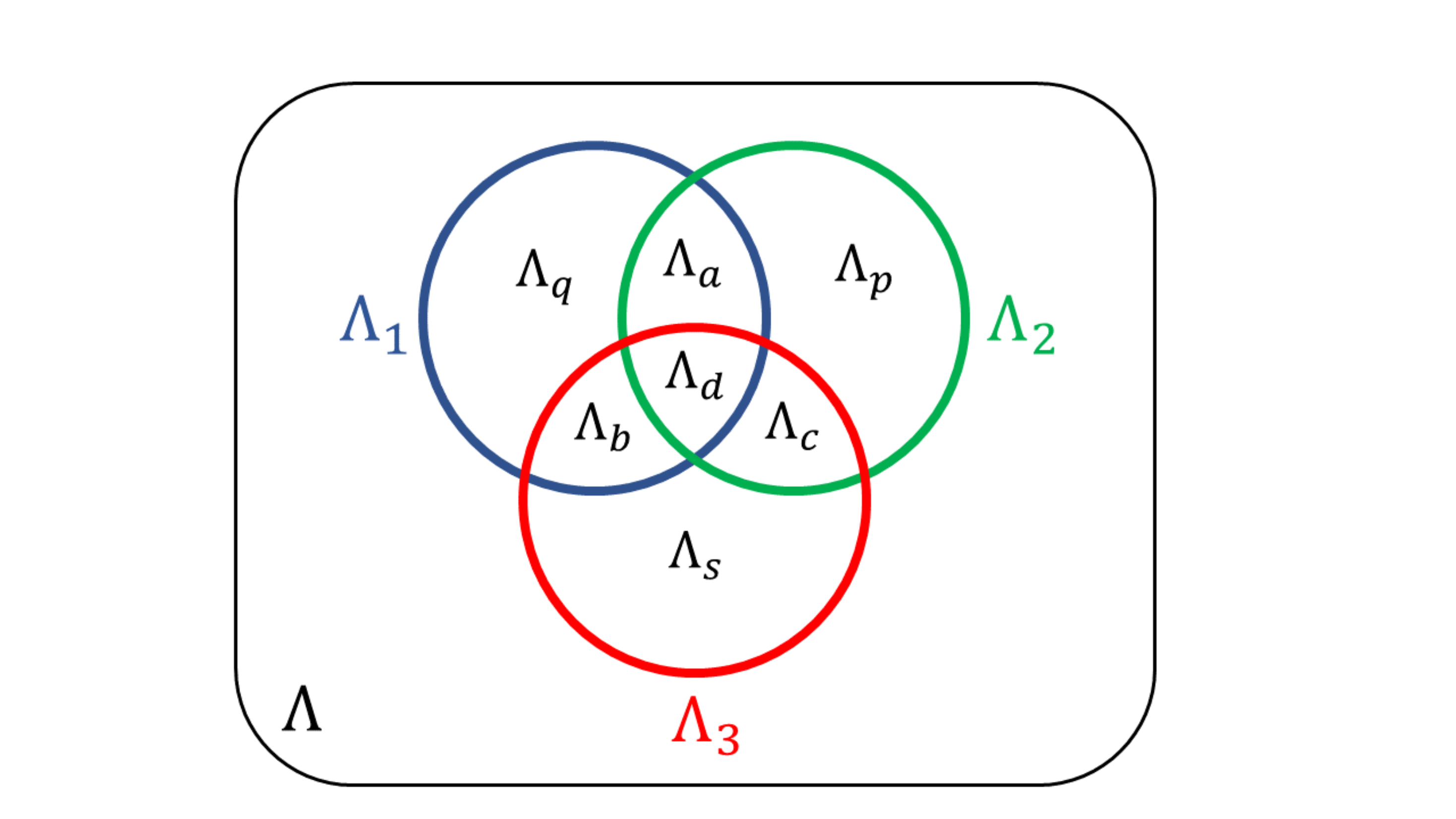}
\caption{The overall `hidden-variable' state space is $\Lambda$. Its subspaces $\Lambda_1,\Lambda_2,\Lambda_3$ for which the photon is detected at $t_1,t_2,t_3$ are respectively represented by blue, green and red circles. All the detectors are assumed to have the same detection efficiency. Different subspaces within $\Lambda_1 \cup \Lambda_2 \cup \Lambda_3$ are denoted by the symbols $q,p,s,a,b,c,d$. }
\label{fig:ptf}
\end{figure}

In an experiment, usually all the events for which the photon is not detected by the inefficient detectors used are rejected and it is assumed that the collected data is a faithful representation of the data that would have been recorded for perfect detectors. This is known as the fair-sampling assumption. If the efficiency of the detectors is sufficiently low, then exploiting this fact it is possible to reproduce the quantum mechanical violation of a macrorealist inequality using a non-invasive realist model. Here, for addressing this loophole, we proceed as follows. We assume that the fate of the photon, whether it will be detected or not, is predefined by its `hidden-variable' state $\lambda$ spanned over the `hidden-variable' state space $\Lambda$, and it depends on the measurement time $t_i \ (i=1,2,3)$. 
Photons generated from the source are assumed to be prepared with the distribution $\rho(\lambda)$, where $\int_{\lambda\in\Lambda}\rho(\lambda)d\lambda=1$. The subspace of $\Lambda$ which corresponds to detection at time $t_i$ may, in general, be different from the subspace that corresponds to detection at time $t_j$. Let us assume that $\Lambda_i$ is the subspace of overall space $\Lambda$ for which the photon is detected at $t_i$. We take the detection efficiency of all the detectors to be $\eta \in (0,1]$, which implies that 
\begin{equation}\label{eq:eta}   \int_{\lambda\in\Lambda_{i}}\rho(\lambda) d\lambda=\eta,~~\forall i=1,2,3  
\end{equation}
The distributions of $\rho(\lambda)$ for different sub-spaces are shown in Fig. \ref{fig:ptf}. For example, $\Lambda_q$ is the subspace for which photons are detected at $t_1$ but not detected at $t_2$ and $t_3$; $\Lambda_a$ is the subspace for which photons are detected at $t_1$ and $t_2$ but not detected at $t_3$. Similarly, other subspaces $\Lambda_p,\Lambda_s,\Lambda_b,\Lambda_c,\Lambda_d$ correspond to distinct possible fates of the photons as depicted in Fig. \ref{fig:ptf}.

According to macrorealism, the joint probabilities of the measurement outcomes at times $t_1,t_2,t_3$ are predefined by the state $\lambda$. In each of the seven subspaces, we denote the joint probabilities of eight possible joint measurement outcomes by the subscript $1,\dots,8$ corresponding to three different times. For instance, the joint probabilities within $\Lambda_q$ are as follows
\begin{equation}
\label{eq:hvm1}
\begin{split}
&q_1=\int_{\lambda \in \Lambda_q}  P_{t_{1},t_{2},t_{3}}(+,+,+|\lambda)\rho(\lambda)d\lambda,~~ q_2=\int_{\lambda \in \Lambda_q}  P_{t_{1},t_{2},t_{3}}(+,+,-|\lambda)\rho(\lambda)d\lambda,\\ 
&q_3=\int_{\lambda \in \Lambda_q}  P_{t_{1},t_{2},t_{3}}(+,-,+|\lambda)\rho(\lambda)d\lambda,~~
q_4=\int_{\lambda \in \Lambda_q}  P_{t_{1},t_{2},t_{3}}(+,-,-|\lambda)\rho(\lambda)d\lambda, \\ 
&q_5=\int_{\lambda \in \Lambda_q}  P_{t_{1},t_{2},t_{3}}(-,+,+|\lambda)\rho(\lambda)d\lambda,~~
q_6=\int_{\lambda \in \Lambda_q}  P_{t_{1},t_{2},t_{3}}(-,+,-|\lambda)\rho(\lambda)d\lambda,\\ 
&q_7=\int_{\lambda \in \Lambda_q}  P_{t_{1},t_{2},t_{3}}(-,-,+|\lambda)\rho(\lambda)d\lambda,~~
q_8=\int_{\lambda \in \Lambda_q}  P_{t_{1},t_{2},t_{3}}(-,-,-|\lambda)\rho(\lambda)d\lambda . \\
\end{split}
\end{equation} 
Similarly, we define $p_i,s_i,a_i,b_i,c_i,d_i$ for the respective subspaces $\Lambda_p,\Lambda_s,\Lambda_a,\Lambda_b,\Lambda_c,\Lambda_d$. Further, we denote the contribution arising from $\rho(\lambda)$ on the subspace $\Lambda_q$ by
\begin{equation}
    q = \int_{\lambda \in \Lambda_q}  \rho(\lambda)d\lambda = \sum_{i=1}^{8}q_{i} , 
\end{equation} and similarly, for the other subspaces
\begin{equation}\label{eq:pqab}
    p = \sum_{i=1}^{8}p_{i}, \quad s = \sum_{i=1}^{8}s_{i} ,\quad a = \sum_{i=1}^{8}a_{i} ,\quad b =  \sum_{i=1}^{8}b_{i} , \quad c = \sum_{i=1}^{8}c_{i} , \quad d = \sum_{i=1}^{8}d_{i} .
\end{equation}
Based on the above considerations, we now present the analyses pertaining to the two different experimental
scenarios: (1) setup involving only detectors, and (2) setup involving detectors and blockers which we have implemented.

\subsection{The setup involving only detectors}
\label{appendix:a1}
First, we consider the scenario where three detectors are used to observe the joint probabilities in our setup. For example, if we want to measure $P_{t_{1},t_{2}}(+,+)$ using negative result measurement, we place a detector D1 in the `-1' arm at $t_1$ and two detectors, say, D2 and D3, in the arms `+1' and `-1'  at $t_2$. 
The joint probability is given by
\begin{equation}
\label{eq:hvm2}
P_{t_{1},t_{2}}(+,+)=\frac{\text{no. of photons detected at D2}}{\text{total no. of photons detected at three detectors}} .
\end{equation}
In terms of a realist model, we can write the above as
\begin{equation}
\label{eq:hvm3}
\begin{split}
P_{t_{1},t_{2}}(+,+) &= \frac{\int\limits_{\lambda \in \Lambda_2} P_{t_{1},t_{2}}(+,+|\lambda)\rho(\lambda)d\lambda}{\int\limits_{\lambda \in \Lambda_2} P_{t_{1},t_{2}}(+,+|\lambda)\rho(\lambda)d\lambda+\int\limits_{\lambda \in \Lambda_2} P_{t_{1},t_{2}}(+,-|\lambda)\rho(\lambda)d\lambda+\int\limits_{\lambda \in \Lambda_1} P_{t_{1}}(-|\lambda)\rho(\lambda)d\lambda}\\
&= \frac{\int\limits_{\lambda \in \Lambda_2} P_{t_{1},t_{2}}(+,+|\lambda)\rho(\lambda)d\lambda}{\int\limits_{\lambda \in \Lambda_2} P_{t_{1}}(+|\lambda)\rho(\lambda)d\lambda+\int\limits_{\lambda \in \Lambda_1} P_{t_{1}}(-|\lambda)\rho(\lambda)d\lambda} .
\end{split}
\end{equation}
Similarly, by obtaining the other joint probabilities, we get the correlation $\langle Q_{t_{1}} Q_{t_{2}}\rangle$ in terms of the above introduced quantities in \eqref{eq:hvm1}-\eqref{eq:pqab}, given by the following expression
\begin{equation}
\label{eq:hvm4}
\begin{split}
\langle Q_{t_{1}} Q_{t_{2}}\rangle &=P_{t_{1},t_{2}}(+,+)-P_{t_{1},t_{2}}(+,-)-P_{t_{1},t_{2}}(-,+)+P_{t_{1},t_{2}}(-,-)\\
&=\frac{\int\limits_{\lambda \in \Lambda_2} P_{t_{1},t_{2}}(+,+|\lambda)\rho(\lambda)d\lambda-\int\limits_{\lambda \in \Lambda_2} P_{t_{1},t_{2}}(+,-|\lambda)\rho(\lambda)d\lambda}{\int\limits_{\lambda \in \Lambda_2} P_{t_{1}}(+|\lambda)\rho(\lambda)d\lambda+\int\limits_{\lambda \in \Lambda_1} P_{t_{1}}(-|\lambda)\rho(\lambda)d\lambda}+\frac{\int\limits_{\lambda \in \Lambda_2} P_{t_{1},t_{2}}(-,-|\lambda)\rho(\lambda)d\lambda-\int\limits_{\lambda \in \Lambda_2} P_{t_{1},t_{2}}(-,+|\lambda)\rho(\lambda)d\lambda}{\int\limits_{\lambda \in \Lambda_2} P_{t_{1}}(-|\lambda)\rho(\lambda)d\lambda+\int\limits_{\lambda \in \Lambda_1} P_{t_{1}}(+|\lambda)\rho(\lambda)d\lambda}\\
&=\frac{\sum\limits_{i=1,2}(a_i+c_i+d_i+p_i)-\sum\limits_{i=3,4}(a_i+c_i+d_i+p_i)}{a+d+\sum\limits_{i=1,2,3,4} (c_i+p_i) +\sum\limits_{i=5,6,7,8} (b_i+q_i)}+\frac{\sum\limits_{i=7,8}(a_i+c_i+d_i+p_i)-\sum\limits_{i=5,6}(a_i+c_i+d_i+p_i)}{a+d+\sum\limits_{i=5,6,7,8} (c_i+p_i) +\sum\limits_{i=1,2,3,4} (b_i+q_i)} .
\end{split}
\end{equation}
Subsequently, the LGI expression given by \eqref{eq:lgiApx} reduces to
\begin{equation}
\label{eq:hvm5}
\begin{split}
& \langle Q_{t_{1}} Q_{t_{2}}\rangle + \langle Q_{t_{2}} Q_{t_{3}}\rangle - \langle Q_{t_{1}} Q_{t_{3}}\rangle \\ 
 = &\frac{\sum\limits_{i=1,2}(a_i+c_i+d_i+p_i)-\sum\limits_{i=3,4}(a_i+c_i+d_i+p_i)}{a+d+\sum\limits_{i=1,2,3,4} (c_i+p_i) +\sum\limits_{i=5,6,7,8} (b_i+q_i)}+\frac{\sum\limits_{i=7,8}(a_i+c_i+d_i+p_i)-\sum\limits_{i=5,6}(a_i+c_i+d_i+p_i)}{a+d+\sum\limits_{i=5,6,7,8} (c_i+p_i) +\sum\limits_{i=1,2,3,4} (b_i+q_i)}\\
& +\frac{\sum\limits_{i=1,5}(b_i+c_i+d_i+s_i)-\sum\limits_{i=2,6}(b_i+c_i+d_i+s_i)}{c+d+\sum\limits_{i=1,2,5,6} (b_i+s_i) +\sum\limits_{i=3,4,7,8} (a_i+p_i)}+\frac{\sum\limits_{i=4,8}(b_i+c_i+d_i+s_i)-\sum\limits_{i=3,7}(b_i+c_i+d_i+s_i)}{c+d+\sum\limits_{i=3,4,7,8} (b_i+s_i) +\sum\limits_{i=1,2,5,6} (a_i+p_i)}\\
& -\frac{\sum\limits_{i=1,3}(b_i+c_i+d_i+s_i)-\sum\limits_{i=2,4}(b_i+c_i+d_i+s_i)}{b+d+\sum\limits_{i=1,2,3,4} (c_i+s_i) +\sum\limits_{i=5,6,7,8} (a_i+q_i)}-\frac{\sum\limits_{i=6,8}(b_i+c_i+d_i+s_i)-\sum\limits_{i=5,7}(b_i+c_i+d_i+s_i)}{b+d+\sum\limits_{i=5,6,7,8} (c_i+s_i) +\sum\limits_{i=1,2,3,4} (a_i+q_i)}.
\end{split}
\end{equation}
Now, in order to obtain the maximum macrorealist value of the LGI expression, the right-hand-side of the above equation is maximized under two conditions:\\
$(i)$ the normalization condition, that is,
\begin{equation}\label{c1}
    \sum_{i=1}^8 \left( p_i +  q_i + s_i +a_i + b_i + c_i + d_i \right)  \leq 1;
\end{equation} and $(ii)$ the condition \eqref{eq:eta} expressed as follows using \eqref{eq:hvm1} 
\begin{equation}\label{c2}
    \sum_{i=1}^8 (p_i +  a_i + b_i +d_i ) = \sum_{i=1}^8 ( q_i +  a_i + c_i + d_i) = \sum_{i=1}^8 (s_i +  b_i + c_i + d_i) = \eta . 
\end{equation}
For $\eta < 2/3$, 
the maximum value of the right-hand-side of \eqref{eq:hvm5} is 8/3, which is larger than the maximum quantum value of LGI expression. 
This maximum value is obtained when the parameters
\begin{equation}\label{parameters1}
    a_7=b_5=c_1=\frac{\eta}{2},
\end{equation} and all other parameters are zero.
On the other hand, for $\eta \geq 2/3$, this maximum value occurs for 
\begin{equation}
    a_1=b_4=c_1=1-\eta,\ d_1=3\eta -2,
\end{equation} 
 and when other parameters are zero, so that the LGI expression in \eqref{eq:hvm5} simplifies to $2/\eta - \eta$. Thus, taking into account the detection efficiency loophole, the LGI given by \eqref{eq:lgiApx} is modified to the following form
 \begin{eqnarray}\label{eq:mlgi}
\langle Q_{t_{1}}Q_{t_{2}}\rangle + \langle Q_{t_{2}}Q_{t_{3}}\rangle - \langle Q_{t_{1}}Q_{t_{3}}\rangle\leq 
\begin{cases}
 \frac83 , & \text{ for } \eta < \frac23 \\
 \frac{2}{\eta} - \eta , & \text{ for } \eta \geq \frac23 .
\end{cases} 
\end{eqnarray}
Now, considering the maximum quantum value $3/2$ of the LGI expression, it follows from \eqref{eq:mlgi} that a realist model reproduces all quantum predictions up to that maximum value whenever
\begin{equation}
     \frac{2}{\eta} - \eta \leq \frac32 \implies \eta \leq 0.8508.
\end{equation}
Thus, in order to close the detection efficiency loophole in this case for testing LGI, one requires $\eta > 0.851$. 

Next, considering the WLGI expression given by \eqref{eq:wlgiApx}, it takes the following form
\begin{equation}
\label{eq:hvm6}
\begin{split}
&P_{t_{1},t_{3}}(-,+)-P_{t_{1},t_{2}}(-,+)-P_{t_{2},t_{3}}(-,+)\\
 =& \frac{\int\limits_{\lambda \in \Lambda_3} P_{t_{1},t_{3}}(-,+|\lambda)\rho(\lambda)d\lambda}{\int\limits_{\lambda \in \Lambda_3} P_{t_{1}}(-|\lambda)\rho(\lambda)d\lambda+\int\limits_{\lambda \in \Lambda_1} P_{t_{1}}(+|\lambda)\rho(\lambda)d\lambda} - \frac{\int\limits_{\lambda \in \Lambda_2} P_{t_{1},t_{2}}(-,+|\lambda)\rho(\lambda)d\lambda}{\int\limits_{\lambda \in \Lambda_2} P_{t_{1}}(-|\lambda)\rho(\lambda)d\lambda+\int\limits_{\lambda \in \Lambda_1} P_{t_{1}}(+|\lambda)\rho(\lambda)d\lambda}\\
&-\frac{\int\limits_{\lambda \in \Lambda_3} P_{t_{2},t_{3}}(-,+|\lambda)\rho(\lambda)d\lambda}{\int\limits_{\lambda \in \Lambda_3} P_{t_{2}}(-|\lambda)\rho(\lambda)d\lambda+\int\limits_{\lambda \in \Lambda_2} P_{t_{2}}(+|\lambda)\rho(\lambda)d\lambda}\\
=& \frac{\sum\limits_{i=5,7}(b_i+c_i+d_i+s_i)}{b+d+\sum\limits_{i=5,6,7,8} (c_i+s_i) +\sum\limits_{i=1,2,3,4} (a_i+q_i)}-\frac{\sum\limits_{i=5,6}(a_i+c_i+d_i+p_i)}{a+d+\sum\limits_{i=5,6,7,8} (c_i+p_i) +\sum\limits_{i=1,2,3,4} (b_i+q_i)}\\
&-\frac{\sum\limits_{i=3,7}(b_i+c_i+d_i+s_i)}{c+d+\sum\limits_{i=3,4,7,8} (b_i+s_i) +\sum\limits_{i=1,2,5,6} (a_i+p_i)} .
\end{split}
\end{equation}
It can be verified that whenever $\eta < 2/3$, the above WLGI expression is 1 for the choice of parameter values given in \eqref{parameters1}.
For $\eta \geq 2/3$,  the maximum value of the right-hand-side of \eqref{eq:hvm6} is obtained when all other parameters are zero except
\begin{equation}
    a_7=b_5=c_1=1-\eta,\ d_1=3\eta -2 ,
\end{equation} 
so that the expression in \eqref{eq:hvm6} simplifies to
$ (1-\eta)/(2\eta-1)$. Subsequently, we have the modified form of WLGI given by
\begin{eqnarray}\label{eq:mwlgi}
P_{t_{1},t_{3}}(-,+)-P_{t_{1},t_{2}}(-,+)-P_{t_{2},t_{3}}(-,+)\leq
\begin{cases}
 1 , & \text{ for } \eta < \frac23 \\
 \frac{1-\eta}{2\eta-1} , & \text{ for } \eta \geq \frac23 .
\end{cases} 
\end{eqnarray}  Taking into account the maximum quantum value 0.4034 of the WLGI expression, it follows from \eqref{eq:mwlgi} that a realist model reproduces all quantum predictions up to that maximum value whenever
\begin{equation}
      \frac{1-\eta}{2\eta-1}  \leq 0.4034 \implies \eta \leq 0.78.
\end{equation}
Thus, in order to close the detection efficiency loophole in this case for testing WLGI, one requires $\eta > 0.78$.

\subsection{The setup involving detectors and blockers}
\label{appendix:a2}
For the modified version of the setup, the detectors used for negative result measurement are replaced by ideal blockers, while detectors are placed only at time $t_{3}$. So, it is the subspace $\Lambda_{3}$ which is essentially relevant for our subsequent analysis. For example, the joint probability $P_{t_{1},t_{3}}(+,+)$ is written in terms of the realist model considered as follows
\begin{equation}
\label{eq:hvm7}
P_{t_{1},t_{3}}(+,+) = \frac{\int\limits_{\lambda \in \Lambda_3} P_{t_{1},t_{3}}(+,+|\lambda)\rho(\lambda)d\lambda}{\int\limits_{\lambda \in \Lambda_3} P_{t_{1}}(+|\lambda)\rho(\lambda)d\lambda+\int\limits_{\lambda \in \Lambda_3} P_{t_{1}}(-|\lambda)\rho(\lambda)d\lambda}=\frac{\int\limits_{\lambda \in \Lambda_3} P_{t_{1},t_{3}}(+,+|\lambda)\rho(\lambda)d\lambda}{\int\limits_{\lambda \in \Lambda_3}\rho(\lambda)d\lambda}=\frac{\int\limits_{\lambda \in \Lambda_3} P_{t_{1},t_{3}}(+,+|\lambda)\rho(\lambda)d\lambda}{\eta} .
\end{equation}
Similar expressions are valid for the other joint probabilities $P_{t_{1},t_{3}}(q_{t_{1}},q_{t_{3}}), P_{t_{2},t_{3}}(q_{t_{2}},q_{t_{3}})$. However, for determining the joint probabilities of the form $P_{t_{1},t_{2}}(q_{t_{1}},q_{t_{2}})$, we have considered an additional step of marginalizing the outcome at $t_3$ using the induction condition at the level of observable probabilities \eqref{eq:AoTApx}. For instance,
\begin{equation}
\begin{split}
P_{t_{1},t_{2}}(+,+) = P_{t_{1},t_{2},t_{3}}(+,+,+)+P_{t_{1},t_{2},t_{3}}(+,+,-)&=\frac{\int\limits_{\lambda \in \Lambda_3} \{P_{t_{1},t_{2},t_{3}}(+,+,+|\lambda)+P_{t_{1},t_{2},t_{3}}(+,+,-|\lambda)\}\rho(\lambda)d\lambda}{\eta}\\
&=\frac{\int\limits_{\lambda \in \Lambda_3} P_{t_{1},t_{2}}(+,+|\lambda)\rho(\lambda)d\lambda}{\eta} .
\end{split}
\end{equation}
Further, the correlation functions are expressed as follows
\begin{equation}
\begin{split}
& \langle Q_{t_{1}} Q_{t_{3}}\rangle =\frac{\int\limits_{\lambda \in \Lambda_3}\{P_{t_{1},t_{3}}(+,+|\lambda)-P_{t_{1},t_{3}}(+,-|\lambda)-P_{t_{1},t_{3}}(-,+|\lambda)+P_{t_{1},t_{3}}(-,-|\lambda)\}\rho(\lambda)d\lambda}{\eta}=\frac{\int\limits_{\lambda \in \Lambda_3}\langle Q_{t_{1}} Q_{t_{3}}\rangle_{\lambda}\rho(\lambda)d\lambda}{\eta} , \\
& \langle Q_{t_{1}} Q_{t_{2}}\rangle = \frac{\int\limits_{\lambda \in \Lambda_3}\langle Q_{t_{1}} Q_{t_{2}}\rangle_{\lambda}\rho(\lambda)d\lambda}{\eta}, ~~ \langle Q_{t_{2}} Q_{t_{3}}\rangle = \frac{\int\limits_{\lambda \in \Lambda_3}\langle Q_{t_{2}} Q_{t_{3}}\rangle_{\lambda}\rho(\lambda)d\lambda}{\eta} .
\end{split}
\end{equation}
Hence, the LGI expression involves integration only on the subspace of $\Lambda_3$ and reduces to the following form 
\begin{equation}
\label{eq:hvmLGI}
\begin{split}
\langle Q_{t_{1}} Q_{t_{2}}\rangle + \langle Q_{t_{2}} Q_{t_{3}}\rangle - \langle Q_{t_{1}} Q_{t_{3}}\rangle
=  \frac{\int\limits_{\lambda \in \Lambda_3}\{\langle Q_{t_{1}} Q_{t_{2}}\rangle_{\lambda}+\langle Q_{t_{2}} Q_{t_{3}}\rangle_{\lambda}-\langle Q_{t_{1}} Q_{t_{3}}\rangle_{\lambda}\}\rho(\lambda)d\lambda}{\eta} ,
\end{split}
\end{equation} 
where the numerator is the macrorealist expression of the left-hand-side of the LGI given by \eqref{eq:lgiApx}. It is then seen that for any 
non-vanishing value of $\eta$, the above expression remains less than or equal to 1.
Similarly, the WLGI expression given by \eqref{eq:wlgiApx} reduces to the following form
\begin{equation}
\label{eq:hvmWLGI}
\begin{split}
P_{t_{1},t_{3}}(-,+)-P_{t_{1},t_{2}}(-,+)-P_{t_{2},t_{3}}(-,+) = \frac{\int\limits_{\lambda \in \Lambda_3}\{P_{t_{1},t_{3}}(-,+|\lambda)-P_{t_{1},t_{2}}(-,+|\lambda)-P_{t_{2},t_{3}}(-,+|\lambda)\}\rho(\lambda)d\lambda}{\eta} ,
\end{split}
\end{equation} 
where the numerator is its macrorealist expression on the subspace of $\Lambda_3$. Thus, the WLGI expression given by \eqref{eq:wlgiApx} retains its same macrorealist upper bound 0 for any non-vanishing value of $\eta$.
Therefore, it follows from \eqref{eq:hvmLGI} and \eqref{eq:hvmWLGI} that any quantum mechanical violation of LGI/WLGI for this setup cannot be reproduced by a macrorealist model, whatever be the detection efficiency.

\section{Closing the multiphoton emission loophole}
\label{appendix:b} 

\subsection{Derivation of 
the modified upper bounds of LGI and WLGI, considering the presence of multiphotons inside the setup}
\label{appendix:b1}
In our analysis, we consider the probability of generating multiple photon pairs to be negligibly small 
so that we can ignore the possibility of more than two photons being present in our setup
. Now, considering the case of LGI, we first show that there exists a realist model of two photons, each of which is present in either of the two different paths,  for which the LGI expression attains its algebraic upper bound.  To describe such a model, let us denote the hidden-variable state $\lambda^1,\lambda^2$ for photons 1 and 2 respectively. Due to the presence of both these photons, they may mutually influence each other on arriving at the beamsplitter simultaneously, which, in turn, can affect their subsequent individual behaviors. 

Now, recall that  we place a blocker at $t_1$ (either in path `$+1$' or path `$-1$') for implementing the negative result measurement to obtain the joint probabilities $P_{t_{1},t_{3}}(q_{t_{1}},q_{t_{3}}) , P_{t_{1},t_{2}}(q_{t_{1}},q_{t_{2}})$, while we do not place any blocker at $t_1$ to obtain the joint probabilities $P_{t_{2},t_{3}}(q_{t_{2}},q_{t_{3}})$. If the two photons respectively are in different paths at $t_1$ and there is no blocker placed at $t_1$, the photons may influence each other at the second beamsplitter just before $t_2$. Hence, in principle, the behavior of the photon which is present at $t_2$ depends on whether a blocker is placed at $t_1$ or not.
Further, to specify the realist model of two photons we are considering, we take the joint probability distributions of the measurement outcomes at $t_1,t_2,t_3$ to be given by the following expressions for all $\lambda^1,\lambda^2,$ 
\begin{eqnarray}
\label{eq:m1}
 & P^{1}_{t_{1},t_{2},t_{3}}(+,+,+|\lambda^1,N_1)= 1,~~P^{2}_{t_{1},t_{2},t_{3}}(-,-,-|\lambda^2,N_1) = 1, \\
& P^{1}_{t_{1},t_{2},t_{3}}(+,+,-|\lambda^1,B_1-)= 1,~~ P^{2}_{t_{1},t_{2},t_{3}}(-,-,+|\lambda^2,B_1+)=1, \label{eq:m2}
\end{eqnarray}
where $P^i$ denotes the probability distribution for the photon $i\in \{1,2\}$, and $N_1$ denotes the configuration in which no blocker is placed at $t_1$, $B_1\pm$ denotes the configuration in which the blocker is placed at $t_1$ in the `$\pm 1$' path. 

Now,  for the realist model specified by the relations \eqref{eq:m1}-\eqref{eq:m2}, let us evaluate the joint probabilities $P_{t_{1},t_{3}}(q_{t_{1}},q_{t_{3}})$, $P_{t_{1},t_{2}}(q_{t_{1}},q_{t_{2}})$, $P_{t_{2},t_{3}}(q_{t_{2}},q_{t_{3}})$ measured in our experimental setup using (\eqref{eq:P23Apx}-\eqref{eq:P12Apx}), as will be explained in Appendix \ref{appendix:c1}. For example, the joint probabilities $P_{t_{2},t_{3}}(q_{t_{2}},q_{t_{3}})$ are obtained from the relations \eqref{eq:m1} for which we do not place any blocker at $t_1$. For this purpose, we place a blocker at $t_2$ and two detectors, say, $D_+$ and $D_-$, at $t_3$ in `$\pm1$' path. When the blocker is placed in the `-1' path, we observe clicks only at $D_+$, since from \eqref{eq:m1} we know that photon 2 is blocked at $t_2$ and photon 1 is in the `+1' path all the time. Likewise, when the blocker is placed in the `+1' path, we observe clicks only at $D_-$, since from \eqref{eq:m1} we know that photon 1 is blocked at $t_2$ and photon 2 is in the `-1' path all the time.\\ 
Next, considering in the context of \eqref{eq:P23Apx} relating the quantities $P_{t_{2},t_{3}}(q_{t_{2}},q_{t_{3}})$ to the actual observed coincidence counts, let $N$ to be the total number of photons in each of the two sets of runs corresponding to each of the two configurations (the blocked being placed in `$\pm1$' path at $t_2$) respectively. Then, in view of what has been explained above using the relation \eqref{eq:m1}, it follows that the relevant coincidence counts are given by
\begin{equation}
    C_{t_{2},t_{3}}(+,+) = C_{t_{2},t_{3}}(-,-)= N, \ C_{t_{2},t_{3}}(+,-)=C_{t_{2},t_{3}}(-.+) =0 .
\end{equation}
Thus, from \eqref{eq:P23Apx} we obtain
\begin{equation}\label{mprob1}
P_{t_{2},t_{3}}(+,+)=P_{t_{2},t_{3}}(-,-)=\frac12, \ P_{t_{2},t_{3}}(+,-)=P_{t_{2},t_{3}}(-,+)=0.
\end{equation}
Similarly, following the above procedure, we can obtain the other joint probabilities $P_{t_{1},t_{3}}(q_{t_{1}},q_{t_{3}}), P_{t_{1},t_{2}}(q_{t_{1}},q_{t_{2}})$ from the relations \eqref{eq:m2} and (\eqref{eq:P13Apx}-\eqref{eq:P12Apx}), which are given as follows
\begin{eqnarray} \label{mprob}
P_{t_{1},t_{3}}(+,+)=P_{t_{1},t_{3}}(-,-)=0, \ P_{t_{1},t_{3}}(+,-)=P_{t_{1},t_{3}}(-,+)=\frac12, \nonumber \\
P_{t_{1},t_{2}}(+,+)=P_{t_{1},t_{2}}(-,-)=\frac12, \ P_{t_{1},t_{2}}(+,-)=P_{t_{1},t_{2}}(-,+)=0. 
\end{eqnarray}
Using these joint probabilities \eqref{mprob1}-\eqref{mprob}, one can readily show that $\langle Q_{t_{1}} Q_{t_{2}}\rangle = \langle Q_{t_{2}} Q_{t_{3}}\rangle = - \langle Q_{t_{1}} Q_{t_{3}}\rangle = 1$. Hence, for the realist model specified by the relations \eqref{eq:m1}-\eqref{eq:m2}, the value of the LGI expression in \eqref{eq:lgiApx} attains its algebraic maximum value 3. 

Now, if $\gamma$ is the fraction of the total number of runs for which multiple photons occur, the modified upper bound for LGI, in this case, can be written as
\begin{equation}
\label{eq:mp_bound}
\text{modified upper bound of LGI}=\gamma\times\text{algebraic maximum value of LGI}\ + \ (1-\gamma)\times\text{macrorealist upper bound of LGI} .
\end{equation}
It then follows from \eqref{eq:mp_bound} that the LGI given in \eqref{eq:lgiApx} is modified to 
\begin{equation}\label{eq:modboundlgiApx}
\langle Q_{t_{1}}Q_{t_{2}}\rangle + \langle Q_{t_{2}}Q_{t_{3}}\rangle - \langle Q_{t_{1}}Q_{t_{3}}\rangle\leq 1 + 2\gamma . 
\end{equation}
Next, by 
considering all possible the two-photon realist models, one can show that the WLGI expression given by \eqref{eq:wlgiApx} attains the maximum value 1/2 for the two-photon realist model specified above by the relations  \eqref{eq:m1}-\eqref{eq:m2}. Subsequently, it follows from \eqref{mprob1}-\eqref{mprob} 
that the WLGI given by \eqref{eq:wlgiApx} is modified to 
\begin{equation}\label{eq:modboundwlgiApx} P_{t_{1},t_{3}}(-,+)-P_{t_{1},t_{2}}(-,+)-P_{t_{2},t_{3}}(-,+)\leq \frac{\gamma}{2} .
\end{equation}


\subsection{Determining the value of $\gamma$ in our experimental setup}
\label{appendix:b2}
In order to formulate the required procedure, we first consider the simplest configuration in which the photons are incident on a beamsplitter, followed by the two detectors in the reflecting and transmitting channels respectively. Here, in evaluating the single and coincidence counts, apart from the single photons, we take into account the possibility of simultaneous occurrence of two photons within the setup, ignoring the negligible possibility of the simultaneous presence of more than two photons for the type of source used in our experiment.

\begin{figure}[!ht]
\centering
\includegraphics[scale=0.3]{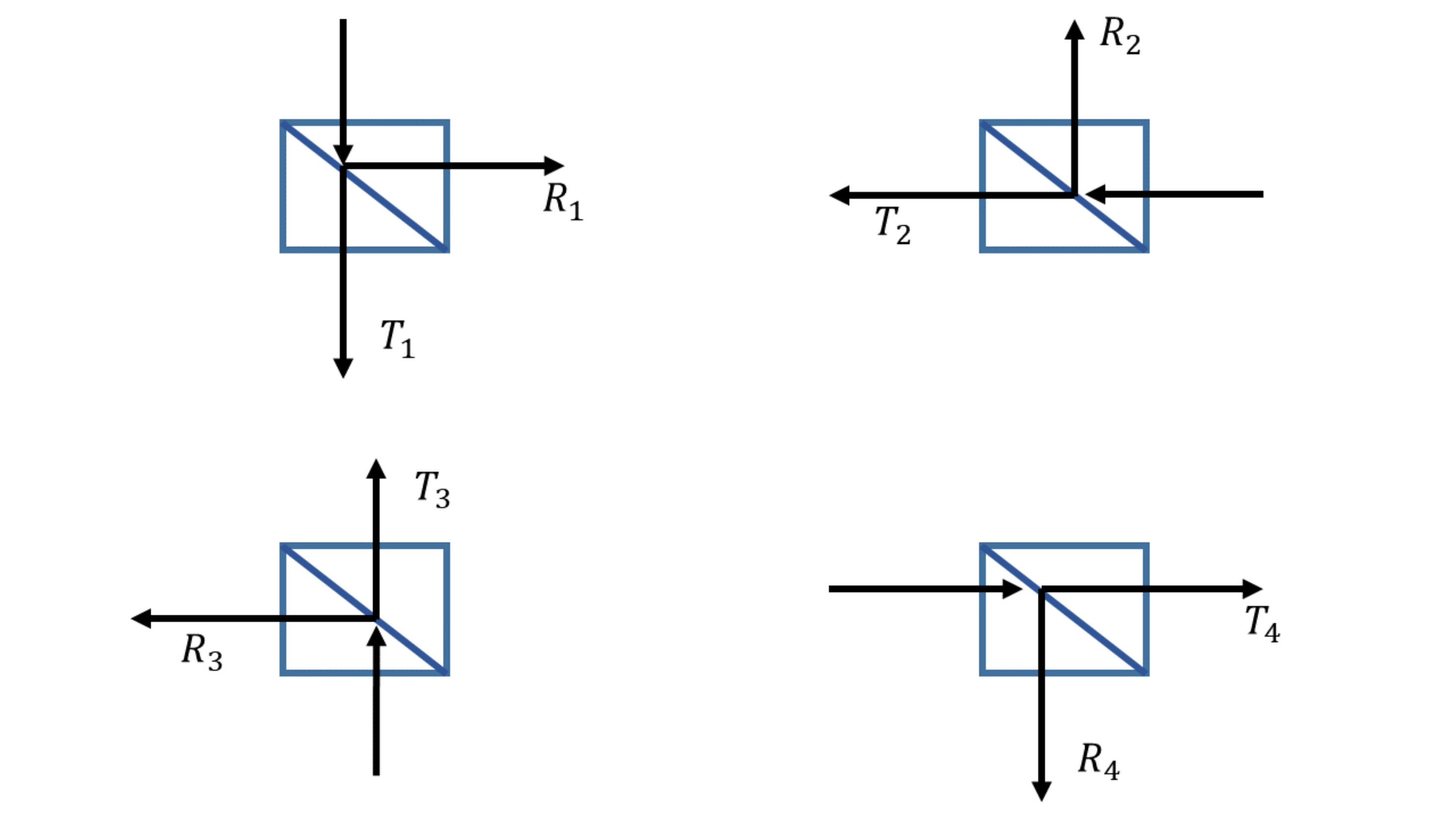}
\caption{Splitting ratios of the non-polarizing beamsplitter (NPBS) are denoted as $T_i:R_i$ for $i=1, 2, 3, 4$, depending on the input port of the NPBS. Here, $T_i+R_i=1, ~\forall~i$.}
\label{fig:bso}
\end{figure}

Let's say for a large time-interval ($T$), $N$ number of photon events are generated, among which $N_{1}$ are the single photon events and $N_{2}$ are the two-photon events. So, $N=N_{1}+N_{2}$ and $\gamma=N_{2}/N=1/(\frac{N_{1}}{N_{2}}+1)$. Let us assume $N$ photons are sent to a beamsplitter (BS) with a splitting ratio ($T:R$). In the transmitting arm of the BS, a single photon detector (D1) is placed whose efficiency is $\eta_{1}$. Similarly, in the reflecting arm detector D2 is placed with efficiency $\eta_{2}$. For the $N_{1}$ single photons, they have $T\eta_{1}$ and $R\eta_{2}$ probabilities of getting detected in D1 and D2, respectively. For the $N_{2}$ two-photon events, the probability that D1 will get a detection is $T^{2}\eta_{1}(2-\eta_{1})+2TR\eta_{1}$. The probability that D2 will get a detection is $R^{2}\eta_{2}(2-\eta_{2})+2TR\eta_{2}$. The probability that both D1 and D2 will get simultaneous detection is $2TR\eta_{1}\eta_{2}$. Consequently, the single counts denoted by $C_1, C_2$ and the coincidence counts between the detectors D1 and D2 denoted by $C_{1,2}$ are then given by,
\begin{equation}
\label{eq:C1}
C_{1}=N_{1}T\eta_{1}+N_{2}\{T^{2}\eta_{1}(2-\eta_{1})+2TR\eta_{1}\} ,
\end{equation}
\begin{equation}
\label{eq:C2}
C_{2}=N_{1}R\eta_{2}+N_{2}\{R^{2}\eta_{2}(2-\eta_{2})+2TR\eta_{2}\} ,
\end{equation}
\begin{equation}
\label{eq:C12}
C_{1,2}=2N_{2}TR\eta_{1}\eta_{2} .
\end{equation}
We can determine the values of of $C_{1}, C_{2}$, and $C_{1,2}$ from experiment, and using \eqref{eq:C1}, \eqref{eq:C2} and \eqref{eq:C12}, we can obtain $\gamma$. This process forms the basis of our procedure to measure $\gamma$ directly from the LGI experimental setup. 
For this purpose, we consider measurements of the single and coincidence counts
in each of the four different configurations used in our experiment, wherein the
two blockers are respectively placed corresponding to each of the instants $t_1$ and
$t_2$, swapped among the two arms of each of the two interferometers, while the two
detectors are respectively placed in the final two output channels at $t_3$. It is for each
of these four configurations (denoted by the set 1, 2, 3 or 4), the expressions for the
single and coincidence counts are given below in terms of the relevant
parameters, similar to the way \eqref{eq:C1}-\eqref{eq:C12} have been derived. \\
Note that in this treatment, set in the context of the setup we have actually used in
our experiment (Fig. \ref{fig:schematic2}), to be empirically relevant, we take the $T:R$ splitting ratio
of the non-polarizing beamsplitter (NPBS) to be varying, dependent upon which input port of the NPBS the photons are impinging on (see Figure \ref{fig:bso}). Splitting ratio due to the HWP and PBS combination (HWP2, PBS2 in Figure \ref{fig:schematic2}) is denoted by ${|\alpha|}^{2}:{|\beta|}^{2}$.

\textit{Set 1: $t_{1}+, t_{2}+$ (blocking arm 2 and arm 3 in the experimental setup shown in Figure \ref{fig:schematic2})}\\
\begin{equation}
\label{eq:mp1}
C_{1}(+,+)=N_{1}{|\alpha|}^{2}T_{1}R_{2}\eta_{1}+N_{2}{|\alpha|}^{4}{(T_{1})}^{2}\{{(R_{2})}^{2}\eta_{1}(2-\eta_{1})+2T_{2}R_{2}\eta_{1}\}+N_{2}({|\alpha|}^{4}2T_{1}R_{1}+2{|\alpha|}^{2}{|\beta|}^{2}T_{1})R_{2}\eta_{1}
\end{equation}
\begin{equation}
\label{eq:mp2}
C_{2}(+,+)=N_{1}{|\alpha|}^{2}T_{1}T_{2}\eta_{2}+N_{2}{|\alpha|}^{4}{(T_{1})}^{2}\{{(T_{2})}^{2}\eta_{2}(2-\eta_{2})+2T_{2}R_{2}\eta_{2}\}+N_{2}({|\alpha|}^{4}2T_{1}R_{1}+2{|\alpha|}^{2}{|\beta|}^{2}T_{1})T_{2}\eta_{2}
\end{equation}
\begin{equation}
\label{eq:mp3}
C_{1,2}(+,+)=2N_{2}{|\alpha|}^{4}{(T_{1})}^{2}T_{2}R_{2}\eta_{1}\eta_{2}
\end{equation}
\textit{Set 2: $t_{1}+, t_{2}-$ (blocking arm 2 and arm 4)}\\
\begin{equation}
\label{eq:mp4}
C_{1}(+,-)=N_{1}{|\alpha|}^{2}R_{1}T_{3}\eta_{1}+N_{2}{|\alpha|}^{4}{(R_{1})}^{2}\{{(T_{3})}^{2}\eta_{1}(2-\eta_{1})+2T_{3}R_{3}\eta_{1}\}+N_{2}({|\alpha|}^{4}2T_{1}R_{1}+2{|\alpha|}^{2}{|\beta|}^{2}R_{1})T_{3}\eta_{1}
\end{equation}
\begin{equation}
\label{eq:mp5}
C_{2}(+,-)=N_{1}{|\alpha|}^{2}R_{1}R_{3}\eta_{2}+N_{2}{|\alpha|}^{4}{(R_{1})}^{2}\{{(R_{3})}^{2}\eta_{2}(2-\eta_{2})+2T_{3}R_{3}\eta_{2}\}+N_{2}({|\alpha|}^{4}2T_{1}R_{1}+2{|\alpha|}^{2}{|\beta|}^{2}R_{1})R_{3}\eta_{2}
\end{equation}
\begin{equation}
\label{eq:mp6}
C_{1,2}(+,-)=2N_{2}{|\alpha|}^{4}{(R_{1})}^{2}T_{3}R_{3}\eta_{1}\eta_{2}
\end{equation}
\textit{Set 3: $t_{1}-, t_{2}+$ (blocking arm 1 and arm 3)}\\
\begin{equation}
\label{eq:mp7}
C_{1}(-,+)=N_{1}{|\beta|}^{2}R_{4}R_{2}\eta_{1}+N_{2}{|\beta|}^{4}{(R_{4})}^{2}\{{(R_{2})}^{2}\eta_{1}(2-\eta_{1})+2T_{2}R_{2}\eta_{1}\}+N_{2}({|\beta|}^{4}2T_{4}R_{4}+2{|\alpha|}^{2}{|\beta|}^{2}R_{4})R_{2}\eta_{1}
\end{equation}
\begin{equation}
\label{eq:mp8}
C_{2}(-,+)=N_{1}{|\beta|}^{2}R_{4}T_{2}\eta_{2}+N_{2}{|\beta|}^{4}{(R_{4})}^{2}\{{(T_{2})}^{2}\eta_{2}(2-\eta_{2})+2T_{2}R_{2}\eta_{2}\}+N_{2}({|\beta|}^{4}2T_{4}R_{4}+2{|\alpha|}^{2}{|\beta|}^{2}R_{4})T_{2}\eta_{2}
\end{equation}
\begin{equation}
\label{eq:mp9}
C_{1,2}(-,+)=2N_{2}{|\beta|}^{4}{(R_{4})}^{2}T_{2}R_{2}\eta_{1}\eta_{2}
\end{equation}
\textit{Set 4: $t_{1}-, t_{2}-$ (blocking arm 1 and arm 4)}\\
\begin{equation}
\label{eq:mp10}
C_{1}(-,-)=N_{1}{|\beta|}^{2}T_{4}T_{3}\eta_{1}+N_{2}{|\beta|}^{4}{(T_{4})}^{2}\{{(T_{3})}^{2}\eta_{1}(2-\eta_{1})+2T_{3}R_{3}\eta_{1}\}+N_{2}({|\beta|}^{4}2T_{4}R_{4}+2{|\alpha|}^{2}{|\beta|}^{2}T_{4})T_{3}\eta_{1}
\end{equation}
\begin{equation}
\label{eq:mp11}
C_{2}(-,-)=N_{1}{|\beta|}^{2}T_{4}R_{3}\eta_{2}+N_{2}{|\beta|}^{4}{(T_{4})}^{2}\{{(R_{3})}^{2}\eta_{2}(2-\eta_{2})+2T_{3}R_{3}\eta_{2}\}+N_{2}({|\beta|}^{4}2T_{4}R_{4}+2{|\alpha|}^{2}{|\beta|}^{2}T_{4})R_{3}\eta_{2}
\end{equation}
\begin{equation}
\label{eq:mp12}
C_{1,2}(-,-)=2N_{2}{|\beta|}^{4}{(T_{4})}^{2}T_{3}R_{3}\eta_{1}\eta_{2}
\end{equation}
We measure $C_{1}$, $C_{2}$, and $C_{1,2}$ from the above four sets of experiments, and numerically fit the 12 equations (\eqref{eq:mp1}-\eqref{eq:mp12}) by varying all input parameters, ${|\alpha|}^{2}$, $T_1$, $T_2$, $T_3$, $T_4$, $\eta_1$, $\eta_2$ and  $\gamma$. We take ten iteration 
of raw data-sets (time-stamps) for each of the four runs of the experiments and measure the average values of the $C_{1}$, $C_{2}$, and $C_{1,2}$, reported in the table \ref{tab:gamma}.\\
\begin{table}[!ht]
\centering
\begin{tabular}{|c|c|}
\hline
$C_{1}(+,+)$ & 9412\\
\hline
$C_{2}(+,+)$ & 36458.33\\
\hline
$C_{1,2}(+,+)$ & 7.67\\
\hline
$C_{1}(+,-)$ & 9589.33\\
\hline
$C_{2}(+,-)$ & 2611.67\\
\hline
$C_{1,2}(+,-)$ & 0.67\\
\hline
\end{tabular}
\quad
\begin{tabular}{|c|c|}
\hline
$C_{1}(-,+)$ & 2206\\
\hline
$C_{2}(-,+)$ & 11286\\
\hline
$C_{1,2}(-,+)$ & 1\\
\hline
$C_{1}(-,-)$ & 32375.33\\
\hline
$C_{2}(-,-)$ & 10656.67\\
\hline
$C_{1,2}(-,-)$ & 7.33\\
\hline
\end{tabular}
\caption{Measured values of singles and coincidences from the experiment.}
\label{tab:gamma}
\end{table}

For the purpose of the numerical fitting, we use the chi-squared test using the formula $\chi^2=\sum_{i}\frac{{(C^E_i-C^T_i)}^2}{C^T_i}$, where $C^E_i$ represents all the 12 coincidence values obtained from the experiment, $C^T_i$ represents all coincidence values from theory (\eqref{eq:mp1}-\eqref{eq:mp12}). We tweak all the fitting parameters to get the minimum value of $\chi^2$. 
We find the best fit
for ${|\alpha|}^{2}=0.48, T_{1}=0.74, T_{2}=0.77, T_{3}=0.81, T_{4}=0.65, \eta_{1}=0.56, \eta_{2}=0.64$ and $\gamma=0.0023$, where the chi-squared value is $7.2$ ($p$-value $> 0.05$). Using the measured value of $\gamma=0.0023$, we get the maximum upper bound for LGI is 1.0046 and for WLGI, it is 0.0012 (using \eqref{eq:mp_bound}). As we can see that the changes in the upper bounds due to the contribution of multi-photons are very small, i. e., in the third decimal place only. So, we can ignore this effect in our experiment.\\

\section{ Calculations and error analyses pertaining to the experimental data}
\label{appendix:c}
In this section, we describe the procedures for calculating the results mentioned in Section \ref{sec:result} of the main text.\\
(1) Details on experimental methods.\\
(2) Estimating the standard deviations in the data statistics obtained from the experiment, for the LGI, WLGI and NSIT.\\
(3) Estimating the range of the measured values of LGI, WLGI and NSIT, considering all the significant imperfections in the optical components.
\subsection{Details on experimental methods}
\label{appendix:c1}
In order to measure LGI and WLGI values from the experimental setup, we perform three experimental runs. In the first run, we measure all four joint probabilities of the form $P_{t_2,t_3}(q_{t_2},q_{t_3})\, \forall\, q_{t_2}=\pm 1\,, q_{t_3}=\pm 1$. In the second run, we measure four joint probabilities of the form $P_{t_1,t_3}(q_{t_1},q_{t_3})$. In the third run, we measure eight joint probabilities of the form $P_{t_1,t_2,t_3}(q_{t_1},q_{t_2},q_{t_3})$. All these sixteen measured joint probabilities are used for calculating the LGI and WLGI expression, given by \eqref{eq:lgiApx}
and \eqref{eq:wlgiApx}. 
For WLGI, we get the three joint probabilities directly from the three experimental runs mentioned above. In case of LGI, we do an additional step of calculating correlation values from measured joint probabilities, by following the equation, $\langle Q_{t_{i}} Q_{t_{j}}\rangle  =P_{t_i,t_j}(+,+)-P_{t_i,t_j}(+,-)-P_{t_i,t_j}(-,+)+P_{t_i,t_j}(-,-)$. To show the maintenance of all relevant NSITs, we perform the fourth run of the experiment where we measure the probabilities $P_{t_3}(+)$ and $P_{t_3}(-)$, and then calculate the following three expressions:
\begin{equation}
\label{eq:nsit12}
 NSIT_{(t_{1})t_{2}}:\,\left|P_{t_{2}}(+)-P_{t_{1},t_{2}}(+,+)-P_{t_{1},t_{2}}(-,+)\right|
\end{equation}
\begin{equation}
\label{eq:nsit13}
NSIT_{(t_{1})t_{3}}:\,\left|P_{t_{3}}(+)-P_{t_{1},t_{3}}(+,+)-P_{t_{1},t_{3}}(-,+)\right| .
\end{equation}
\begin{equation}
\label{eq:nsit23}
 NSIT_{(t_{2})t_{3}}:\,\left|P_{t_{3}}(+)-P_{t_{2},t_{3}}(+,+)-P_{t_{2},t_{3}}(-,+)\right|
\end{equation}
As we have implemented negative result measurement in our experiment for preserving NIM condition, we perform multiple sub-runs under each of the first three experimental runs. For each of the sub-runs, coincidence data is recorded for 10 seconds and repeated for large number of times (300 iterations for the second experimental run, 150 iterations for both the first and the third run), for better averaging. Details of the sub-runs are provided below.

In the first experimental run, we measure the correlation between time $t_2$ and $t_3$. For this purpose, we perform two sub-experimental runs, incorporating negative result measurement. In the first sub-run, a blocker is placed at $-1$ arm at $t_{2}$ (or at arm 3 in the experimental setup provided in Fig. \ref{fig:schematic2}), and coincidence counts are measured between SPAD1, SPAD2+ and between SPAD1 and SPAD2-. We denote the first coincidence as $C_{t_{2},t_{3}}(+,+)$ as any photon that gets detected in SPAD2+ must be in $+1$ arm at $t_{2}$ and $+1$ arm at $t_{3}$. Similarly, the second coincidence count can be denoted as $C_{t_{2},t_{3}}(+,-)$. In the second sub-run of the experiment, a blocker is placed at $+1$ arm at $t_{2}$. So coincidence count between SPAD1 and SPAD2+ is denoted as $C_{t_{2},t_{3}}(-,+)$, and coincidence count between SPAD1 and SPAD2- is denoted as $C_{t_{2},t_{3}}(-,-)$. The total coincidence count from the first run becomes, $C^{T}_{t_{2},t_{3}}=C_{t_{2},t_{3}}(+,+)+C_{t_{2},t_{3}}(+,-)+C_{t_{2},t_{3}}(-,+)+C_{t_{2},t_{3}}(-,-)$. We calculate four joint probabilities of the form $P_{t_{2},t_{3}}(q_{t_{2}},q_{t_{3}})$, where $q_{t_{2}}=\pm 1$, $q_{t_{3}}=\pm 1$, by normalizing the four coincidence values obtained in the first run\\
\begin{equation}
\label{eq:P23Apx}
P_{t_{2},t_{3}}(q_{t_{2}},q_{t_{3}})=\dfrac{C_{t_{2},t_{3}}(q_{t_{2}},q_{t_{3}})}{C^{T}_{t_{2},t_{3}}} .
\end{equation}
Similar strategy is used in the second run of the experiment, where we perform two sub-runs as well. In this case, a blocker is placed in $-1$ arm at time $t_{1}$ in the first sub-run, and in $+1$ arm for the second sub-run. We obtain four coincidence values, i. e., $C_{t_{1},t_{3}}(+,+)$, $C_{t_{1},t_{3}}(+,-)$, $C_{t_{1},t_{3}}(-,+)$, $C_{t_{1},t_{3}}(-,-)$, which are normalized to calculate the joint probabilities of the form $P_{t_{1},t_{3}}(q_{t_{1}},q_{t_{3}})$\\
\begin{equation}
\label{eq:P13Apx}
P_{t_{1},t_{3}}(q_{t_{1}},q_{t_{3}})=\dfrac{C_{t_{1},t_{3}}(q_{t_{1}},q_{t_{3}})}{C^{T}_{t_{1},t_{3}}} .
\end{equation}
The third experimental run, where we measure correlation between time $t_1$ and $t_2$, is slightly different, as we do not place any detectors (SPAD2+ and SPAD2-) at time $t_{2}$; in order to close the detection efficiency loophole. Here, we still fix the detectors at time $t_3$ and use two blockers (instead of one) at both times $t_{1}$ and $t_{2}$. In this case, we perform four sub-runs of the experiment, by changing the positions of the two blockers inside the setup. To give an example, in the first sub-run, we place one blocker in the $-1$ arm at $t_1$ and the other blocker in the $-1$ arm at $t_2$. So, the measured coincidence count between SPAD1 and SPAD2+ is denoted as $C_{t_{1},t_{2},t_{3}}(+,+,+)$, and between SPAD1 and SPAD2- is denoted as $C_{t_{1},t_{2},t_{3}}(+,+,-)$. In this way, we obtain eight coincidence values of the form $C_{t_{1},t_{2},t_{3}}(q_{t_1},q_{t_2},q_{t_3})$ from the third experimental run, and calculate all eight three-time joint probabilities, by using the following normalization
\begin{equation}
\label{eq:P12Apx}
P_{t_{1},t_{2},t_{3}}(q_{t_{1}},q_{t_{2}},q_{t_{3}})=\dfrac{C_{t_{1},t_{2},t_{3}}(q_{t_{1}},q_{t_{2}},q_{t_{3}})}{C^{T}_{t_{1},t_{2},t_{3}}} .
\end{equation}
All four joint probabilities of the form $P_{t_1,t_2}(q_{t_1},q_{t_2})$ are then calculated from these eight three-time joint probabilities by using the induction or arrow of time (AoT) expression, given below
\begin{equation}
\label{eq:AoTApx}
P_{t_{1},t_{2}}(q_{t_{1}},q_{t_{2}})= \sum_{q_{t_{3}}=\pm 1}P_{t_{1},t_{2},t_{3}}(q_{t_{1}},q_{t_{2}},q_{t_{3}}) .
\end{equation}
We also perform another experiment run (fourth run) where we do not place any blocker in any arm of the setup and measure coincidences. In this case, the coincidence count between SPAD1 and SPAD2+ is denoted as $C_{t_3}(+)$ and coincidence count between SPAD1 and SPAD2- is denoted as $C_{t_3}(-)$. So, the probabilities at time $t_3$ are measured as,
\begin{equation}
\label{probat3}
P_{t_3}(+)=\frac{C_{t_3}(+)}{C_{t_3}(+)+C_{t_3}(-)}, \quad  P_{t_3}(-)=\frac{C_{t_3}(-)}{C_{t_3}(+)+C_{t_3}(-)}  .
\end{equation}
These probabilities (\ref{probat3}) are used to calculate the two-time NSIT equations, $NSIT_{(t_1),t_3}$ and $NSIT_{(t_2),t_3}$ (\eqref{eq:nsit23} and \eqref{eq:nsit13}). Details of the experimental methods, as discussed in this section, have been summarized in Table \ref{expsum}.
\begin{table}[!ht]
\centering
\begin{tabular}{|c|c|c|c|c|c|c|}
\hline
\begin{tabular}[c]{@{}c@{}}Experi-\\mental\\ runs\end{tabular} & \begin{tabular}[c]{@{}c@{}}Experi-\\mental\\ sub-runs\end{tabular} & \begin{tabular}[c]{@{}c@{}}No. of\\  blockers\\ used\end{tabular} & \begin{tabular}[c]{@{}c@{}}Blocker\\ placement\end{tabular} & \begin{tabular}[c]{@{}c@{}}Measured\\ coincidences\end{tabular}               & \begin{tabular}[c]{@{}c@{}}Total\\ coincidence\end{tabular} & \begin{tabular}[c]{@{}c@{}}Measured\\ probabilities\end{tabular}                                                                          \\ \hline
\multirow{2}{*}{1} & 1.1 & 1 & at $t_2$ in the $-1$ path (arm 3) & \begin{tabular}[c]{@{}c@{}}$C_{t_2,t_3}(+,+)$\\ $C_{t_2,t_3}(+,-)$\end{tabular} & \multirow{2}{*}{$C^T_{t_2,t_3}$}            & \multirow{2}{*}{\begin{tabular}[c]{@{}c@{}}$P_{t_2,t_3}(+,+)$, $P_{t_2,t_3}(+,-)$,\\ $P_{t_2,t_3}(-,+)$, $P_{t_2,t_3}(-,-)$\end{tabular}} \\ \cline{2-5} & 1.2 & 1 & at $t_2$ in the $+1$ path (arm 4) & \begin{tabular}[c]{@{}c@{}}$C_{t_2,t_3}(-,+)$\\ $C_{t_2,t_3}(-,-)$\end{tabular} &  &  \\ \hline
\multirow{2}{*}{2} & 2.1 & 1 & at $t_1$ in the $-1$ path (arm 2) & \begin{tabular}[c]{@{}c@{}}$C_{t_1,t_3}(+,+)$\\ $C_{t_1,t_3}(+,-)$\end{tabular} & \multirow{2}{*}{$C^T_{t_1,t_3}$} & \multirow{2}{*}{\begin{tabular}[c]{@{}c@{}}$P_{t_1,t_3}(+,+)$, $P_{t_1,t_3}(+,-)$,\\ $P_{t_1,t_3}(-,+)$, $P_{t_1,t_3}(-,-)$\end{tabular}} \\ \cline{2-5} & 2.2 & 1 & at $t_1$ in the $+1$ path (arm 1) & \begin{tabular}[c]{@{}c@{}}$C_{t_1,t_3}(-,+)$\\ $C_{t_1,t_3}(-,-)$\end{tabular} & & \\ \hline
\multirow{4}{*}{3} & 3.1 & 2 & \begin{tabular}[c]{@{}c@{}}at $t_1$ in the $-1$ path (arm 2)\\ at $t_2$ in the -1 path (arm 3)\end{tabular} & \begin{tabular}[c]{@{}c@{}}$C_{t_1,t_2,t_3}(+,+,+)$\\ $C_{t_1,t_2,t_3}(+,+,-)$\end{tabular} & \multirow{4}{*}{$C^T_{t_1,t_2,t_3}$} & \multirow{4}{*}{\begin{tabular}[c]{@{}c@{}}$P_{t_1,t_2,t_3}(+,+,+)$, $P_{t_1,t_2,t_3}(+,+,-)$,\\ $P_{t_1,t_2,t_3}(+,-,+)$, $P_{t_1,t_2,t_3}(+,-,-)$,\\ $P_{t_1,t_2,t_3}(-,+,+)$, $P_{t_1,t_2,t_3}(-,+,-)$,\\ $P_{t_1,t_2,t_3}(-,-,+)$, $P_{t_1,t_2,t_3}(-,-,-)$,\end{tabular}} \\ \cline{2-5} & 3.2 & 2 & \begin{tabular}[c]{@{}c@{}}at $t_1$ in the $-1$ path (arm 2)\\ at $t_2$ in the $+1$ path (arm 4)\end{tabular} & \begin{tabular}[c]{@{}c@{}}$C_{t_1,t_2,t_3}(+,-,+)$\\ $C_{t_1,t_2,t_3}(+,-,-)$\end{tabular} & & \\ \cline{2-5}
 & 3.3 & 2 & \begin{tabular}[c]{@{}c@{}}at $t_1$ in the $+1$ path (arm 1)\\ at $t_2$ in the $-1$ path (arm 3)\end{tabular} & \begin{tabular}[c]{@{}c@{}}$C_{t_1,t_2,t_3}(-,+,+)$\\ $C_{t_1,t_2,t_3}(-,+,-)$\end{tabular} & & \\ \cline{2-5} & 3.4 & 2  & \begin{tabular}[c]{@{}c@{}}at $t_1$ in the +1 path (arm 1)\\ at $t_2$ in the $+1$ path (arm 4)\end{tabular}  & \begin{tabular}[c]{@{}c@{}}$C_{t_1,t_2,t_3}(-,-,+)$\\ $C_{t_1,t_2,t_3}(-,-,-)$\end{tabular} & & \\ \hline 4 & 4.1 & 0 & N/A & \begin{tabular}[c]{@{}c@{}}$C_{t_3}(+)$\\ $C_{t_3}(-)$\end{tabular} & $C^T_{t_3}$ & $P_{t_3}(+)$, $P_{t_3}(-)$ \\ \hline
\end{tabular}
\caption{Details on the various runs and sub-runs of the experiment performed to obtain all relevant probabilities.}
\label{expsum}
\end{table}

Now, in Table \ref{expresulttab} we show the measured coincidence values obtained from the four runs of the experiment, for a representative dataset. 
We also show calculations for getting all the relevant joint probabilities from these coincidence values, and how to calculate the LGI, WLGI and NSIT values.
\begin{table}[!ht]
\centering
\begin{tabular}{|c|c|c|c|}
\hline
\begin{tabular}[c]{@{}c@{}}Experi-\\ mental\\ run\end{tabular} & \begin{tabular}[c]{@{}c@{}}Measured\\ coincidences\end{tabular} & \begin{tabular}[c]{@{}c@{}}Total\\ coincidence\end{tabular} & \begin{tabular}[c]{@{}c@{}}Measured\\ probabilities\end{tabular} \\ \hline
1 & \begin{tabular}[c]{@{}c@{}}$C_{t_2,t_3}(+,+)$=41644.94\\ $C_{t_2,t_3}(+,-)$=10725.33\\ $C_{t_2,t_3}(-,+)$=12334.57\\ $C_{t_2,t_3}(-,-)$=35954.40\end{tabular} & $C^T_{t_2,t_3}$=100659.24 & \begin{tabular}[c]{@{}c@{}}$P_{t_2,t_3}(+,+)$=0.414, $P_{t_2,t_3}(+,-)$=0.107\\ $P_{t_2,t_3}(-,+)$=0.122, $P_{t_2,t_3}(-,-)$=0.357\end{tabular} \\ \hline
2 & \begin{tabular}[c]{@{}c@{}}$C_{t_1,t_3}(+,+)$=22977.20\\ $C_{t_1,t_3}(+,-)$=30456.67\\ $C_{t_1,t_3}(-,+)$=35541.98\\ $C_{t_1,t_3}(-,-)$=19814.86\end{tabular} & $C^T_{t_1,t_3}$=108790.71 & \begin{tabular}[c]{@{}c@{}}$P_{t_1,t_3}(+,+)$=0.211, $P_{t_1,t_3}(+,-)$=0.280\\ $P_{t_1,t_3}(-,+)$=0.327, $P_{t_1,t_3}(-,-)$=0.182\end{tabular} \\ \hline
3 & \begin{tabular}[c]{@{}c@{}}$C_{t_1,t_2,t_3}(+,+,+)$=34430.15\\ $C_{t_1,t_2,t_3}(+,+,-)$=8957.09\\ $C_{t_1,t_2,t_3}(+,-,+)$=2203.34\\ $C_{t_1,t_2,t_3}(+,-,-)$=9067.37\\ $C_{t_1,t_2,t_3}(-,+,+)$=10218.37\\ $C_{t_1,t_2,t_3}(-,+,-)$=1900.90\\ $C_{t_1,t_2,t_3}(-,-,+)$=10171.06\\ $C_{t_1,t_2,t_3}(-,-,-)$=30126.35\end{tabular} & $C^T_{t_1,t_2,t_3}$=107074.63 & \begin{tabular}[c]{@{}c@{}}$P_{t_1,t_2,t_3}(+,+,+)$=0.322, $P_{t_1,t_2,t_3}(+,+,-)$=0.084\\ $P_{t_1,t_2,t_3}(+,-,+)$=0.021, $P_{t_1,t_2,t_3}(+,-,-)$=0.085\\ $P_{t_1,t_2,t_3}(-,+,+)$=0.095, $P_{t_1,t_2,t_3}(-,+,-)$=0.018\\ $P_{t_1,t_2,t_3}(-,-,+)$=0.095, $P_{t_1,t_2,t_3}(-,-,-)$=0.281\\
$P_{t_1,t_2}(+,+)$=0.406, $P_{t_1,t_2}(+,-)$=0.106\\ $P_{t_1,t_2}(-,+)$=0.113, $P_{t_1,t_2}(-,-)$=0.376\end{tabular} \\ \hline 4 & \begin{tabular}[c]{@{}c@{}}$C_{t_3}(+)$=49888.96\\ $C_{t_3}(-)$=42526.46\end{tabular} & $C^T_{t_3}$=92415.42 & $P_{t_3}(+)$=0.540, $P_{t_3}(-)$=0.460\\ \hline
\end{tabular}
\caption{Experimentally measured coincidence values and calculated joint probabilities from a representative experimental datset.}
\label{expresulttab}
\end{table}
From the Table \ref{expresulttab} we calculate the three correlation values, $\langle Q_{t_{1}} Q_{t_{2}}\rangle=0.56$, $\langle Q_{t_{2}} Q_{t_{3}}\rangle=0.54$, $\langle Q_{t_{1}} Q_{t_{3}}\rangle=-0.22$, and get the LGI value of $1.32$ using \eqref{eq:lgiApx}. For the same dataset, WLGI value is obtained from \eqref{eq:wlgiApx} to be 0.09. NSIT values are the following: $NSIT_{(t_1)t_2}$ (\ref{eq:nsit12})=0.002, $NSIT_{(t_2)t_3}$ (\ref{eq:nsit23})=0.004, and $NSIT_{(t_1)t_3}$ (\ref{eq:nsit13})=0.002.

\subsection{Evaluating the standard deviations}
\label{appendix:c2}
We observe time dependent fluctuations in the measured coincidence counts, which can be primarily ascribed to three different reasons; the first being the fluctuations in the intensity of the pump beam itself, the second one being the random statistical fluctuations from the SPDC process, and the third one from the fluctuation in the interference pattern due to the slight unavoidable instability in the interferometer. The intensity of the pump beam coming from a diode laser can oscillate with time due to its sensitivity towards the minute temperature fluctuations inside the lab. We measure these fluctuations in the pump power by placing a power meter just before the BBO crystal and recording input power for around 2 hours. We observe a sinusoidal fluctuation in the power of the laser. The period of this fluctuation is less than 15 minutes. Mean pump power is 13.45 mW, with a standard deviation by mean (SD/M) value of 0.95$\%$.

The random fluctuation present in the coincidence count is due to the randomness of the single-photon pair generation process in spontaneous parametric down-conversion. To understand these fluctuations, we record the coincidence counts between any one of the detectors and the heralding detector (coincidence between SPAD1 and SPAD2+) for 10 seconds and 1000 iterations. 
We observe singles count of $4.87\times10^{5}$ cps (SPAD1), and $1.73\times10^{5}$ cps (SPAD2+), and the coincidence of the order of $10^{4}$ MHz for a time-window of 8192 ps. In order to understand the effect of the instability in the Sagnac interferometer on the temporal fluctuations in coincidence count, we record coincidences for two different scenarios. First, for the non-interference scenario, we placed two blockers, one in arm 1 and another in arm 3 (see Fig. \ref{fig:schematic2}) and measure coincidences between SPAD1 and SPAD2+. For the second case, which we call the interference case, we place one blocker only in arm 1 and measure coincidences between SPAD1 and SPAD2+. As we do not place any blocker at time $t_2$ inside the Sagnac interferometer, we allow interference to take place for the second case. For both cases, coincidence counts were recorded for 10 seconds at each iteration and repeated for 1000 iterations. From the statistical distribution of these 1000 iterations, we observe higher fluctuations in the coincidence counts from the interference condition than the non-interference condition. 

\begin{figure}[!ht]
\centering
\includegraphics[scale=0.5]{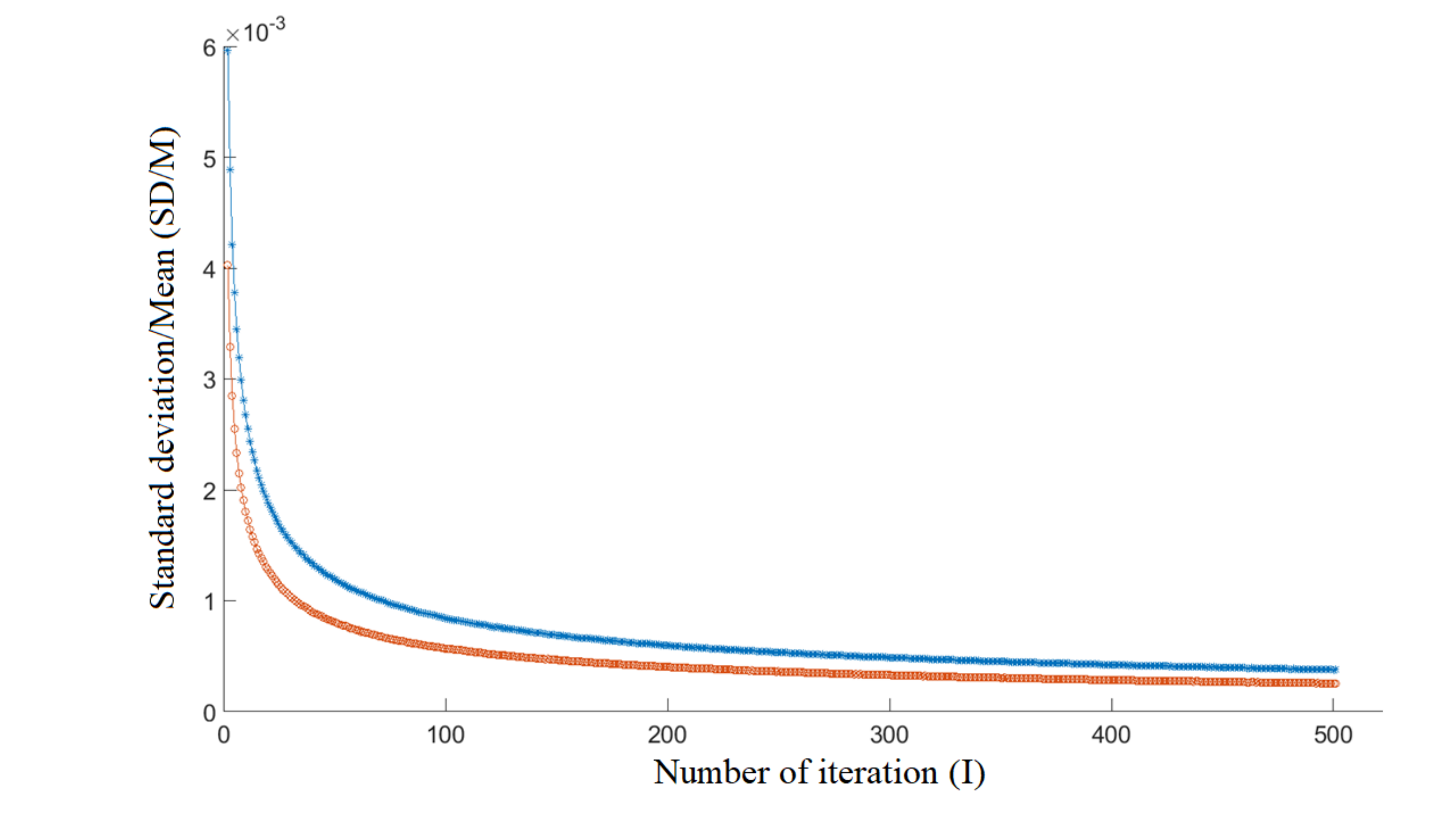}
\caption{Plot for SD/M vs. the number of iterations. The blue curve represents the interference case, whereas the red curve represents the non-interference case.}
\label{fig:bootstrap}
\end{figure}

To counter the random fluctuations described in the previous paragraph, we measure average coincidence values from a representative sample of datasets. The first step is to find the minimum number of datasets (or the number of iteration) necessary for the averaging process. For this purpose, we apply a bootstrapping strategy described as follows: we measure the number of coincidence detections in a time interval $T$, for a significantly large number of iteration ($I_{L}$), where we assume that the measured samples are a good representation of the actual sample space. We select any integer value $I$ where $I<I_L$, randomly select $I$ number of samples from $I_{L}$ number of total samples, and measure the average value, $\mu_{I}$. We repeat this process with re-sampling (with replacement) for $K$ number of times and get a list of $K$ number of $\mu_{I}$, e. g., $\mu^{1}_{I}$, $\mu^{2}_{I}$,.., $\mu^{K}_{I}$. We calculate the average of this list, i. e., $\overline{\mu}_{I}=\frac{1}{K}\sum_{k=1}^{K}\mu^{k}_{I}$, and the standard deviation, $\sigma^{\mu}_{I}=\sqrt{\frac{1}{K-1}\sum_{k=1}^{K}(\mu^{k}_{I}-\overline{\mu}_{I})^{2}}$. In case $I$ number of iterations is sufficient for the averaging, we would expect a negligibly small $\sigma^{\mu}_{I}$ as compared to $\overline{\mu}_{I}$ ($\sigma^{\mu}_{I}<<\overline{\mu}_{I}$). If we plot the ratio between the standard deviation and the mean (SD/M), that is, $\frac{\sigma^{\mu}_{I}}{\overline{\mu}_{I}}$ as a function of iteration ($I$), we will see an exponential decay as $I$ increases. So, we can select a value of $I$, for which the SD/M is negligibly small or below a reasonable preferred lower threshold value. We record coincidence value in a time interval $T=10$ seconds and for $I_L=1000$ iterations for both the interference and the non-interference conditions. We apply a bootstrapping algorithm with $K=10^5$; to find the standard deviation by mean (SD/M) value for the different number of iterations of the experiment. We pick a representative SD/M value of 0.05$\%$ obtained for $I=150$ iterations in the non-interference case, as we can observe from the plot (see figure \ref{fig:bootstrap}) that SD/M curve quickly flattens after 150 iterations. The same value of SD/M requires $I=300$ iterations in the interference case. This observation implies that we have to take at least 300 datasets to calculate the average coincidences, where the coincidences are dependant on interference in the Sagnac interferometer. Otherwise, we have to take at least 150 datasets for averaging. 

To quote the values for LGI, WLGI, and NSITs in the result section (tables in Fig. \ref{tab:resultTab1} and \ref{tab:resultTab2}), we calculate the average value and the standard deviation from all the datasets obtained from the experiment. 
We repeat all experimental runs for a statistically significant number of times, where we record coincidence counts for 10 seconds in different runs. We take 300 iterations while measuring the correlation between time $t_{1}, t_{3}$ (interference case) and 150 iterations for correlation between time $t_{1}, t_{2}$ and between $t_{2}, t_{3}$ (non-interference case). For LGI, we measure three average values ${\langle Q_{t_{1}}Q_{t_{2}}\rangle}_{\mu}, {\langle Q_{t_{2}}Q_{t_{3}}\rangle}_{\mu}, {\langle Q_{t_{1}}Q_{t_{3}}\rangle}_{\mu}$ along with their respective standard deviations denoted as $\sigma_{1,2}, \sigma_{2,3}, \sigma_{1,3}$. Here, the subscript $\mu$ is used to identify these correlation values as experimentally measured averages of large iterations. The measured LGI value which is $LGI={\langle Q_{t_{1}}Q_{t_{2}}\rangle}_{\mu}+{\langle Q_{t_{2}}Q_{t_{3}}\rangle}_{\mu}-{\langle Q_{t_{1}}Q_{t_{3}}\rangle}_{\mu}$ has a maximum error of $\pm\Delta$, where $\Delta=\sigma_{1,2}+\sigma_{2,3}+\sigma_{1,3}$. The reason for defining the maximum error as the summation of three standard deviations is that we measured all the three average values from three separate experimental runs performed one after another and combined together. So, 
we consider the worst-case scenario where we assume that all the three experimental runs are independent of each other, and hence the error from each run adds up to form the maximum error. Any violation with a magnitude much higher than the maximum error obtained for the worst-case scenario can then be considered as a significant violation. Each of the three sets of the experiment also has two or more subsets of experimental runs since we are using negative result measurements. For example, while measuring ${\langle Q_{t_{1}}Q_{t_{3}}\rangle}_{\mu}$ we perform two sub-experiments, where in the first one we block $'+1'$ arm at $t_{1}$, and in the second run, we block $'-1'$ arm at $t_{1}$. Although we take data for the same amount of time duration ($T=10$ seconds) for both cases, due to statistical fluctuations, the input number of photons to the setup itself can be different for the two cases. Then the four joint probabilities measured from the two different experiments may be self-inconsistent and introduce error in the average value ${\langle Q_{t_{1}}Q_{t_{3}}\rangle}_{\mu}$. For this, we apply a strategy where we compare all 300 iterations of the first experiment with all 300 iterations of the second, generating 300$\times$300 combinations of $\langle Q_{t_{1}}Q_{t_{3}}\rangle$ values. We measure average ${\langle Q_{t_{1}}Q_{t_{3}}\rangle}_{\mu}$ and standard deviation $\sigma_{1,3}$ of this distribution. Similarly, we measure average value ${\langle Q_{t_{2}}Q_{t_{3}}\rangle}_{\mu}$ and standard deviation $\sigma_{2,3}$ from a combination of 150$\times$150 possibilities. For ${\langle Q_{t_{1}}Q_{t_{2}}\rangle}_{\mu}$, however we measure all tree-times joint probabilities rather than two-times joint probabilities, which requires four sub-experiment instead of two, leading to a distribution of $150^{4}$ possible values of $\langle Q_{t_{1}}Q_{t_{2}}\rangle$, from which we measure average and standard deviation.

For WLGI, we measure average violation, that is, ${\{P_{t_{1},t_{3}}(-,+)\}}_{\mu}-{\{P_{t_{1},t_{2}}(-,+)\}}_{\mu}-{\{P_{t_{2},t_{3}}(-,+)\}}_{\mu}$ with maximum error of $\pm\Delta$, where $\Delta=\sigma_{1,2}+\sigma_{2,3}+\sigma_{1,3}$, and $\sigma_{i,j}$ represents the standard deviation of the distribution $P_{t_{i},t_{j}}(q_{t_{i}},q_{t_{j}})$. We measure the three averages and standard deviations in the same procedure as implemented for LGI. For NSITs, we measure three expressions, $NSIT_{(t_{1})t_{2}}=|\{P_{t_{2}}(+)\}_{\mu}-\{P_{t_{1},t_{2}}(+,+)+P_{t_{1},t_{2}}(-,+)\}_{\mu}|$, $NSIT_{(t_{2})t_{3}}=|\{P_{t_{3}}(+)\}_{\mu}-\{P_{t_{2},t_{3}}(+,+)+P_{t_{2},t_{3}}(-,+)\}_{\mu}|$, and $NSIT_{(t_{1})t_{3}}=|\{P_{t_{3}}(+)\}_{\mu}-\{P_{t_{1},t_{3}}(+,+)-P_{t_{1},t_{3}}(-,+)\}_{\mu}|$. Each of the NSIT equations requires only two separate experimental runs rather than three used in LGI and WLGI. 
\subsection{Estimating the quantum mechanical predicted values of LGI, WLGI and NSIT by incorporating experimental non-idealities}
\label{appendix:c3}
We write the expression for LGI and WLGI, using various experimental parameters, as the following:
\begin{equation}
\label{eq:lgi2Apx}
LGI:~\langle Q_{t_{1}} Q_{t_{2}}\rangle + \langle Q_{t_{2}} Q_{t_{3}}\rangle - \langle Q_{t_{1}} Q_{t_{3}}\rangle = 1-4R^{2}+4TR~cos(\theta_2)
\end{equation}
\begin{equation}
\label{eq:wlgi2Apx}
WLGI:~P_{t_{1},t_{3}}(-,+)-P_{t_{1},t_{2}}(-,+)-P_{t_{2},t_{3}}(-,+) = 2{|\beta|}^{2}TR~cos(\theta_2) - R^{2}
\end{equation}
where $T:R$ is the splitting ratio of the NPBS (see Fig. \ref{fig:schematic2}) and ${|\alpha|}^{2}:{|\beta|}^{2}$ is the splitting ratio due to the HWP2 and PBS2 combination. 
$\theta_2$ represents the overall phase difference between the two arms of the Sagnac interferometer. Hence, the $cos(\theta_2)$ term can be considered as a quantifier of the interferometric visibility obtained in the setup
.

However, while deriving \eqref{eq:lgi2Apx} and \eqref{eq:wlgi2Apx}, we assume an ideal case where the beamsplitting ratio  $T:R$ for the NPBS has a fixed value. But, as we observe experimentally, the splitting ratio varies based on the input polarization and the input port of the NPBS. Even if we consider that all photons impinging on the NPBS are vertically polarized, they experience different $T:R$ values based on which of the four input ports of the NPBS they are impinging on. We follow the same notation for the splitting ratios $T_{i}:R_{i},~\forall i=\{1,2,3,4\}$ as mentioned in Figure \ref{fig:bso}. With this modification, LGI and WLGI expressions become,
\begin{equation}
\label{eq:lgimod}
\begin{split}
LGI:~~&\langle Q_{t_{1}} Q_{t_{2}}\rangle + \langle Q_{t_{2}} Q_{t_{3}}\rangle - \langle Q_{t_{1}} Q_{t_{3}}\rangle \\& = {|\alpha|}^{2}\{R_1(T_3-3R_3)+T_1+2\sqrt{T_1T_2R_1R_3}~cos(\theta_2)+2\sqrt{T_1T_3R_1R_2}~cos(\theta_2)\}\\
&+{|\beta|}^{2}\{R_4(T_2-3R_2)+T_4+2\sqrt{T_2T_4R_3R_4}~cos(\theta_2)+2\sqrt{T_3T_4R_2R_4}~cos(\theta_2)\} \\
\end{split}
\end{equation}
\begin{equation}
\label{eq:wlgimod}
WLGI: ~~P_{t_{1},t_{3}}(-,+)-P_{t_{1},t_{2}}(-,+)-P_{t_{2},t_{3}}(-,+) = 2{|\beta|}^{2}\sqrt{T_2T_4R_3R_4}~cos(\theta_2)-{|\alpha|}^{2}R_1R_3- {|\beta|}^{2}R_2R_4\\
\end{equation}
$NSIT_{(t_1)t_2}=|P_{t_2}(+)-P_{t_1,t_2}(+,+)-P_{t_1,t_2}(-,+)|=0$ and $NSIT_{(t_1)t_3}=|P_{t_3}(+)-P_{t_1,t_3}(+,+)-P_{t_1,t_3}(-,+)|=0$; independent of the values of the splitting ratios ($T_i:R_i$). However, $NSIT_{(t_2)t_3}$ is dependant on the splitting ratio of the beamsplitter.
\begin{equation}
\label{eq:nsitmod}
\begin{split}
NSIT_{(t_2)t_3}:~~&|P_{t_3}(+)-P_{t_2,t_3}(+,+)-P_{t_2,t_3}(-,+)| \\& = \left|2{|\alpha|}^{2}\sqrt{T_1T_2R_1R_3}~cos(\theta_2)-2{|\beta|}^{2}\sqrt{T_2T_4R_3R_4}~cos(\theta_2)\right| \\
\end{split}
\end{equation}
Using the ideal values of ${|\alpha|}^{2}={|\beta|}^{2}=0.5$, $cos(\theta_2)=1$, and experimentally characterized values, $T_{1}=0.80$, $T_{2}=0.79$, $T_{3}=0.82$, $T_{4}=0.82$, in \eqref{eq:lgimod}, \eqref{eq:wlgimod} and \eqref{eq:nsitmod}, the quantum mechanical estimated values of LGI, WLGI and $NSIT_{(t_2)t_3}$ are obtained as $1.47$, $0.11$ and $0.006$, respectively. In the experimental setup, these values of the input parameters may change slightly, depending on the precision of various optical components. So, instead of the quantum mechanical estimated fixed values for LGI and WLGI expressions, we obtain a range of the quantum mechanical predicted values, considering most of the experimental imperfections along with their respective least count errors. The value ${|\alpha|}^{2}$ depends on the rotation angle of the HWP2, where we consider a typical least count error of $\pm 1^{\circ}$ for the rotation angle. Also, for all the $T_{i}$ values, we consider a least count of $\pm 2\%$. Considering these experimental error ranges, the range of the quantum mechanical predicted LGI value is calculated to be from $1.45$ to $1.49$, while the range of the WLGI value is calculated to be from $0.09$ to $0.13$. For $NSIT_{(t_2)t_3}$, this range is from $0$ to $0.03$. 

A point to be noted is that so far we have considered a simplified model for these calculations, 
where we assume that the interference visibility of the displaced Sagnac interferometer (dSI) reaches the optimal value in the experiment, or $cos(\theta_2)=1$. However, we observe from the experimental setup that the interferometric visibility value can only be reached up to 85\%, with our best effort in the alignment process, and also fluctuates between 70\% to 85\% along with time. The reasons for this phenomenon can be attributed to various observations. First of all, we simultaneously maintain optimal interference for both the two input arms of the beamsplitter (NPBS in Fig. \ref{fig:schematic2} in the main text). For achieving this condition, both the input beams should be collinear (on top of each other) throughout the whole interference process. This is not easy to maintain, given that the reflection angle from the NPBS is sensitive to the polarization of the input beam. Although we maintain both the input beams to have the same polarization (vertical polarization) with appropriate polarization control, a slight difference can lead to an observable change in the relative angle between the two beams as they pass through many reflecting mirrors inside the interferometer. Another issue arises due to the geometry of the dSI, where the two interfering beams hit the mirrors placed inside the interferometer at two different spots. So, any surface imperfections in the mirrors lead to a path difference (of the order of micrometers) between the two beams, causing a reduction of the visibility. We also observe that the visibility value slowly drifts between 85\% and 70\% due to the minute changes inside the lab environment, such as temperature, mechanical vibrations, etc.

From the above discussion, it is evident that the ideal assumption of $cos(\theta_2)=1$ is not suitable for the experimental setup that we used. So, we calculate a modified range of quantum mechanical prediction using Eqs. (\ref{eq:lgimod}-\ref{eq:nsitmod}), where we vary the value of $cos(\theta_2)$ from 0.7 to 0.85, along with the imperfections stemming from other experimental parameters and their respective least count errors, as mentioned earlier. We obtain the modified quantum mechanical predicted range for LGI to be from 1.28 to 1.40, for WLGI, it is from 0.05 to 0.11, and for $NSIT_{(t_2)t_3}$, from 0 to 0.026.

\section{Closing the coincidence loophole and the preparation state loophole by adjusting the coincidence time window for different measurement settings}
\label{appendix:d}
In 
experiments performing violation of Bell inequalities
, the coincidence loophole arises from the fact that different measurement settings may introduce various time delays between the two time-correlated entangled photons, which results in an asymmetric detection of correlated photon pairs within a predefined coincidence time window. Here, one can show that by adjusting the delay between the two photons of the same entangled pair for different measurement settings, a local hidden variable theory can produce a fake violation of the Bell inequalities. For the experiments testing macrorealism, it is necessary to use single photons. So, the coincidence loophole does not directly apply to this scenario. However, one can argue that a similar type of loophole may arise when different measurement settings introduce time variation in the arrival time of the individual photons. In this case, one can control which photon will be detected in which detector; by introducing different time delays to the single photons based on the choice of the measurement settings. Our use of heralded photon or time-correlated photon has been discussed later in this section but, before understanding the same, it is important to point out why non-heralded photons are amenable to this loophole. 

In the experimental setup, all single photon detectors (SPAD2+, SPAD2-) are fixed at time $t_3$, and different measurement settings 
are 
realized by changing the position of the blockers only. Now, consider a general scenario where these different measurement settings introduce different path lengths for the photons to traverse inside the setup before getting detected in SPAD2+ or SPAD2-. For example, let us consider two measurement settings where the blocker is placed at time $t_1$ in the +1 arm for the first setting and in the -1 arm for the second setting. If there is a difference in path length between the +1 and -1 arm, then a photon will take different time (say $T$ for +1 arm, $T+\tau$ for -1 arm) to reach the detectors. 

Suppose one uses non-heralded single photons for the experiment. In that case, a detection time window needs to be preset in order to reduce the background noise, as non-heralded single photons do not have any timing references. In the case of different path lengths introduced by different measurement settings, a preset detection time window is disadvantageous. For example, if one fixes the time window around $T$, then photons traversing +1 arm have higher chances of getting detected than photons traveling -1 arm. This may introduce inaccurate joint probability distribution such that even a macrorealist theory may produce a false violation of LGI/WLGI.

In light of the above discussion, the usage of heralded single photon for the experimental test of macrorealism has a certain advantage, as will be discussed 
next
. Single photons generated from an SPDC source always come as a pair, or in other words, two photons from each pair are time-correlated. We use the arrival time of the heralding photon (in SPAD1) as the timing reference for the heralded single photon that goes to the experimental setup. So, instead of any preset detection time window, we measure the coincidence between the heralding and the heralded photon (or between SPAD1 and SPAD2+/SPAD2-) in the post-processing stage. Now, going back to the two different measurement settings mentioned earlier. In the case of all the photons that go through the +1 arm, they form a coincidence peak around time $T$ 
with the heralding photon
. So, we place a coincidence window at full-width-half-maxima (FWHM) around the peak position. On the other hand, the photons passing through the -1 arm form the coincidence peak 
with the heralding photon
, at a shifted position $T+\tau$. We place the FWHM coincidence window around $T+\tau$ for the second case. So, by adjusting the coincidence window for different measurement settings, we have nullified any error introduced by the photon-time shift occurring due to different settings.

Another important 
consideration 
is the way we select the coincidence window for individual data sets such that the contribution of background noise is minimized. This step is essential to close the preparation state loophole as it ensures that the post-selected sample of single photons is generated from the SPDC process only; hence we can safely assume that all the selected photons have the same preparation state, 
within permissible errors
. Once we select the coincidence time window, only the coincidence events occurring within the selected time window are considered as valid coincidence counts. In most cases, the coincidence window is chosen 
as 
a time interval slightly larger than the temporal 
width 
of the coincidence peak, such that 
the area under the peak 
is 
properly 
included. This strategy is appropriate if the background noise is significantly lower than the signal itself. However, in the presence of significant noise, a large coincidence window also captures background noise, which may increase the errors in the experimental results. For example, if we measure the ratio between two coincidence counts $C_{1}$ and $C_{2}$, and if the large window size also captures additional noise counts $N_{1}$ and $N_{2}$ as well, then the ratio becomes $\frac{C_{1}+N_{1}}{C_{1}+C_{2}+N_{1}+N_{2}}$, which is significantly different than $\frac{C_{1}}{C_{1}+C_{2}}$ if $\frac{C_{1}}{N_{1}}\neq\frac{C_{2}}{N_{2}}$.

We follow the strategy to set a coincidence window that is smaller than the temporal width of the coincidence peak 
such that the noise is minimized. We select two positions in the x-axis of the coincidence curve around the peak position and adjust them such that the signal-to-noise ratio is optimal. We can also adjust the window such that it covers the full-width at half-maxima (FWHM) of the coincidence peak, or we may also apply the standard $\frac{1}{e^{2}}$ width value. In our case, we select the FWHM size as our coincidence window due to the observation that it provides an optimal signal-to-noise ratio as well as a higher data collection rate for our datasets. In this way, we reduce the contribution of unnecessary background noise in the experiment. Also, there is a flatline on both sides of the coincidence peak in the coincidence plot, which is due to the accidental coincidences coming from background noise. So, to correct for these accidental coincidences, we first calculate the total background coincidences within the time window as, $\text{background}=\text{average coincidence value in the flatline} \times \text{size of the coincidence window}$. We then subtract this background value from the coincidence value to get the background-corrected coincidences. Interested readers may look at Ref. \cite{PhysRevApplied.14.024036} 
in the main text, to find more detailed analysis regarding the selection and adjustment of the coincidence time window based on various experimental requirements in the context of the quantum key distribution experiment. 

\end{document}